\documentclass{aa}
\usepackage[utf8]{inputenc}

\usepackage{graphicx}
\usepackage[breaklinks, colorlinks, citecolor=blue, linkcolor=blue]{hyperref}
\usepackage{threeparttable} 
\usepackage{tabularx}
\usepackage{siunitx}
\usepackage{natbib}
\usepackage{longtable}
\usepackage{lscape} 
\usepackage{xcolor}

\makeatletter
\renewcommand*\aa@pageof{, page \thepage{} of \pageref*{LastPage}}
\makeatother

\let\orgautoref\autoref
\renewcommand{\autoref}
        {\def\equationautorefname{Eq.}%
         \def\figureautorefname{Fig.}%
         \def\sectionautorefname{Sect.}%
         \def\subsectionautorefname{Sect.}%
         \def\subsubsectionautorefname{Sect.}%
         \orgautoref}

\newcommand*\samethanks[1][\value{footnote}]{\footnotemark[#1]}

\makeatletter

\def\instrefs#1{{\def\scsep{\def\scsep{,}}\@for\w:=#1\do{\scsep\ref{inst:\w}}}}
\renewcommand{\inst}[1]{\unskip$^{\instrefs{#1}}$}

\begin{document}

\title{Gaussian processes for radial velocity modeling}

\titlerunning{Gaussian processes for RV modeling}

\subtitle{{Better rotation periods and planetary parameters with the quasi-periodic kernel and constrained priors}}

\author{Stephan~Stock\inst{lsw}\thanks{{Both authors contributed equally to the manuscript.}}
    \and {Jonas~Kemmer\inst{lsw}}\samethanks{}
    \and {Diana~Kossakowski\inst{mpia}}
    \and {Silvia~Sabotta\inst{lsw}}
    \and {Sabine~Reffert\inst{lsw}}
    \and {Andreas~Quirrenbach\inst{lsw}}
}

\institute{
\label{inst:lsw}Landessternwarte, Zentrum für Astronomie der Universität Heidelberg, Königstuhl 12, 69117 Heidelberg, Germany\\ \email{sstock@lsw.uni-heidelberg.de or sreffert@lsw.uni-heidelberg.de}
\and \label{inst:mpia}Max-Planck-Institut f\"{u}r Astronomie, K\"{o}nigstuhl  17, 69117 Heidelberg, Germany
}

\authorrunning{S. Stock et~al.}

\date{Received <day month year> / Accepted <day month year>}

\abstract
{Instrumental radial velocity (RV) precision has {reached} a level where the detection of planetary signals is limited by the ability to understand and simultaneously model stellar astrophysical ``noise.'' {A common method for mitigating the effects of stellar activity is Gaussian process (GP) regression.}}
{In this study we present an analysis of the performance and properties of the quasi-periodic (QP) GP kernel, which is the multiplication of the squared-exponential kernel by the exponential-sine-squared kernel, based on an extensive set of synthetic RVs, into which the signature of activity was {injected}.}{The stellar activity within our synthetic data sets was simulated using astrophysically motivated models with different spot distributions and spot lifetimes rotating on the surface of a modeled late-type star. We used dynamic nested sampling to fit different model sets, including QP-GPs, Keplerian models, white noise models, and combinations of these, to synthetic RV time series data that in some cases included additional injected planetary signals.}{We find that while the QP-GP rotation parameter matches the simulated rotation period of the star, the length scale cannot be directly connected to the spot lifetimes on the stellar surface. Regarding the {setup of the priors for} the QP-GP, we find that it can be advantageous to constrain the QP-GP hyperparameters in different ways depending on the application and the goal of the analysis. We find that a constraint on the length scale of the QP-GP can lead to a significant improvement in identifying the correct rotation period of the star, while a constraint on the rotation hyperparameter tends to lead to improved planet detection efficiency and more accurately derived planet parameters. Even though for most of the simulations the Bayesian evidence performed as expected, we identified not far-fetched cases where a blind adoption of this metric would lead to wrong conclusions.}{We conclude that modeling stellar astrophysical noise by using a QP-GP considerably improves detection efficiencies and leads to precise planet parameters. Nevertheless, there are also cases in which the QP-GP does not perform optimally, for example {RV variations dynamically evolving on short timescales} or a mixture of a very stable activity component and random variations. Knowledge of these limitations is essential for {drawing} correct conclusions from observational data.}

\maketitle

\section{Introduction}
The search for and characterization of low-mass exoplanets, in particular in the habitable zone of their host stars, is in full swing. Indispensable to the mass determination of these exoplanets is the radial velocity (RV) method.
Current and next-generation high-resolution spectrographs searching for exoplanet signals in the spectra of stars via the RV method {are reaching precisions of around or below} 1\,$\mathrm{m\,s^{-1}}$, {such as} HARPS \citep{Mayor2003}, CARMENES \citep{Quirrenbach2014}, ESPRESSO \citep{Pepe2014}, MAROON-X \citep{Seifahrt2018}, {NEID \citep{Gupta2021}, EXPRES \citep{Petersburg2020, Zhao2022}, and KPF \citep{Gibson2020}.}
However, {the stellar activity induced by host stars} can be a significant {obstacle} to the detection and precise characterization {of these low-mass, low-amplitude} ($\sim1\,\mathrm{m\,s^{-1}}$) planetary signals, as even the quietest stars show intrinsic RV variations on this order of magnitude \citep{Robertson2014, Anglada2015}.
For the detection and characterization of planets with the RV method, it is therefore crucial to understand the consequences of stellar activity in order to find optimal ways to mitigate its impact.

The stellar ``noise'' affecting RV measurements of stars is mainly dominated by four different effects that occur on different timescales: oscillations, granulation, active regions, and magnetic cycles \citep{Dumusque2016}.
The most problematic effects, especially for later-type stars, are caused by active regions and magnetic cycles, since the timescales and RV amplitudes can be similar to those of potential planetary signals.
Active regions on the stellar surface have limited lifetimes and are constantly evolving dynamically as they move across the observed stellar hemisphere due to stellar rotation. This leads to correlated noise that can manifest itself as quasi-periodic (QP) variations in the measured RVs. This noise can mask, mimic, or influence planetary signals in the RV data \citep[e.g.,][]{Saar1998, Queloz2001, Boisse2011, Baluev2013, Haywood2014, Hatzes2016, Stock2020b}. In this work we focus on stellar spots and neglect faculae. In particular, for the astrophysical modeling of the correlated noise, we use an M dwarf of spectral class M2.0V. M dwarfs can show a high degree of magnetic activity and are in general spot dominated \citep{Jeffers2022}. Furthermore, no stellar oscillations, for example p-modes, have been observed for M dwarfs thus far, and the magnitude of any stellar oscillations is expected to be in the $\mathrm{cm\,s^{-1}}$ range \citep{Rodriguez2019} and can therefore be neglected in our simulated RV data set. Although we use a specific spectral type for our simulations, the general findings of this work can also be {quite beneficial} for modeling the RVs of other spectral types.

One way to mitigate correlated time variations is to apply noise models in the framework of Gaussian process (GP) regression models, and one particular kernel that is often used in such cases is the QP kernel \citep[e.g.,][]{Haywood2014, Rajpaul2015, Ribas2018, Stock2020c}. These stochastic models fit the stellar and instrumental noise contribution in RVs or light curves. However, the use of these models has raised some questions and concerns regarding the flexibility of these GPs (e.g., whether GPs are prone to overfitting or to absorbing planetary signals into the noise model), their application (e.g., selecting the best priors), and the correct interpretation of the results (e.g., the interpretation of hyperparameters and the implications for derived planetary parameters).

The main objective of this work is to perform a systematic analysis based on numerical simulations with known conditions in order to investigate these important questions. We created synthetic RV time series data that include the stellar astrophysical noise due to active regions. For this purpose, we used \texttt{StarSim} \citep{Herrero2016} to create RV data sets based on different spot patterns and distributions. We then investigated the properties of the QP-GP applied to these data sets in order to derive possible ways of optimally constraining the QP-GP priors. In addition, we injected Keplerian signals into the synthetic RV time series data to investigate the effect of the stellar activity and the GP modeling on the derived parameters of the planets. Investigations based on numerical simulations, such as those presented in this work, are critical for understanding how boundary conditions, properties of the data, or properties of the model may affect the results of an analysis involving a GP fit.

In Sect.~\ref{Sect: Methods} we describe the methods as well as the setup of the simulations and models that are used in this work. Section~\ref{Sect: Results1} presents and visualizes our results for GP fits to simulated activity data without additional Keplerian components, while Sect.~\ref{Sect: Results2} includes simulations with injected Keplerian signals. In Sect.~\ref{Sect: Discussion} we provide a discussion of our results as well as useful guidelines for using GP models for photometric and spectroscopic time series data. Finally, we give our conclusions and a {comprehensive} summary of this work in Sect.~\ref{Sect: Conclusion}.

\section{Setup and methods}
\label{Sect: Methods}

\begin{table*}
    \centering
    \begin{threeparttable}
        \caption{Relevant fixed parameters used to simulate the stellar activity within \texttt{StarSim}. }
        \label{Tab: Priors_starsim}
        \begin{tabular}{lccc}
            \hline
            \hline
            \noalign{\smallskip}
            Parameter                        & Value & Units     & Description                                       \\
            \noalign{\smallskip}
            \hline
            \noalign{\smallskip}
            \multicolumn{4}{c}{ \em general}                                                                         \\
            \noalign{\smallskip}
            ~~~$\lambda_{\mathrm{min}}$      & $380$ & nm        & Lower boundary of spectral range of RV instrument \\
            ~~~$\lambda_{\mathrm{max}}$      & $690$ & nm        & Upper boundary of spectral range of RV instrument \\
            ~~~$t_{\mathrm{Obs}}$            & $220$ & d         & Time span of simulations                          \\
            ~~$\Delta t$                     & $60$  & min       & Time cadence                                      \\
            \noalign{\smallskip}
            \multicolumn{4}{c}{\em radial velocities}                                                                \\
            \noalign{\smallskip}
            ~~~$\Delta v_{\textnormal{CCF}}$ & 15    & km/s      & CCF velocity range                                \\
            ~~~Mask                          & M2    & \ldots    & Spectral mask (G2 or K5 or M2)                    \\
            \noalign{\smallskip}
            \multicolumn{4}{c}{ \em star}                                                                            \\
            \noalign{\smallskip}
            ~~~$T_{\mathrm{eff}}$            & 3500  & K         & Stellar effective temperature                     \\
            ~~~$\Delta T_{\mathrm{Spot}}$    & 400   & K         & Spot temperature contrast$^{1}$                   \\
            ~~~$P_{\mathrm{rot}}$            & 22    & d         & Stellar rotation period                           \\
            ~~~$R$                           & 0.36  & $R_\odot$ & Stellar radius                                    \\
            ~~~$\log(g)$                     & 5     & dex       & Stellar surface gravity                           \\
            ~~~[Fe/H]                        & 0     & dex       & Stellar metallicity                               \\
            ~~~$[\alpha/\mathrm{Fe}]$        & 0     & dex       & Stellar alpha element abundance                   \\
            ~~~$i$                           & 90    & deg       & Axis inclination                                  \\
            ~~~$k_{\mathrm{rot}}$            & 0     & deg/d     & Stellar differential rotation                     \\
            \noalign{\smallskip}
            \multicolumn{4}{c}{ \em spots}                                                                           \\
            \noalign{\smallskip}
            ~~~$\sigma_{t_{\mathrm{spot}}}$  & 0     & d         & Spot lifetime standard deviation                  \\
            ~~~$\bar{N}_{\mathrm{Spot}}$     & 9     & \ldots    & Average number of spots on surface at any time    \\

            \noalign{\smallskip}

            \hline
        \end{tabular}
        \begin{tablenotes}              \item $^1$ The spot temperature contrast for a dwarf star with $T_{\mathrm{eff}}=3500$\,K {is based on the results of \cite{Anderson2015} provided by their Fig.~2}.            \item
        \end{tablenotes}
    \end{threeparttable}
\end{table*}

\subsection{Simulation of activity-induced radial velocity signals}

In order to produce synthetic activity-contaminated RV data, we made use of the tool \texttt{StarSim}\footnote{ \url{https://github.com/rosich/starsim-2}} \citep{Herrero2016}. \texttt{StarSim} models RVs by dividing the stellar surface into a grid of different effective temperatures depending on whether there is an active surface feature {(spot) at the grid element position or not (quiet photosphere).} A synthetic Phoenix spectrum \citep{Husser2013} based on the effective temperature, {the surface gravity} and other relevant stellar parameters, is assigned to each of the two different surface regions. In addition, each grid element is Doppler-shifted according to its position on the rotating surface. \texttt{StarSim} then calculates the cross-correlation function (CCF) for each of the two different spectra, and integrates the entire grid of surface elements in terms of the CCFs instead of spectra to speed up the calculations. The effects of limb darkening, convection and limb brightening are included within \texttt{StarSim}. The RVs are computed from the CCFs for a user-defined grid of epochs.

\begin{table*}
    \centering
    \begin{threeparttable}
        \caption{Grid of investigated stellar activity configurations. }
        \label{Tab: Priors_starsim_I}
        \begin{tabular}{lccc}
            \hline
            \hline
            \noalign{\smallskip}
            spot lifetime [d]         & spot {radius} [deg] & spot size evolution rate [deg/d] & $\overline{\sigma}_{\text{model}}$ [m/s] \\
            \noalign{\smallskip}
            \hline
            \noalign{\smallskip}
            \multicolumn{4}{c}{ \em random spot distribution}                                                                             \\
            \noalign{\smallskip}
            11  ($0.5P_{\text{rot}}$) & 6                   & 1.2                              & 1.9                                      \\
            22  ($1P_{\text{rot}}$)   & 6                   & 0.6                              & 1.6                                      \\
            44  ($2P_{\text{rot}}$)   & 6                   & 0.3                              & 1.4                                      \\
            110 ($5P_{\text{rot}}$)   & 6                   & 0.12                             & 1.5                                      \\
            220 ($10P_{\text{rot}}$)  & 6                   & 0.06                             & 1.1                                      \\
            \noalign{\smallskip}
            \multicolumn{4}{c}{ \em two active longitudes}                                                                                \\
            \noalign{\smallskip}
            11  ($0.5P_{\text{rot}}$) & 4.6                 & 0.92                             & 2.2                                      \\
            22  ($1P_{\text{rot}}$)   & 4.6                 & 0.46                             & 2.0                                      \\
            44  ($2P_{\text{rot}}$)   & 4.6                 & 0.23                             & 1.6                                      \\
            110 ($5P_{\text{rot}}$)   & 4.6                 & 0.092                            & 1.6                                      \\
            220 ($10P_{\text{rot}}$)  & 4.6                 & 0.046                            & 1.0                                      \\
            \noalign{\smallskip}

            \hline
        \end{tabular}

    \end{threeparttable}

\end{table*}

\subsection{Setup of the investigated stellar activity configurations}

Within our investigations, we focus on the effect of different spot sizes, spot distribution, spot lifetimes, spot number, and the stellar rotation period on the stellar astrophysical noise, and hence on the derived properties of the QP-GP.
We neglect instrumental jitter for most of our simulations. {Nevertheless, we allow for a small instrumental uncertainty
of 30\,$\mathrm{cm\,s^{-1}}$, which corresponds to the performance of current ultra-stable spectrographs (e.g., shown for ESPRESSO by \citealt{}{Suarez2020}).}  To introduce the RV uncertainty into the modeled stellar activity signal, we perturbed the simulated \texttt{StarSim} RVs with a Gaussian distribution with mean zero and standard deviation of 30\,$\mathrm{cm\,s^{-1}}$, if not stated otherwise.

The fixed parameters of the simulated star by \texttt{StarSim} to model the RV variations due to stellar activity are given in Table \ref{Tab: Priors_starsim}.
The stellar parameters, for example the stellar mass, radius, surface gravity, and the effective temperature, are based on our M2.0V example star. {The spectral range of the simulated RV instrument is 380\,nm--690\,nm, typical for modern optical spectrographs}. We assumed a temperature contrast of 400\,K for an effective temperature of 3500\,K based on the investigations of \cite{Anderson2015}. This temperature contrast reflects roughly the result of both the linear and polynomial fit in Fig.\,2 provided in \cite{Anderson2015}. Simulations with different temperature contrasts have been carried out and showed that the influence of this parameter, as expected, mainly affects the RV amplitude \citep{Reiners2013, Bauer2018}.

We fixed the rotation period of our example star to $P_{\text{rot}}=22$\,d, which is within the range of common values found for M2 stars \citep{Newton2018, Jeffers2018, Popinchalk2021}. To keep the investigations computationally feasible, the total time baseline of the RV observations, if not stated otherwise, is set to $10\,P_{\text{rot}}=220$\,d, in which 11 RV measurements per rotation period were assumed (one measurement every two days). This results in 111 RV measurements over 220\,d, close to a best-case scenario for the RV observations and our GP analysis.

\begin{figure*}
    \centering
    \includegraphics[width=9cm]{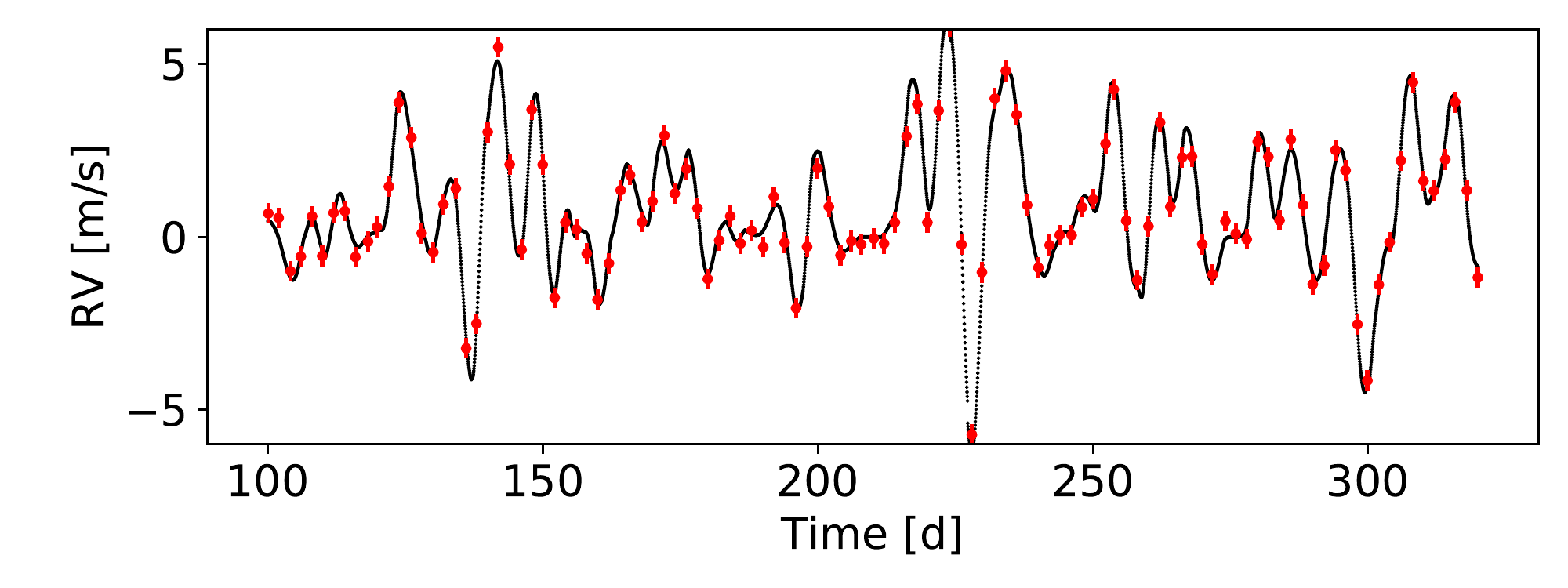}
    \includegraphics[width=9cm]{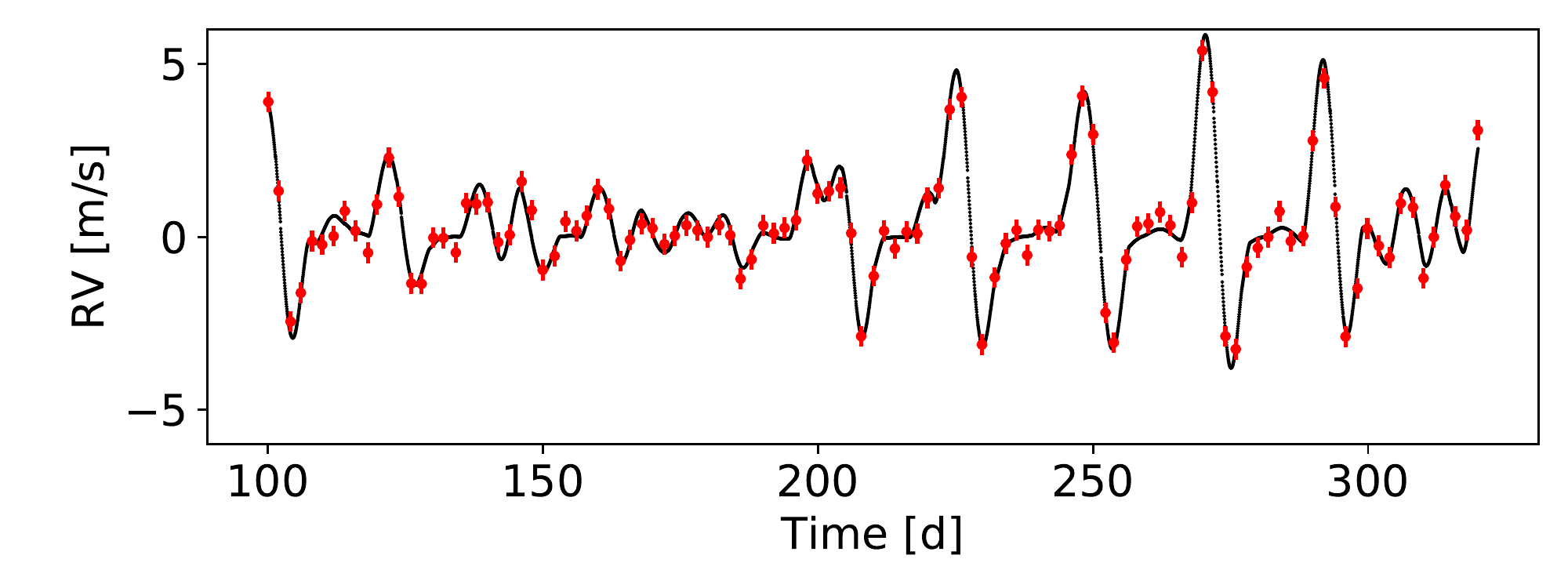}\\
    \includegraphics[width=9cm]{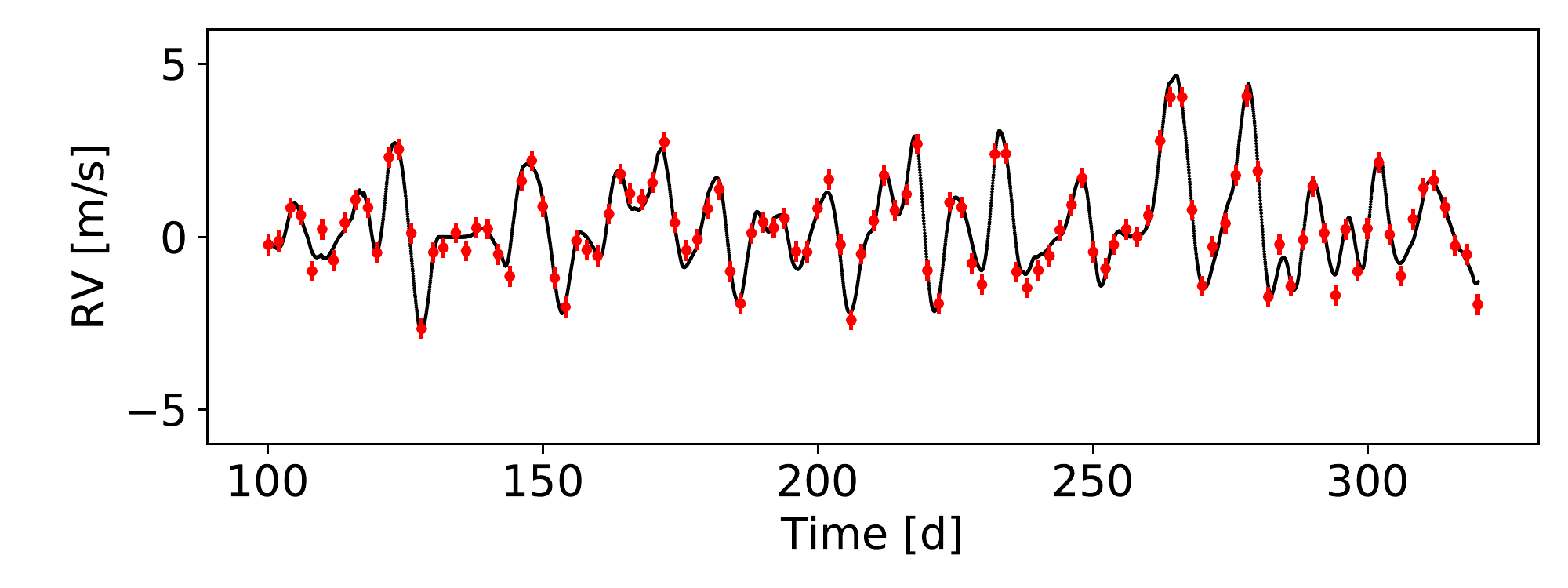}
    \includegraphics[width=9cm]{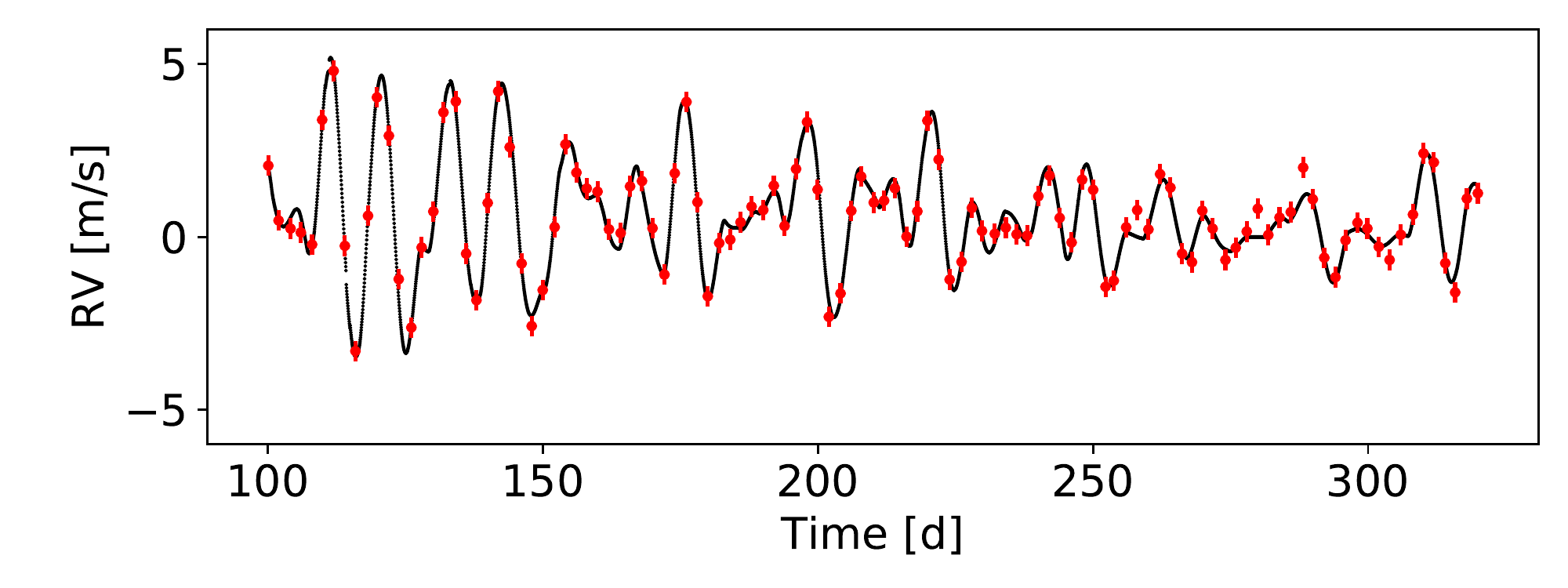}
    \caption{Four simulated example data sets. The black points represent the modeled \texttt{StarSim} RVs, and the red points represent the simulated observations with their error bars. \textit{Top left:} Random spot distribution, $t_{\text{spot}}=P_{\text{rot.}}=22$\,d. \textit{Top right:} Two active longitudes, $t_{\text{spot}}=P_{\text{rot.}}=22$\,d. \textit{Bottom left:} Random spot distribution, $t_{\text{spot}}=5P_{\text{rot.}}=110\,$. \textit{Bottom right:} Two active longitudes, $t_{\text{spot}}=5P_{\text{rott}}=110$\,d.  }
    \label{Fig: dataset_examples}
\end{figure*}{}

We systematically investigated different ratios of the spot lifetimes with respect to the stellar rotation period for spots either distributed randomly or centered at two active longitudes, resulting in a grid of ten
different cases whose details are provided in Table~\ref{Tab: Priors_starsim_I}.
Regarding the spot distributions, we partly followed the approach in \cite{Perger2021}. We assumed a case where spots appear completely randomly on the stellar surface and a case where spots preferentially appear at longitudes that have a distance of 180\,deg (so-called active longitudes).
The random case can be characteristic of active, fast rotating stars \citep{Morin2010, Perger2021}. It may result in a more difficult and less favorable case for the QP-GP to pick up any correlated stellar activity signal.
The case of two active longitudes has been observed for our Sun \citep{Zhang2011}. Here we assumed that the spots can only appear at longitudes drawn from the sum of two normal distributions with means of $x$\,deg and $x$+180\,deg, where $x$ is randomly chosen between 0 and 180\,deg. The standard deviation of these normal distributions has been set to $\sigma_{\text{act.\,long.}}=45$\,deg.
Since we used different discrete spot lifetimes and distributions and had fixed the rotation period, we tweaked the spot sizes and evolution rates to get comparable activity signals of about $\sigma_{RV}=1$ -- $2\,\mathrm{m\,s^{-1}}$. {For each of the activity configurations under investigation, we generated 100 random realizations, taking into account random birth times and positions of spots.}

The spot birth times are drawn from a uniform distribution ranging from {[$t_0-t_{\text{spot}},t_0+t_\text{obs}+t_{\text{spot}}$], where $t_{\text{spot}}$ is the spot lifetime and $t_\text{obs}$ the time span of the observations. This accommodates for the fact that spots appearing before the first RV observation at $t_0$ takes place can still be visible.} In addition, the random birth dates imply that there will always be periods of time when more or fewer star spots are present on the simulated stellar surface. The average number of star spots on the stellar disk visible to the observer per time interval is one of the parameters affecting the RV amplitude. We used the definition of \cite{Basri2020}, which is based on the average number of star spots present on the whole stellar surface at a random time. We show four example data sets of different configurations in Fig.~\ref{Fig: dataset_examples}.

\texttt{StarSim} assumes that spots have a circular shape and a linear growth and decay law of the spot radius (in degrees) that is based on the spot evolution rate (in deg/day). For the grid analysis we assumed that spots grow until they reach their maximum spot size defined as $R_{\text{spot,max}}$, then stay about 10\% of their lifetime at that radius, and decay thereafter, all within the assumed spot lifetime. We also performed simulations with different spot evolution rates. However, we are aware that the assumption of a linear and symmetric growth and decay of the spot radii does not represent the properties of observed Sun spots \citep{Petrovay1997, vanDriel2015, Forgacs-Daika2021}. This prescription is a fixed
property of \texttt{StarSim} that has not been changed for the investigations in this work.

The maximum star spot radius, $R_{\text{spot,max}}$, is set in such a way that the desired median RV scatter is obtained, which is between $1$--$2\,\mathrm{m\,s^{-1}}$ for the grid analysis. However, due to the {randomness} regarding spot distribution and birth times of spots within our simulations, it is not possible to perfectly control the RV scatter caused by the stellar activity simulations. We also note that due to the growth and decay of the spots, the fixed maximum star spot radius only represents an upper boundary. In general, a variety of different spot sizes are apparent within our simulations at any time.

\subsection{The quasi-periodic Gaussian process (QP-GP)}
A full text search for ``exoplanet'' and ``Gaussian process'' in the peer-reviewed astronomical literature results in more than 300 publications\footnote{As of July 2022.}. A significant fraction of them mention the ``quasi-periodic'' GP or kernel.
Generally, a QP kernel is a family of GP covariance functions that are able to model both a periodic signal and a correlation length of that signal simultaneously. The term does not define a particular unique kernel function; however, in the astronomical literature it is common to denote the kernel resulting from the multiplication of an exponential-sine-squared kernel with a squared-exponential kernel as a ``quasi-periodic kernel'' \citep{Haywood2014, Rajpaul2015}. This kernel has the form
\begin{equation}
    \label{Eq: kernel}
    k\left(\tau\right)=\sigma^2_{\textnormal{GP}}\exp\left(-\frac{\tau^2}{2l^2}-\Gamma\sin^2{\left( \frac{\pi\tau}{P_{\text{rot}}} \right)} \right),
\end{equation}
where $\sigma_{\text{GP}}$ is the amplitude of the GP component given in units of the data, $\Gamma$ defines the relative weight between the GP sine-squared component and squared-exponential component and is dimensionless, $l$ is the correlation length scale of the GP squared-exponential component, $P_{\text{rot}}$ is the period of the GP sine-squared component, and $\tau$ 
is the time lag. $l$, $P_{\text{rot}}$, and $\tau$ are commonly given in units of time.
A very common extension of the kernel function in Eq.~\ref{Eq: kernel} is to add a jitter term to the diagonal terms of the covariance function, allowing for a white noise contribution; this is the form we use and refer to in this work as the QP kernel. From Eq.~\ref{Eq: kernel} it is evident that the epithet QP for this kernel comes from its sinusoidal component. However, there are many other forms of kernels that will have a QP behavior, for example the quasi-periodic cosine-kernel \citep[QPC;][]{Perger2021}.

The particular choice of a GP kernel represents a significant part of the prior knowledge on which the modeling is based. Here, we focus on the modeling of the effects from star spots and stellar activity on RV data, where the QP-GP kernel in Eq.~\ref{Eq: kernel} has several advantages, as it was specifically designed for this purpose and is motivated by characteristics of stellar activity in time series data \citep{Aigrain2012, Rajpaul2015, Angus2018}. Furthermore, it is infinitely differentiable and thus leads to well-behaved representations of the data.

\subsection{The QP-GP hyperparameters}
\label{Sect: hyperparams}
The physical motivation of the QP-GP hyperparameters can be difficult: whether the parameters can be connected to real physical processes on the stellar surface and whether it may be possible to calibrate them is currently debated in the literature \citep{Rajpaul2015, Angus2018, Perger2021}. With this work we also intend to improve the general understanding of the QP-GP hyperparameters.

The amplitude hyperparameter $\sigma_{\text{GP}}^2$ defines the absolute strength of the covariance. This hyperparameter is therefore related to the amplitude of the activity signal. In this regard, it can be considered as an outcome of a number of physical processes on the star, such as the spot size, number of visible spots on the stellar surface, or temperature contrast. Many different combinations of these processes can lead to the same value of $\sigma_{\text{GP}}^2$; thus, it is unfeasible to connect this hyperparameter to any single physical {parameter} on the star.

The length scale $l$ defines the correlation length of the signal. Large values of the length scale suggest a strong correlation for data points separated in time, while small values lead to weak or no correlation. This parameter is generally related to the lifetime of activity features on the star, for example stellar spots, and some studies have tried to calibrate the length scale parameter of kernels similar to the QP-GP kernel in attempts to better physically fit this lifetime \citep[see for example][]{Perger2021, Nicholson2022}.

The hyperparameter $P$ correlates data points that are roughly one period apart from each other. The QP-GP has been designed in such a way that this hyperparameter should be directly connectable to the stellar rotation period, which is responsible for active magnetic regions moving across the observed hemisphere of the stellar disk.

The $\Gamma$ hyperparameter describes the harmonic complexity. For small values of $\Gamma$, points with a lag other than a multiple of $P$, are more highly correlated than for large values of $\Gamma$, where points are less correlated if their lag is not close to a multiple of $P$. This hyperparameter can be related to the number of variations (or local maxima and minima) during one full rotation of the star, and by that include information about the average distribution of active regions on the stellar surface.

\subsection{QP-GP priors}
Regarding the choice of priors for the GP model, there are two {distinct} main approaches that are often used in the literature. One is to use wide and unconstrained priors for the QP-GP to allow maximum flexibility of the GP model \citep[see for example][]{Espinoza2020, Pinamonti2022}. The other one is to curtail the flexibility of the GP model as much as possible to reduce ``overfitting-like'' behavior. This is achieved by applying more constrained, in most cases physically motivated, priors \citep[see for example][]{Nava2020, Stock2020c}. The idea behind using tighter GP priors is to
incorporate all the information that is available in the fit, which as a result can lead to better precision of the planetary signals. An additional benefit is a shorter run time of the sampler.

Typically, the QP-GP is constrained as follows.
The amplitude parameter $\sigma_{\text{GP}}^2$ is challenging to constrain due to the large degeneracy of parameters affecting it. Nevertheless, there is a physically meaningful maximum, namely the scatter of the RV data of the star under consideration, which combines all the parameters of activity influencing the RV amplitude. The distribution of this hyperparameter is often chosen to be positive and uniform, or log-uniform. {For our analysis, we adopted a universal uniform range between \SI{0}{\meter\per\second} and \SI{40}{\meter\per\second}, which generously considers the range of scatter that we expected for our simulated activity.}

The length scale parameter, $l$, is essential when it comes to modeling a meaningful periodic signal. It is a common approach to use a uniform or log-uniform prior spanning a large range of possible values. However, there are some caveats: for example, a small length scale parameter leads to a dominating squared-exponential kernel, and the signal decays before one rotation has finished. {One aim of our investigation was to see how the choice of this prior affects the results. We therefore tested different priors in different situations, as will be described later.}

For the GP rotation period, {$P$}, it is common to use a uniform or log-uniform prior that spans the range of physically possible stellar rotation periods. These can range from less than a day for very young stars to hundreds of days for giant stars in their post main sequence evolutionary phase. In cases where an estimate of the rotation period is available, for example due to auxiliary data from photometry and/or activity indicators, it is possible to use this estimate as a physical prior to further constrain the GP rotation hyperparameter. {In our analysis, we use both uninformed uniform priors and constrained priors to investigate how the results differ between the two.}

Large values of the harmonic complexity parameter, $\Gamma$, allow the GP to become very flexible. {The reason is that the fitted curve drawn from the GP may include a higher number of turning points, which in turn can allow for more frequencies to be fitted that may not necessarily be related to the RV variations by the active regions.} Therefore, {it can be useful} to impose an informative prior or an upper limit on $\Gamma$. \cite{Jeffers2009} have shown that, to first approximation, any light curve can be explained by two unequally large spots that are 180 degress apart, independent of the complexity of the spot distribution on the stellar surface. This roughly corresponds to two active longitudes on the stellar surface, and its RV signal can be modeled by one to three local maxima of the GP within one full rotation period. Therefore, to reduce the flexibility of the GP at least an upper limit of $\Gamma$ should be set to allow for a maximum of two to three inflection points within one rotation period, which concretely means that within this work we use a log-uniform prior on $\Gamma$ ranging from 0.01 to 10. Within the literature, there are cases where even more informative priors were applied to $\Gamma$, for example informative normal priors \citep[see, for example,][and references therein]{Nava2020}.

\subsection{Modeling approach}

Within this work, we used \texttt{juliet} \citep{Espinoza2018} to fit the synthetic RV data that include the simulated stellar activity signals and injected planetary signals. For the modeling of Keplerian signals, \texttt{juliet} uses the publicly available package \texttt{radvel} \citep{Fulton2018}. The QP-GP kernel investigated within this work is computed by the publicly available package \texttt{george} \citep{Ambikasaran2015}. The evaluation of the GP likelihood with \texttt{george} is computationally expensive and scales as $N \ln{N}$, where $N$ is the number of data points \citep{Ambikasaran2015}. Nevertheless, \texttt{george} is one of the most widely used packages.

\begin{figure*}
    \centering
    \includegraphics[width=6cm]{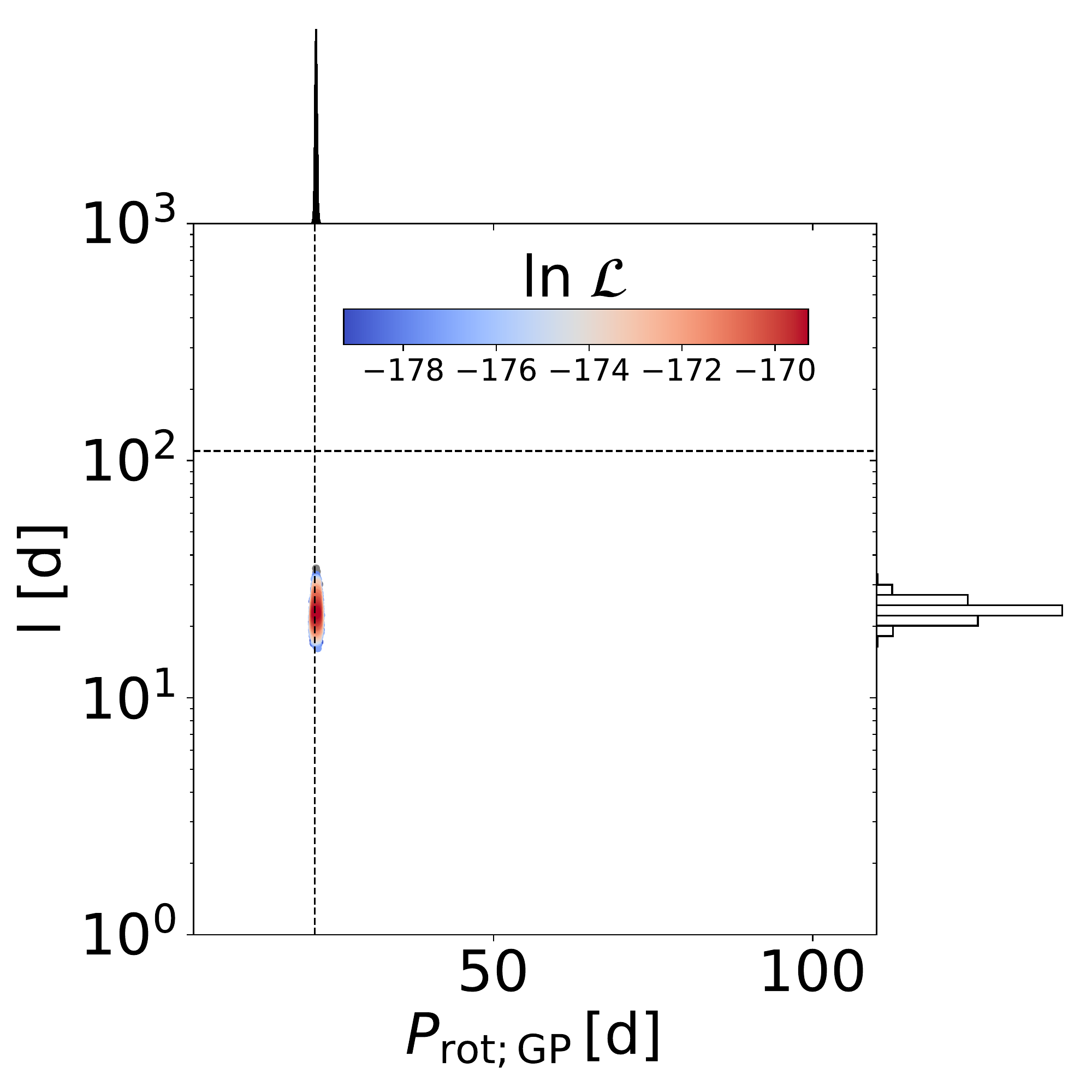}
    \includegraphics[width=6cm]{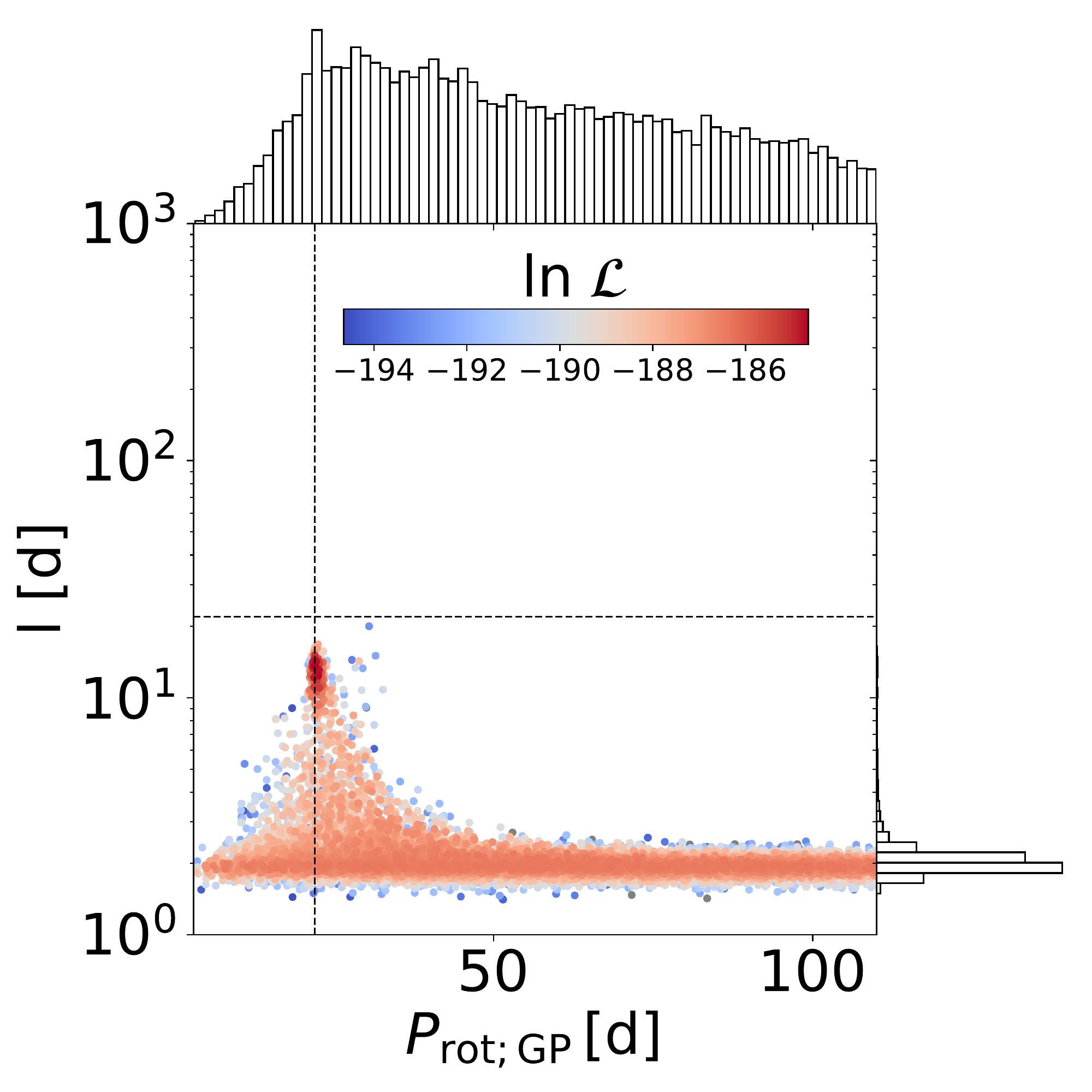}
    \includegraphics[width=6cm]{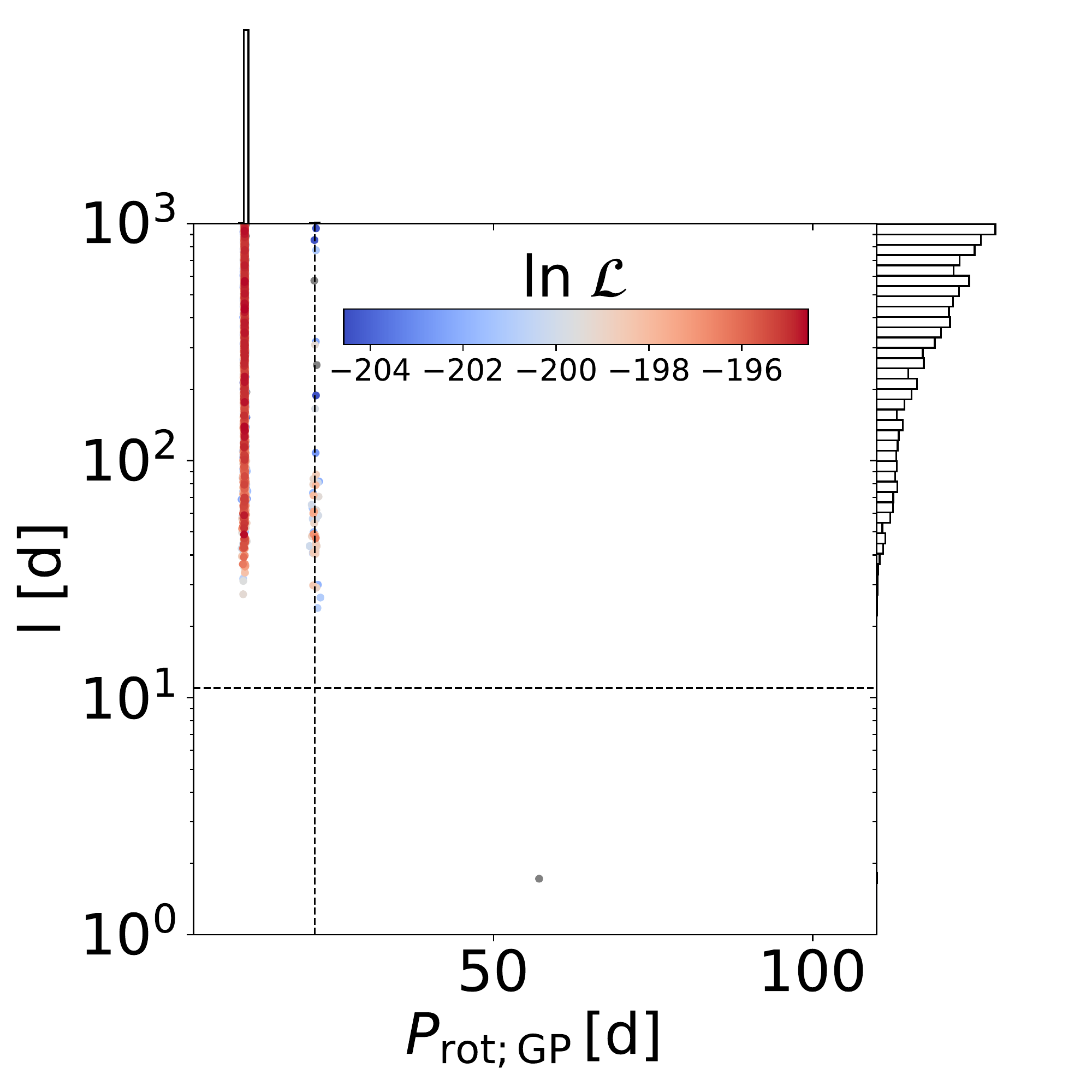}
    \caption{$l$ versus $P$ diagrams showing the posterior samples of a single GP fit with uninformed priors to one example simulation. The horizontal dashed line indicates the simulated spot lifetime, and the vertical dashed line indicates the simulated rotation period. \textit{Left}: Posterior samples fitted to a star with random spot distribution, $P_{\rm rot}=22$\,d, and $P_{\rm life}=110$\,d. A well-behaved peak is found where both the rotation period and the GP length scale are well constrained. \textit{Middle}: Posterior samples fitted to a star with random spot distribution, $P_{\rm rot}=22$\,d, and $P_{\rm life}=22$\,d. A posterior plateau with a short length scale spans the entire posterior volume. The posterior of the rotation period is not well constrained due to this plateau; however, some posterior samples form a ``triangle'' shape at the rotation period. \textit{Right}: Posterior samples fitted to a star with two active longitudes, $P_{\rm rot}=22$\,d, and $P_{\rm life}=11$\,d. The spots that appear on a similar longitude mimic a coherent signal whose length scale cannot be constrained by the GP, resulting in an ``I-shaped'' posterior distribution that reaches the upper boundary of the posterior volume.  }
    \label{Fig: shapes}
\end{figure*}{}

Within \texttt{juliet}, we used nested sampling, in particular, dynamic nested sampling based on the \texttt{dynesty} package \citep{Speagle2019} to fit the models to the data and derive the posterior distributions of the model parameters.
Dynamic nested sampling is better suited than standard Markov chain Monte Carlo procedures for dealing with multi-modal posterior distributions, which are common for the GP hyperparameters, since it is not as susceptible to problems caused by local minima \citep{Feroz2009,Feroz2013,Speagle2019}. While Markov chain Monte Carlo samplers generally require a good initial parameter vector, nested sampling efficiently samples the complete user-defined prior volume. An additional advantage of using nested-sampling algorithms is the fact that the marginal log-likelihood (hereafter referred to as evidence) is a natural outcome of the sampling process \citep{Feroz2009}. Bayesian model comparison, which uses the Bayesian evidence, is a powerful method for comparing distinct or nested models with different numbers of parameters and complexity against each other. Following \citet{Trotta2008}, we consider a model as a significant improvement compared to another model if the log-evidence difference is larger than $\Delta \ln{\mathcal{Z}}>5$, while $\Delta \ln{\mathcal{Z}}>2.5$ is moderate evidence in favor of one model.

\section{Results of QP-GP fits to synthetic RV activity-only data}
\label{Sect: Results1}

\begin{table*}
    \centering
    \caption{Different {model components} that were investigated in this work.}
    \label{Tab: Priors_GP_RV}
    \begin{tabularx}{\textwidth}{lccX}
        \hline
        \hline
        \noalign{\smallskip}
        Parameter name                                       & Prior                                         & Units                & Description                                                         \\
        \noalign{\smallskip}
        \hline

        \noalign{\smallskip}
        \multicolumn{4}{c}{White Noise}                                                                                                                                                                   \\
        \noalign{\smallskip}
        \noalign{\smallskip}
        ~~~$\gamma_{\textnormal{SIM}}$                       & $\mathcal{U}(-10,10)$                         & $\mathrm{m\,s^{-1}}$ & Velocity zero-point for simulated data set                          \\
        ~~~$\sigma_{\textnormal{SIM}}$                       & $\mathcal{J}(0.01,10)$                        & $\mathrm{m\,s^{-1}}$ & Extra jitter term for simulated data set                            \\

        \noalign{\smallskip}
        \multicolumn{4}{c}{GP Prior I (wide priors)}                                                                                                                                                      \\
        \noalign{\smallskip}
        \noalign{\smallskip}
        ~~~$\sigma_{\textnormal{GP, RV}}$                    & $\mathcal{U}(0,40)$                           & $\mathrm{m\,s^{-1}}$ & Amplitude of GP component for RVs                                   \\
        ~~~$\Gamma_{\textnormal{GP, RV}}$                    & $\mathcal{J}(10^{-2},10^1)$                   & \ldots               & Amplitude of GP sine-squared component for RVs                      \\
        ~~~$l_{\textnormal{GP, RV}}$                         & {$\mathcal{J}(1, 10^5)$}                      & {d}                  & Length scale of GP exponential component for RVs\tablefootmark{(a)} \\
        ~~~$P_{\textnormal{rot, GP,RV}}$                     & $\mathcal{U}(3,110)$                          & d                    & Period of the GP quasi-periodic component for RVs                   \\
        \noalign{\smallskip}
        \multicolumn{4}{c}{GP Prior II (length scale constrained)}                                                                                                                                        \\
        \noalign{\smallskip}
        \noalign{\smallskip}
        ~~~$\sigma_{\textnormal{GP, RV}}$                    & $\mathcal{U}(0,40)$                           & $\mathrm{m\,s^{-1}}$ & Amplitude of GP component for RVs                                   \\
        ~~~$\Gamma_{\textnormal{GP, RV}}$                    & $\mathcal{J}(10^{-1},10^1)$                   & \ldots               & Amplitude of GP sine-squared component for RVs                      \\
        ~~~$l_{\textnormal{GP, RV}}$                         & {$\mathcal{J}(1/\sqrt{5\cdot10^{-2}}, 10^5)$} & {d}                  & Length scale of GP exponential component for RVs\tablefootmark{(a)} \\
        ~~~$P_{\textnormal{rot, GP,RV}}$                     & $\mathcal{U}(3,110)$                          & d                    & Period of the GP quasi-periodic component for RVs                   \\
        \noalign{\smallskip}
        \noalign{\smallskip}
        \multicolumn{4}{c}{GP Prior III (period constrained)}                                                                                                                                             \\
        \noalign{\smallskip}
        \noalign{\smallskip}
        ~~~$\sigma_{\textnormal{GP, RV}}$                    & $\mathcal{U}(0,40)$                           & $\mathrm{m\,s^{-1}}$ & Amplitude of GP component for RVs                                   \\
        ~~~$\Gamma_{\textnormal{GP, RV}}$                    & $\mathcal{J}(10^{-1},10^1)$                   & \ldots               & Amplitude of GP sine-squared component for RVs                      \\
        ~~~$l_{\textnormal{GP, RV}}$                         & {$\mathcal{J}(1, 10^5)$}                      & {d}                  & Length scale of GP exponential component for RVs\tablefootmark{(a)} \\
        ~~~$P_{\textnormal{rot, GP,RV}}$                     & $\mathcal{N}(22,2.2)$                         & d                    & Period of the GP quasi-periodic component for RVs                   \\
        \noalign{\smallskip}
        \noalign{\smallskip}
        \multicolumn{4}{c}{GP Prior IV (length scale and period constrained)}                                                                                                                             \\
        \noalign{\smallskip}
        \noalign{\smallskip}
        ~~~$\sigma_{\textnormal{GP, RV}}$                    & $\mathcal{U}(0,40)$                           & $\mathrm{m\,s^{-1}}$ & Amplitude of GP component for RVs                                   \\
        ~~~$\Gamma_{\textnormal{GP, RV}}$                    & $\mathcal{J}(10^{-1},10^1)$                   & \ldots               & Amplitude of GP sine-squared component for RVs                      \\
        ~~~$l_{\textnormal{GP, RV}}$                         & {$\mathcal{J}(1/\sqrt{5\cdot10^{-2}}, 10^5)$} & {d}                  & Length scale of GP exponential component for RVs\tablefootmark{(a)} \\
        ~~~$P_{\textnormal{rot, GP,RV}}$                     & $\mathcal{N}(22,2.2)$                         & d                    & Period of the GP quasi-periodic component for RVs                   \\
        \noalign{\smallskip}
        \noalign{\smallskip}

        \multicolumn{4}{c}{K$X$ (where $X$ is given in days)}                                                                                                                                             \\
        \noalign{\smallskip}
        \noalign{\smallskip}
        ~~~$P_b$                                             & $\mathcal{U}(X-0.1X,X+0.1X)$                  & d                    & Period                                                              \\
        ~~~$t_{0,b} - 2450000$                               & $\mathcal{U}(2458620.,2458620.+1.1X)$         & d                    & Time of transit center                                              \\
        ~~~$K_{b}$                                           & $\mathcal{U}(0,40)$                           & $\mathrm{m\,s^{-1}}$ & RV semi-amplitude                                                   \\
        ~~~$\mathcal{S}_{1,b} = \sqrt{e_{b}}\sin \omega_{b}$ & $\mathcal{U}(-1,1)$                           & \dots                & Parametrization for $e$ and $\omega$                                \\
        ~~~$\mathcal{S}_{2,b} = \sqrt{e_{b}}\cos \omega_{b}$ & $\mathcal{U}(-1,1)$                           & \dots                & Parametrization for $e$ and $\omega$                                \\

        \hline
    \end{tabularx}
    \tablefoot{\tablefoottext{a}{{In \texttt{juliet}, the length scale parameter is parameterized by its inverse as described in \autoref{app:lengthscale}. For better understanding, however, we give the direct length scale.}}

        {The prior labels $\mathcal{U}$, $\mathcal{N,}$ and $\mathcal{J}$ represent uniform, Normal and Jeffrey's distributions \citep{Jeffreys1949}. The $x$ represents either 5.12\,d, 30.35\,d or 44\,d regarding the period of the Kepler signal. The white noise model is always applied on top of the other models in this table.}}
\end{table*}

{In this section we present our investigation of the relationship between the GP hyperparameters and the physical properties of the simulated stellar activity. Our large grid of different activity configurations not only allowed us to investigate the correlation between the parameters of the stellar activity and the GP hyperparameters, but also how the determination of the hyperparameters is affected by different states of activity. Our starting point for this was to use wide, uninformative priors for the GP hyperparameters (see also Table~\ref{Tab: Priors_GP_RV}; referred to as GP Prior I hereafter). The focus was on the question whether the QP-GP can robustly derive the simulated stellar rotation period and spot lifetimes by connecting these values to the physically motivated QP-GP hyperparameters as described in \autoref{Sect: hyperparams}. An analysis based on the Bayesian evidence of how the different conditions affect the probability of finding the (correct) rotation period in practice is presented in \autoref{app:modelcomparison_onlyact}.}

\subsection{QP-GP properties and the $l_{\textnormal{GP}}$ versus $P_{\textnormal{GP}}$ diagram}
\label{Sect: alpha_vs_prot}

{As can be seen in the middle panel of \autoref{Fig: shapes}, the GP rotation period and the length scale parameter can be highly correlated with each other, which is not surprising, given Eq.~\ref{Eq: kernel}.} Both hyperparameters are as an exponential product interconnected with each other and should therefore not be considered in isolation. Shorter QP-GP length scales mean that the signal of the rotation period is less coherent over the time of observations.

    {Within the QP-GP fits to our simulated RV data sets, we find three predominant classes of possible distributions in the $l_{\textnormal{GP}}$ versus $P_{\textnormal{GP}}$ plane: the ``o,'' ``triangle,'' and ``bar'' shapes presented in Fig.~\ref{Fig: shapes} \citep[see also][]{Stock2020b, Stock2020c, Bluhm2021, Kossakowski2021}. We note that real-world scenarios of active stars have however also shown mixtures of the posterior distributions introduced here. For instance, TOI-1201 \citep[][their Fig.~A.1]{Kossakowski2021} was observed to exhibit a mixture of a triangle shape together with a bar shape.}

The left plot shows an o-shaped posterior distribution, resulting in a well-defined peak within the $l_{\textnormal{GP}}$ versus $P_{\textnormal{GP}}$ parameter plane. {Only} in such a case can both the length scale and the rotational period of the QP-GP be well determined from the RV data. However, we only observe o-shaped posterior distributions for spot lifetimes that are at least five times the rotation period. A real-world target for which such a 2D posterior distribution of the GP length scale and rotation period has been observed is the nearby star Lalande~21185 \citep[see Fig.~22 in][]{Stock2020c}.

The middle plot shows a triangle shape that was also commonly identified in our simulations. This 2D posterior distribution is characterized by a broad plateau of solutions at small length scales that are consistent with all allowed rotation periods given the prior volume, and a triangle of posterior solutions (often with higher likelihood) sitting on top of the observed plateau indicating solutions with larger GP length scales that are also consistent with the data. Based on the fits to our simulations, the position of the triangle is often found to be related to the simulated stellar rotation period. The plateau of posterior solutions with small $l_{\textnormal{GP}}$ and no preference for any rotation period within the $l_{\textnormal{GP}}$ versus $P_{\textnormal{GP}}$ parameter space is a consequence of the fact that the rotation hyperparameter within the exponential-sine-squared kernel is undefined if the signal significantly decays on timescales smaller than one rotation, resulting in a dominant exponential-squared term. {This is for example the case if the spot lifetime of the star is close to or smaller than the simulated stellar rotation period.} Real-world examples of such triangle-shaped posterior  distributions have been identified for the RV time series data of YZ~Ceti, GJ~251, and TOI-1201 \citep{Stock2020b, Stock2020c, Kossakowski2021}.

The right diagram in Fig.~\ref{Fig: shapes} shows the third class of frequently observed {relations} between the GP length scale and the GP period. For some of our simulations, the posterior solutions form a bar shape that extends toward the upper boundary of the prior volume of $l_{\textnormal{GP}}$ at a period related to the stellar rotation. Such a shape is caused by the fact that the decay of the signal cannot be constrained given the data or time of observations. As a result, the length scale is estimated to ``infinity'' resulting in valid solutions for any large $l_{\textnormal{GP}}$. In the $l_{\textnormal{GP}}$ versus $P_{\textnormal{GP}}$ diagram, this distribution can be interpreted as a signal that is to some extent coherent over the entire time of observations. A real-world example where such a posterior distribution has been observed is HD~238090 \citep{Stock2020c}.

\subsection{Detection of stellar rotation periods by the QP-GP}
\begin{figure*}
    \centering
    \includegraphics[width=0.48\textwidth]{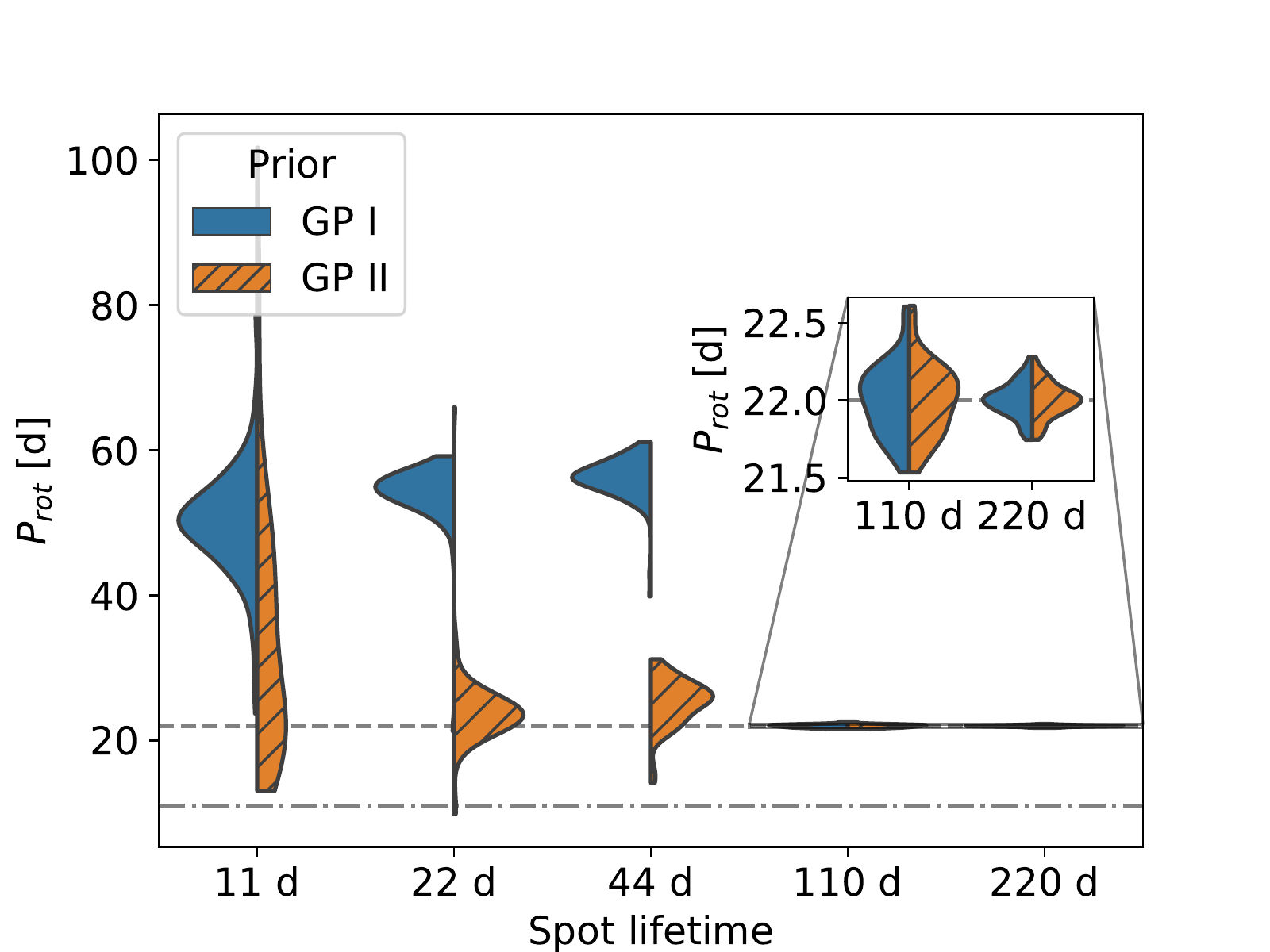}
    \includegraphics[width=0.48\textwidth]{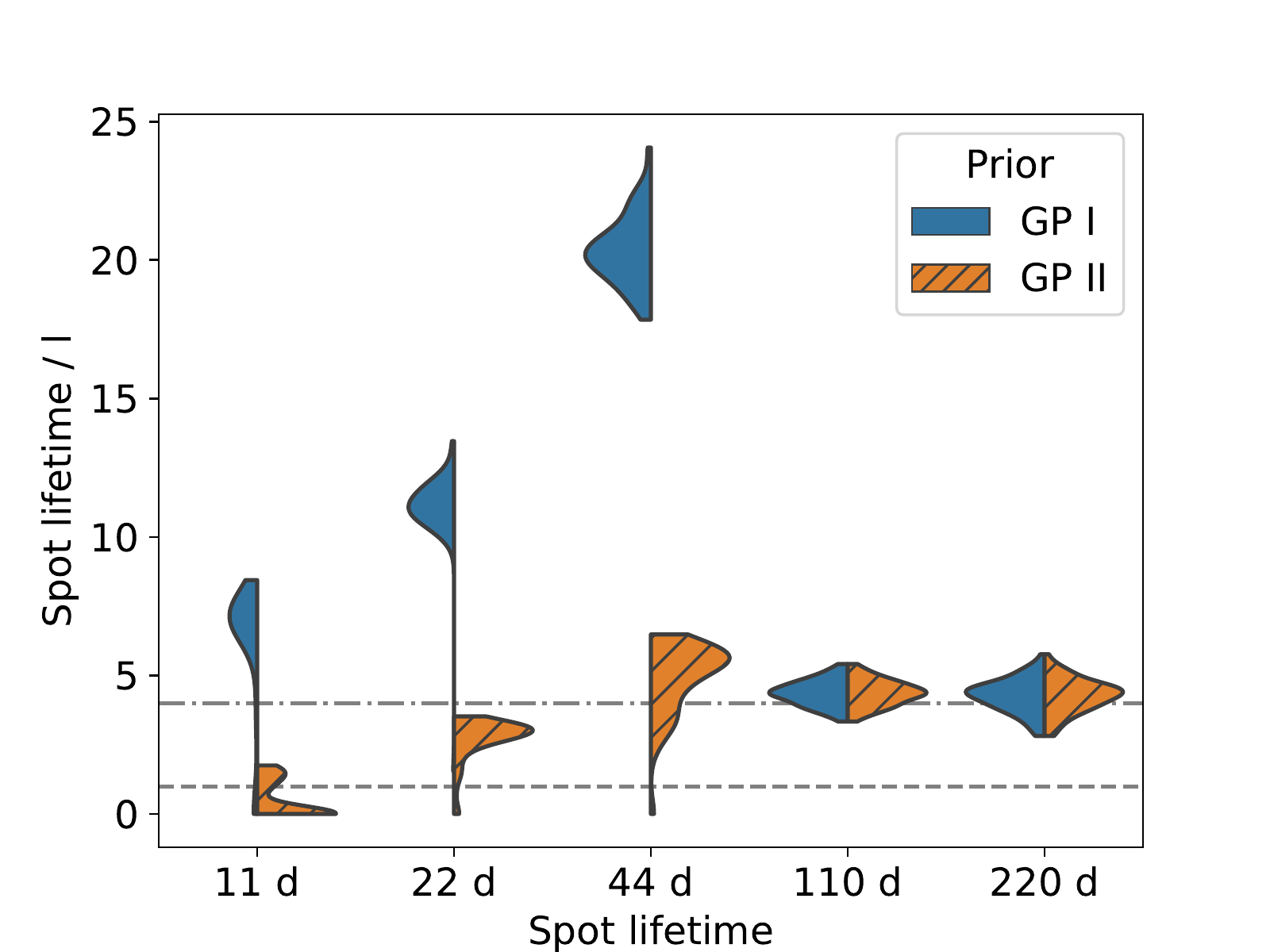}\\
    \includegraphics[width=0.48\textwidth]{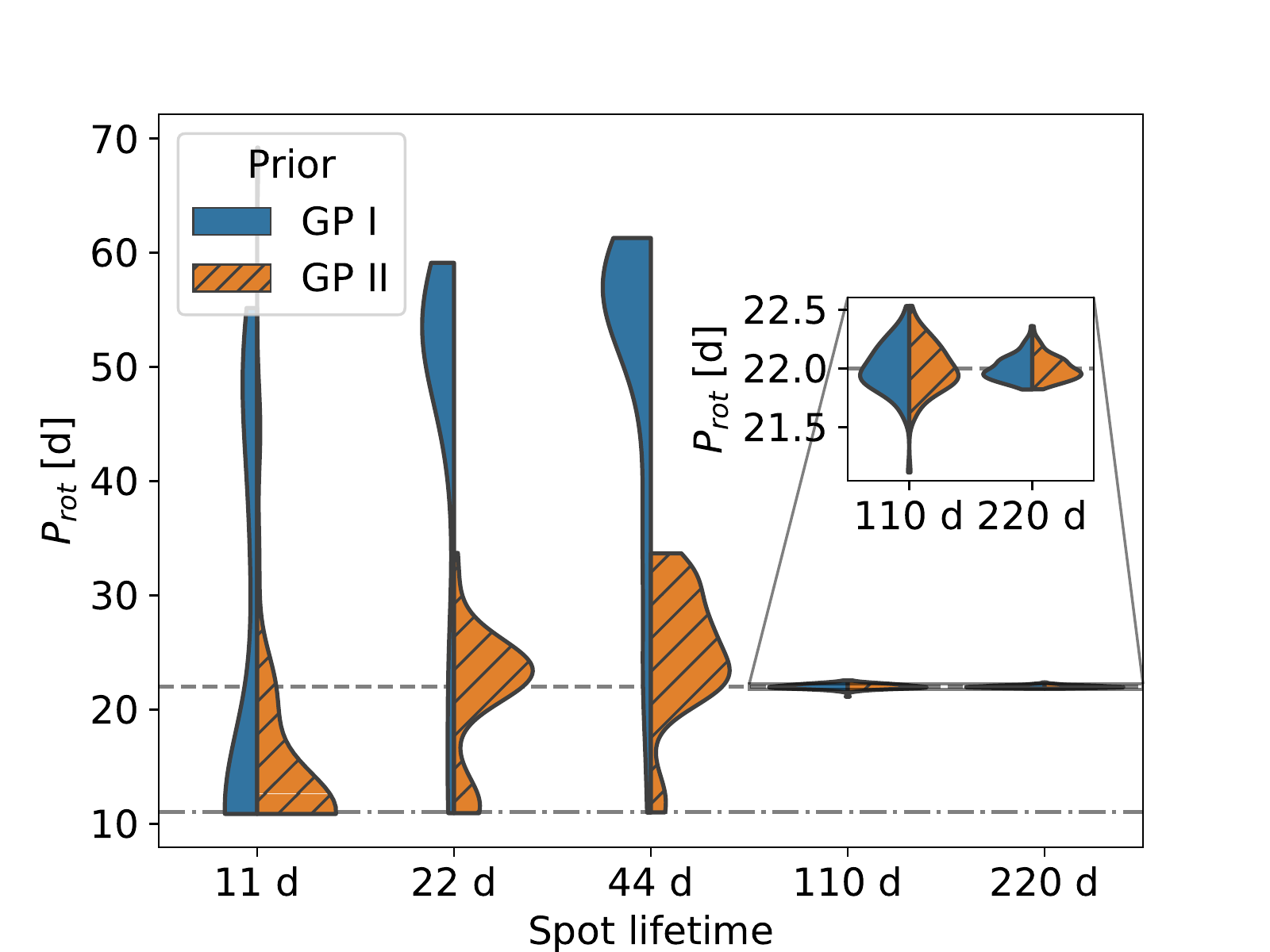}
    \includegraphics[width=0.48\textwidth]{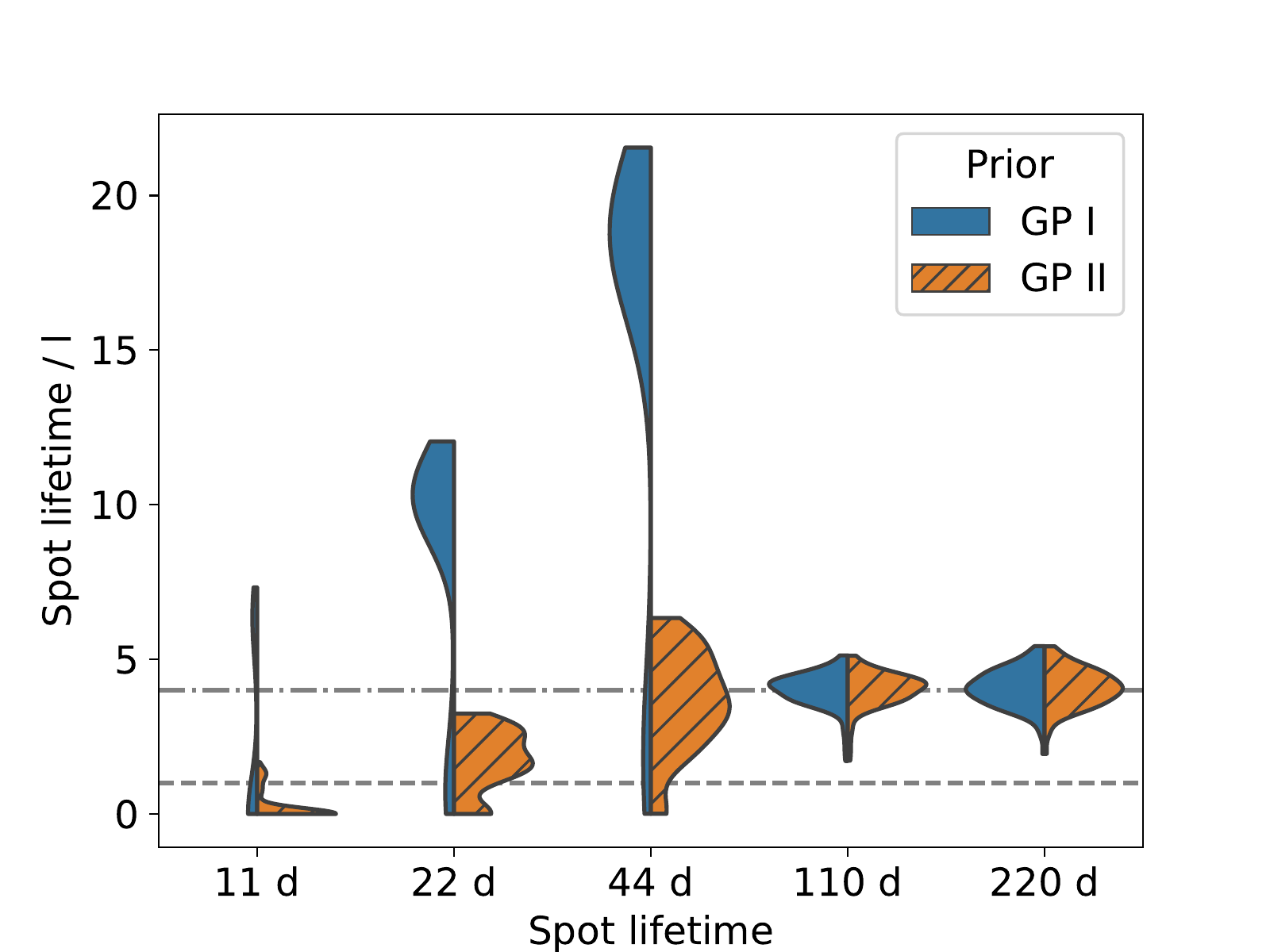}\\
    \caption{{Violin plots showing the distribution of the derived median rotation periods and length scale hyperparameters of the QP-GP based on 100 simulations per spot lifetime. The blue distribution shows the results obtained from using GP Prior I (wide prior), and the orange distribution shows the results obtained by using GP Prior II (length scale constrained). For the period, the dashed line shows the simulated rotation period and the dash-dotted line the first harmonic at $P_{\text{rot}}/2$. For the length scale, we show the simulated spot lifetime divided by the determined GP length scale so that the dashed line marks a 1:1 correspondence and the dash-dotted line a factor of 4 between the two. \textit{Top:} For a random spot distribution on the stellar surface. \textit{Bottom}: With two active longitudes on the stellar surface.}}
    \label{Fig: GP_period_violin}
\end{figure*}

{In Fig.~\ref{Fig: GP_period_violin} we show the comparison between the derived medians of the QP-rotation hyperparameters and the simulated period for five different spot lifetimes and two different spot distributions. In doing so, we consider the unconstrained GP Prior I, as well as a length scale constrained prior (GP Prior II in Table~\ref{Tab: Priors_GP_RV} and hereafter), which was motivated by the triangle shape distribution described in the previous section. The idea behind this is to exclude the plateau with small length scales and unconstrained rotational modulation in order to force the GP to model a QP signal.}

With both GP priors, the rotation period was derived very accurately and precisely by the QP-GP for the majority of cases in which the spot lifetime was at least a few times larger than the rotation period. For such cases, there was also no significant difference between GP Prior I or II regarding the derived median value of the stellar rotation period. In cases where the spot lifetime is close to the stellar rotation timescale ($P_{\mathrm{life}}\leq2P_{\mathrm{rot}}$), the estimates of the rotation period by the QP-GP were less precise. This is expected, as the signal's coherence decreases if the positions, size, and number of stellar spots on the surface change significantly within one rotation.
The determination of the rotation period is particularly hampered if the spot distribution is, in addition to the short spot lifetime, positionally uncorrelated, which means for random spot distributions. {In this case}, we see that the unconstrained GP is not able to recover the simulated rotation period. A feature that can be seen directly from Fig.~\ref{Fig: GP_period_violin} is that fits using GP Prior I for the RV data sets with short spot lifetimes favor solutions with rotation periods that lie exactly in the middle of the predefined prior volume. The reason is the plateau of posterior solutions that dominates the posterior distribution over the entire period prior volume, as shown in the example plot in Fig.~\ref{Fig: shapes}. {For those instances, the median rotation periods derived by the length scale constrained GP Prior II, which excludes this plateau, are much more consistent and closer to the simulated rotation period.}
However, even when applying a QP-GP with Prior II we find that the derived rotation period is sensitive to the spot distribution and spot lifetimes. In the case of two active longitudes and with decreasing spot lifetimes, there are more and more cases where the median of the determined rotation period is half of the simulated one and represents therefore the second harmonic of the true rotation period. It is straight-forward to understand how two active longitudes can lead to such an estimate, as it is very difficult for a correlated noise model, such as the QP-GP, to distinguish between two almost equally spotted sides of the stellar surface. In the case of shorter spot lifetimes and with two active longitudes the much faster dynamic evolution of the spot pattern observed at shorter spot lifetimes complicates the distinction of the active longitudes by the GP noise model, explaining the higher number of solutions that are then found at the second harmonic.

\subsection{Limitations in correctly identifying the star spot lifetimes}

{In \autoref{Sect: alpha_vs_prot}, we showed already that only one of the posterior distributions occurring in the $l_{\textnormal{GP}}$ versus $P_{\textnormal{GP}}$ diagram results in an unambiguous measurement for the length scale hyperparameter of the GP. For a more in-depth analysis of the {connection} between the determined length scale parameter and the simulated spot lifetime, we investigated the outcome from the previous section with focus on the length scale.}

{We observed a clear relation in most of the cases: longer simulated spot lifetimes result in longer correlation length scales of the QP-GP. However, as can be seen in \autoref{Fig: GP_period_violin}, we found no 1:1 correspondence but a  dependence of the derived length scale of the QP-GP that varies with the simulated spot lifetime and spot pattern. As already seen for the rotation period, the constrained GP Prior II had a clear advantage for short spot lifetimes.}

{Figure\,\ref{Fig: GP_period_violin} shows that the largest deviation occurred for the unconstrained GP Prior I and short spot lifetimes (0.5\,$P_{\text{rot.}}$ to 2\,$P_{\text{rot.}}$) and the resulting plateau with indefinite rotation period discussed in \autoref{Sect: alpha_vs_prot}. The length scale hyperparameters determined from this were generally much smaller than the simulated spot lifetime -- on the order of a few days at most.} Strikingly, for the sot lifetimes of 22\,d and 44\,d the gross of the determined length scales in those cases is centered around ratios of $\sim10$ and $\sim20,$ respectively, which corresponds to a length scale of 2\,d and matches the sampling of our data. For a less regular sampling, as in the real observations in \cite{Stock2020b} and \cite{Stock2020c} this relation is however not apparent and the plateau occurs for length scales between 10--100\,d, which highlights the importance of the visual inspection of the posterior using the $l_{\textnormal{GP}}$ versus $P_{\textnormal{GP}}$ diagram as presented in \autoref{Sect: alpha_vs_prot}. Fitting a QP-GP with GP Prior II{, which} excludes the observed posterior plateau in $l_{\textnormal{GP}}${,} reduced the offset between $l$ and the simulated spot lifetime, {but still shows a dependence on the spot lifetime}. For the cases with spot lifetimes of 5\,$P_{\text{rot.}}$ or 10\,$P_{\text{rot.}}$ we observed no difference in the results based on the prior choice as these result in well-behaved o-shaped 2D posterior distributions{, with an approximate factor of 4 to the simulated spot lifetime.}

    {Especially for the shortest spot lifetimes and the configuration of two active longitudes, we recognized a lot of samples that resulted in length scale parameters much larger than the simulated spot lifetime (i.e., ratios close to 0). These are the cases with a bar shape in the 2D posterior distribution. A possible explanation is that due} to our definition of an average number of spots over a specific time interval, such cases are much more dynamic than our simulated cases with longer spot lifetimes. A large number of spots forms and decays over the time of the simulations, and the fast reoccurrence of spots with short lifetimes that are smaller than the rotation period tend to lead to a coherent signal that the QP-GP cannot distinguish from that of a single long-lived signal. {Because the spots occurring on two active longitudes are positionally correlated on opposite sides of the stellar disk, the resulting determined rotation period is often found at the second harmonic (0.5\,$P_{\text{rot.}}$) of the simulated rotation period. For a random spot distribution and spot lifetimes shorter than the rotation period, we find samples with determined length scale parameters longer than the simulated spot lifetimes and measured rotation periods completely unrelated to the rotation period.}

\subsection{Dependence of QP-GP length scale and period on jitter and RV uncertainty}
{So far, we only considered scenarios in which we adopted a fixed instrumental uncertainty of 30\,$\mathrm{cm\,s^{-1}}$ and no additional white-noise contribution by the star or instrument. In the following we discuss the sensitivity of the derived GP hyperparameters with respect to an added jitter, which in reality may result from unresolved stellar oscillations of the star itself or from an unstable RV instrument.} For these investigations we used the stellar activity models with random spot distribution and a simulated spot lifetime of 110\,d, as this is one of the cases where the QP-GP kernel is particularly effective, as we show in the previous section.

We applied different RV uncertainties, different white noise contributions, or a combination of both to the simulated data and then fitted a QP-GP with Prior I to the data. We note the difference between RV uncertainties and jitter: for the RV uncertainties, the shift of the measured value is taken into account by the larger error bars of the data, while this is not the case for the white noise contribution caused by jitter.

\begin{figure}
    \centering
    \includegraphics[width=8.15cm]{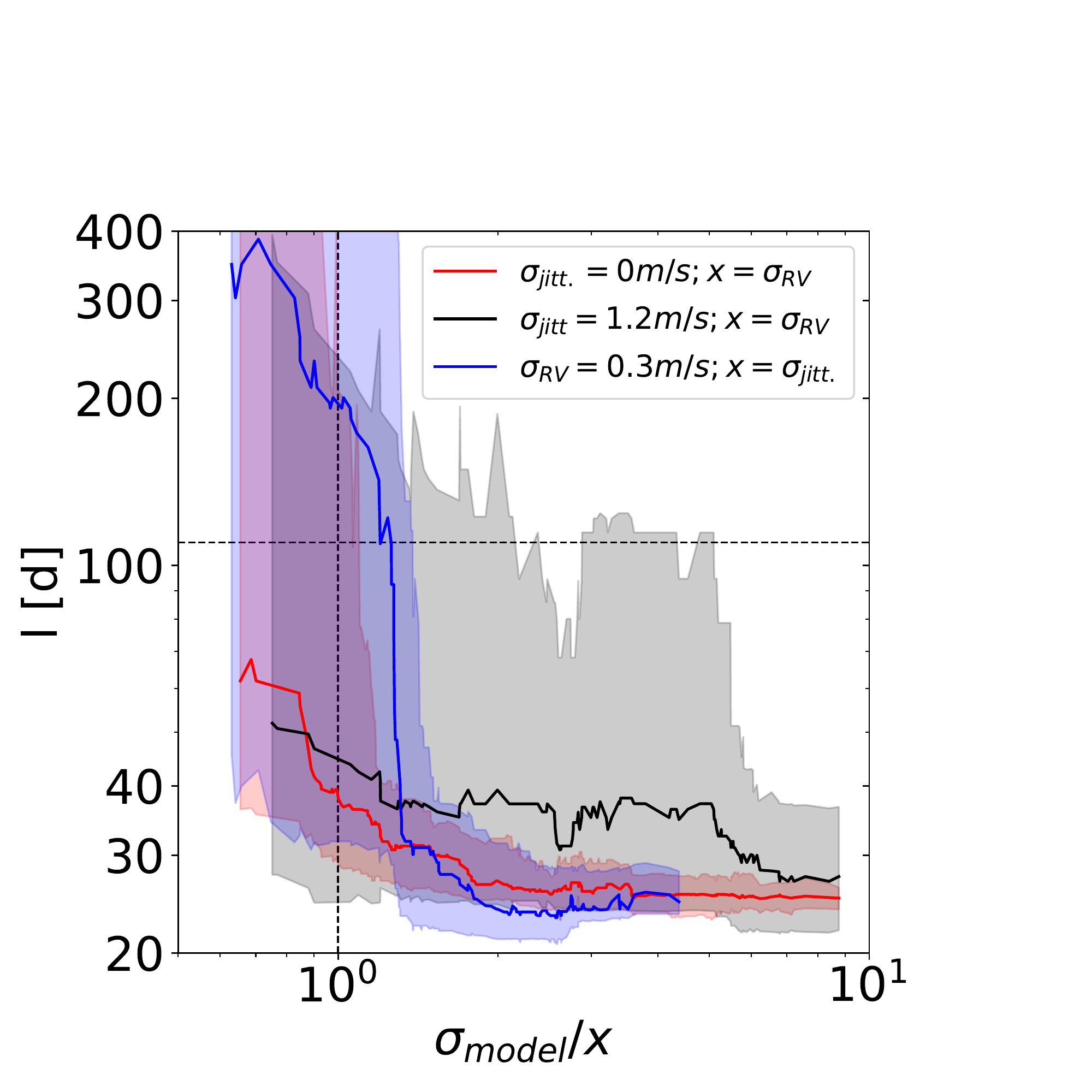}\\
    \includegraphics[width=8.15cm]{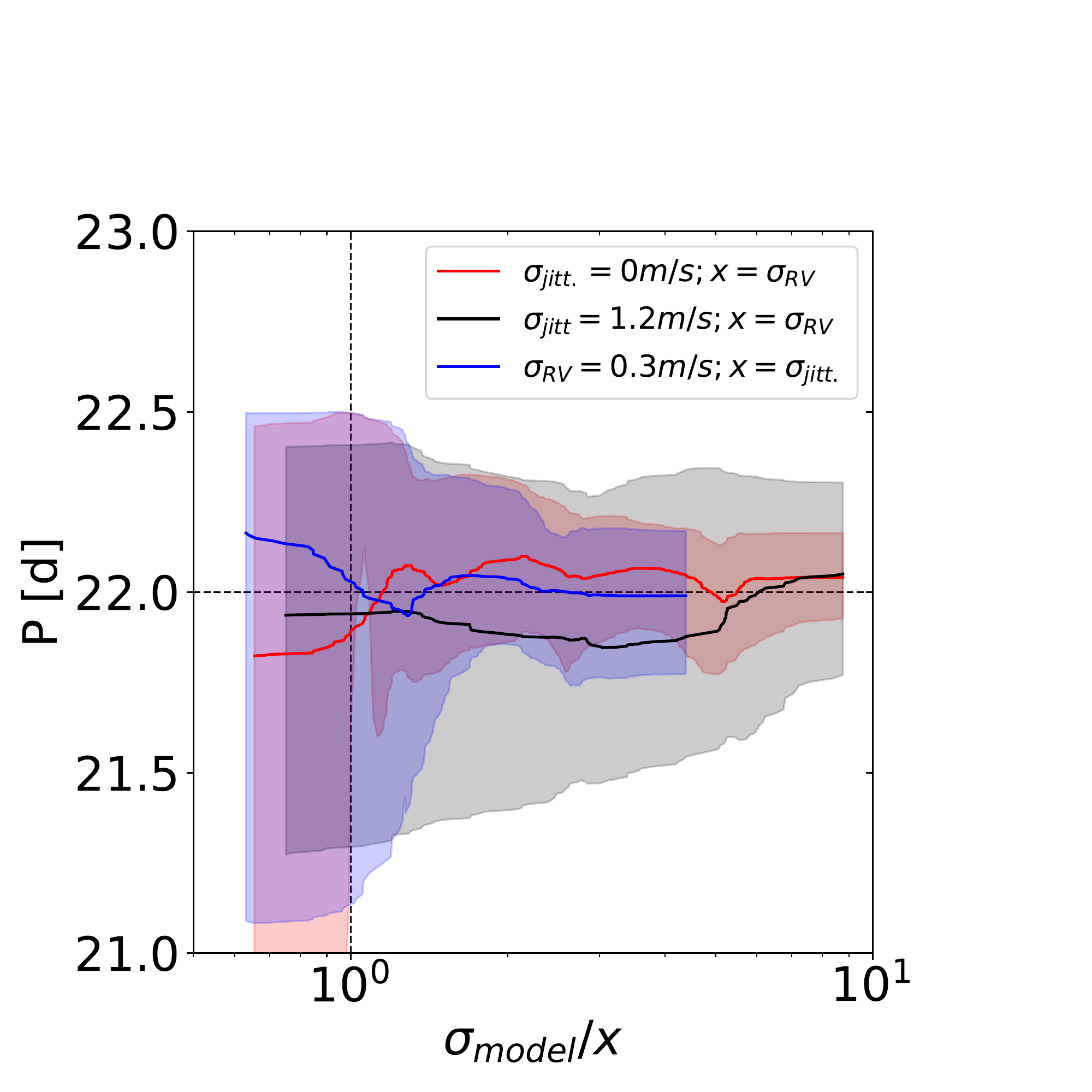}\\
    \caption{Relations of the derived GP parameters to the RV uncertainty. \textit{Top:}
    Derived GP length scale as a function of the ratio of the standard deviation of the stellar activity signal over the RV uncertainty or jitter. {Note that only GP-only fits that had at least moderate significance (with $\Delta\ln \mathcal{Z}>2.5$) compared to a white noise model that includes only jitter have been used.} The curves were computed by applying a median filter with a window size incorporating 50 simulations. The red curve assumes a fixed jitter of $\sigma_{\text{jitter}}=0\,\mathrm{m\,s^{-1}}$ and changes the ratio $\sigma_{\text{model}}/\sigma_{\text{RV}}$, the black curve assumes a fixed jitter of $\sigma_{\text{jitter}}=1.2\,\mathrm{m\,s^{-1}}$ and changes the ratio $\sigma_{\text{model}}/\sigma_{\text{RV}}$, and the blue curve assumes a fixed uncertainty of $\sigma_{\text{RV}}=0.3\,\mathrm{m\,s^{-1}}$ and changes the ratio $\sigma_{\text{model}}/\sigma_{\text{jitter}}$. The colored shaded area shows the appropriate inter-quartile range. The vertical dotted line represents a ratio of 1. The horizontal dashed line represents the simulated spot lifetime.  \textit{Bottom:} Analogous to the top plot, but for the GP rotation period.}
    \label{Fig: GP_hyperparam_jitter_unc}
\end{figure}

Figure~\ref{Fig: GP_hyperparam_jitter_unc} shows the derived GP hyperparameters $l_{\textnormal{GP}}$ and $P_{\textnormal{GP}}$ as a function of the ratio between the simulated stellar activity amplitude, $\sigma_{\text{model}}$, {and} the applied jitter, $\sigma_{\text{jitter}}$, {as well as} the RV uncertainty $\sigma_{\text{RV}}$. As can be seen, the derived length scale by the QP-GP is {sensitive} to this ratio. If the ratio between {the} activity amplitude {and} the RV uncertainty or jitter, respectively, is close to or below one, the GP length scale is generally determined to be larger. {However, when this ratio becomes greater than one, $l_{\textnormal{GP}}$ approaches a lower boundary whose absolute value is only about 25\% of the simulated lifetime.} In comparison, the derived GP rotation period for the same simulations show a rather stable behavior, independent of the ratio of the {simulated stellar activity amplitude $\sigma_{\text{model}}$ to the applied jitter $\sigma_{\text{jitter}}$ or to the RV uncertainty $\sigma_{\text{RV}}$. This showcases that the derived QP-GP length scale is rather sensitive to the data quality.}

{}

\section{Results of QP-GP fits to combined simulated activity and planetary signals}
\label{Sect: Results2}

{It is common practice to use GP components in fits to improve planetary parameters in the presence of stellar activity. However, there are different approaches to the treatment of the GP kernel, such as a free prior on the GP rotation period versus a prior constrained to the possibly known rotation period of the star. In the following, we investigate how the choice of GP prior affects the precision and accuracy of the planetary parameters.}

{In order to do so,} the grid of investigations is extended by injecting Keplerian signals with periods $P$ of 5.12\,d, 30.35\,d and 44\,d, hereafter referred to as KI5, KI30, and KI44. We note that KI44 has a period twice of the rotation period to investigate whether these signals can be recovered, and to which extent. The RV semi-amplitudes, $K$, of the injected Keplerian signals have been set to specific ratios to the standard deviation of the modeled stellar activity to investigate the effect of planetary signals that have higher or smaller amplitudes than the stellar activity signal. The investigated ratios ($K:\sqrt{2}*\sigma_{\text{activity}}$) include 2:1, 1:1, 1:2, and 1:5. The eccentricity $e$, mean anomaly $M$ and argument of periastron $\omega$ of the injected Keplerian were set to zero. {We used five different model types for the fits: (i) a Keplerian model that contains only white noise (K + W); (ii) a Keplerian model plus an unconstrained GP (K + GP I); (iii)  Keplerian model plus a length-scale-constrained GP (K + GP II), motivated by the results from \autoref{Sect: Results1}; (iv) a Keplerian model plus a GP with a constrained rotation period (K + GP III), assuming that the stellar rotation period is known through auxiliary measurements; and (v) a Keplerian model plus a GP that is constrained in terms of the length scale parameter and the rotation period (K + GP IV), which is the combination of the previous two assumptions.

        The priors used for the different model components are given in \autoref{Tab: Priors_GP_RV}.} Using four different stellar activity configurations, each having 12 different planet configurations and with 100 simulated data sets per final configuration, we end up with \num{4800} different data sets that need to be investigated by fitting five different combinations of models to them, resulting in a total of \num{24000} fits. {As for the activity-only analysis, we also used our results for an extensive model comparison, which is presented in \autoref{app:modelcomparison_activity+kepler}}

\subsection{{Accuracy}}

\begin{table*}
    \centering
    \begin{threeparttable}
        \caption{MSA of the derived orbital periods and RV semi-amplitudes and median of the derived eccentricities for different stellar activity configurations or all simulations. {Bold font highlights the best models.}}
        \label{Tab: MSA}
        \begin{tabular}{lccccc}
            \hline
            \hline
            \noalign{\smallskip}
            Configuration                                            & K+W    & K+GP I & K+GP II & K+GP III        & K+GP IV        \\
            \noalign{\smallskip}
            \hline
            \noalign{\smallskip}
            \multicolumn{6}{c}{ \em Orbital periods (MSA ($\xi$), smaller is better)}                                               \\
            \noalign{\smallskip}
            random, $t_{\text{spot}}=P_{\text{rot}}=22\,d$           & 0.468  & 0.403  & 0.498   & \textbf{0.402}  & 0.413          \\
            random, $t_{\text{spot}}=5P_{\text{rot}}=110\,d$         & 0.399  & 0.140  & 0.143   & \textbf{0.098}  & 0.099          \\
            two act. long., $t_{\text{spot}}=P_{\text{rot}}=22\,d$   & 0.479  & 0.356  & 0.405   & \textbf{0.320}  & 0.331          \\
            two act. long., $t_{\text{spot}}=5P_{\text{rot}}=110\,d$ & 0.381  & 0.181  & 0.180   & 0.173           & \textbf{0.168} \\
            all \num{24000} simulations                              & 0.425  & 0.239  & 0.261   & \textbf{0.206}  & 0.211          \\
            \noalign{\smallskip}
            \multicolumn{6}{c}{ \em RV semi-amplitudes (MSA ($\xi$), smaller is better)}                                            \\
            \noalign{\smallskip}
            random, $t_{\text{spot}}=P_{\text{rot}}=22\,d$           & 11.090 & 8.984  & 8.937   & \textbf{8.638}  & \textbf{8.638} \\
            random, $t_{\text{spot}}=5P_{\text{rot}}=110\,d$         & 47.044 & 35.578 & 36.087  & \textbf{30.939} & 31.046         \\
            two act. long., $t_{\text{spot}}=P_{\text{rot}}=22\,d$   & 10.429 & 7.497  & 7.986   & 7.491           & \textbf{7.255} \\
            two act. long., $t_{\text{spot}}=5P_{\text{rot}}=110\,d$ & 17.932 & 3.791  & 3.677   & \textbf{3.531}  & 3.547          \\
            all \num{24000} simulations                              & 17.932 & 9.747  & 10.054  & \textbf{9.183}  & 9.210          \\
            \noalign{\smallskip}
            \multicolumn{6}{c}{ \em eccentricity (median, true value is zero)}                                                      \\
            \noalign{\smallskip}
            random, $t_{\text{spot}}=P_{\text{rot}}=22\,d$           & 0.240  & 0.220  & 0.237   & \textbf{0.213}  & 0.221          \\
            random, $t_{\text{spot}}=5P_{\text{rot}}=110\,d$         & 0.323  & 0.076  & 0.077   & \textbf{0.060}  & \textbf{0.060} \\
            two act. long., $t_{\text{spot}}=P_{\text{rot}}=22\,d$   & 0.240  & 0.215  & 0.225   & \textbf{0.212}  & 0.216          \\
            two act. long., $t_{\text{spot}}=5P_{\text{rot}}=110\,d$ & 0.223  & 0.095  & 0.093   & \textbf{0.089}  & 0.090          \\
            all \num{24000} simulations                              & 0.251  & 0.143  & 0.146   & \textbf{0.121}  & 0.124          \\
            \noalign{\smallskip}

            \noalign{\smallskip}

            \hline
        \end{tabular}

    \end{threeparttable}

\end{table*}

{To investigate the accuracy of the derived planetary parameters we used} the median symmetric accuracy \citep[MSA; see for example chapter 4 in][]{Morley2018}. It is defined as

\begin{equation}
    \xi=100\,\,\biggl(\exp\Bigl(\mathcal{M}\bigl(\bigl|\ln(Q_i)\bigr|\bigr)\Bigr)-1\biggr)~~,
\end{equation}where $\mathcal{M}(\dots)$ is the median function and $\ln(Q_i)$ is the natural logarithm of the accuracy ratio \citep{Kitchenham2001, Tofallis2015}

\begin{equation}\label{Eq: acc_ratio}
    Q_i=\frac{x_i}{y_i},
\end{equation}

\noindent with $x_i$ being the estimated (predicted) value, which in our case is the median of the posterior distribution for a fit to simulation $i$, and $y_i$, which is the actual value of simulation $i$.

The MSA can intuitively be interpreted as a percentage error. This is because if the relative error is defined as always having the same direction, the value of the MSA is the same as the uncertainty of the median percentage \citep[][see in particular Eq. 12 and 13]{Morley2018}.

Table ~\ref{Tab: MSA} shows the MSA that we derived for the orbital periods and RV semi-amplitudes based on different activity configurations, as well as aggregated over all \num{24000} investigated fits. Given the configurations we investigated and all our fit results, we find that using a Keplerian simultaneously with a QP-GP does provide the best results in terms of the MSA metric. A Keplerian model without additionally taking into account the stellar activity does generally perform much worse and results in less accurate planet parameters. While the largest positive effect is observed by adding a QP-GP to model the activity signal, constraining the QP-GP to the rotation period (K + GP III) benefits the accuracy of the derived planet parameters in most of our investigated cases. On the other hand, constraining only the length scale of the QP-GP (K + GP II) does actually lead to less accurate planet parameters based on the MSA metric, even when compared to a fully unconstrained QP-GP (K + GP I). However, constraining both the length scale and the rotation period (K + GP IV) is almost indistinguishable from a prior that only constrains the rotation period.

While Table~\ref{Tab: MSA} provides a general summary of the MSA metric for different combined sets of simulations not taking into account the different amplitude ratios and period ratios between activity and planet signal, we show the MSA of the orbital period and RV semi-amplitude distinguishing between all simulated configurations graphically in Fig.~\ref{Fig: msa_periods} and Fig.~\ref{Fig: msa_amplitudes} in the appendix. Examining the MSA of the fitting results as a function of the period ratio between the stellar rotation period and the injected orbital period, we find that the MSA is generally smaller when the planet's orbital period is shorter than the rotation period. An additional but obvious result is that the larger the amplitude ratio between planet and activity, the more accurately the planet parameters are determined.

    {Using the MSA (or any other metric based on relative accuracy) to investigate the eccentricity for our circular ($e=0$) injected Keplerian signals is problematic {because it would lead to a division by zero in \autoref{Eq: acc_ratio}. Furthermore,} in contrast to the orbital period and RV semi-amplitude, the eccentricity is a parameter that is constrained between 0 and 1, and therefore the expected variance rather small.}
We {therefore} decided to investigate the median of the median eccentricities instead of the MSA for the eccentricity, since RV-semi amplitude and eccentricity are correlated anyway (i.e., a higher fit eccentricity results in higher RV semi-amplitudes). Furthermore, the derived eccentricity should be as close to zero as possible for any of our simulated cases, which makes it easy to compare the different simulations in a similar fashion as for the MSA metric. We visualize the median eccentricity for the 100 simulations of each configuration in Fig.\ref{Fig: med_ecc} while Table~\ref{Tab: MSA} shows the median eccentricities derived for each model class. The {main results from the investigation of} the eccentricity parameter basically reflect what has been observed for the MSA on the period and RV semi-amplitude, which is that K + GP III performs better in terms of planet parameter accuracy than the QP-GP models using the other prior distributions.

    {Besides, w}e find that the median eccentricity increases for an increasing amplitude of the stellar activity with respect to the planet RV semi-amplitude. {This highlights} that the eccentricity is rather difficult to constrain and very sensitive to the correlated noise in the data. {Since we find significantly smaller eccentricities when simultaneously fitting for the activity by using the QP-GP, } it is therefore especially helpful to use an activity model to account for the stellar noise. Nevertheless, even with the QP-GP we retrieve too high eccentricities for the simulations that consist of very prominent activity signals.

An additional effect that can be observed from  Fig.\ref{Fig: med_ecc} is that the derived eccentricity is generally the highest for the orbital period of 44\,d, which is twice that of the stellar rotation period (2:1). This particular period ratio was chosen to investigate how the activity affects the eccentricity of a planet signal when the period of the stellar rotation overlaps with the second harmonic of the injected planet. {For example,} \cite{Stock2020b} stated that the derived eccentricity of YZ~Cet~b is largely influenced by the stellar activity due to this effect. \cite{Anglada2010} showed that eccentric solutions for a planet can hide a second planet in a 2:1 mean motion resonance. Analogously, {our results show that this is} also possible for another periodic signal{, in our case the stellar rotation period,} overlapping with the second harmonic of the planet.

\subsection{{Precision}}

\begin{table*}
    \centering
    \begin{threeparttable}
        \caption{Median value of the standard deviation distance between the derived and simulated orbital periods, RV semi-amplitudes, and eccentricities
            for different stellar activity configurations or all simulations. {Bold font highlights the best models.}}
        \label{Tab: standard}
        \begin{tabular}{lccccc}
            \hline
            \hline
            \noalign{\smallskip}
            Configuration                                            & K+W            & K+GP I         & K+GP II        & K+GP III       & K+GP IV        \\
            \noalign{\smallskip}
            \hline
            \noalign{\smallskip}
            \multicolumn{6}{c}{ \em Orbital periods (median of standard deviation distance, smaller is better)}                                           \\
            \noalign{\smallskip}
            random, $t_{\text{spot}}=P_{\text{rot}}=22\,d$           & 1.120          & 0.679          & 0.915          & \textbf{0.654} & 0.800          \\
            random, $t_{\text{spot}}=5P_{\text{rot}}=110\,d$         & 5.667          & \textbf{0.794} & 0.846          & 0.803          & 0.808          \\
            two act. long., $t_{\text{spot}}=P_{\text{rot}}=22\,d$   & 1.057          & \textbf{0.672} & 0.777          & 0.730          & 0.704          \\
            two act. long., $t_{\text{spot}}=5P_{\text{rot}}=110\,d$ & 0.982          & 0.812          & 0.856          & 0.773          & \textbf{0.754} \\
            all \num{24000} simulations                              & 2.207          & \textbf{0.739} & 0.849          & 0.740          & 0.766          \\
            \noalign{\smallskip}
            \multicolumn{6}{c}{ \em RV semi-amplitudes (median of standard deviation distance, smaller is better)}                                        \\
            \noalign{\smallskip}
            random, $t_{\text{spot}}=P_{\text{rot}}=22\,d$           & 0.875          & 0.731          & 0.721          & \textbf{0.692} & 0.696          \\
            random, $t_{\text{spot}}=5P_{\text{rot}}=110\,d$         & \textbf{3.370} & 10.245         & 10.268         & 13.489         & 13.485         \\
            two act. long., $t_{\text{spot}}=P_{\text{rot}}=22\,d$   & 0.883          & 0.647          & \textbf{0.646} & 0.679          & 0.688          \\
            two act. long., $t_{\text{spot}}=5P_{\text{rot}}=110\,d$ & 0.748          & 0.723          & 0.742          & 0.709          & \textbf{0.706} \\
            all \num{24000} simulations                              & \textbf{1.469} & 3.086          & 3.094          & 3.892          & 3.894          \\
            \noalign{\smallskip}
            \multicolumn{6}{c}{ \em eccentricity (median of standard deviation distance, smaller is better)}                                              \\
            \noalign{\smallskip}
            random, $t_{\text{spot}}=P_{\text{rot}}=22\,d$           & 2.546          & 2.519          & 2.586          & \textbf{1.938} & 2.275          \\
            random, $t_{\text{spot}}=5P_{\text{rot}}=110\,d$         & 4.590          & 2.187          & 2.247          & \textbf{1.589} & \textbf{1.589} \\
            two act. long., $t_{\text{spot}}=P_{\text{rot}}=22\,d$   & 2.536          & 2.305          & 2.336          & \textbf{2.081} & 2.352          \\
            two act. long., $t_{\text{spot}}=5P_{\text{rot}}=110\,d$ & 2.692          & 2.267          & 2.679          & 2.106          & \textbf{1.909} \\
            all \num{24000} simulations                              & 3.091          & 2.320          & 2.462          & \textbf{1.929} & 2.031          \\
            \noalign{\smallskip}

            \noalign{\smallskip}

            \hline
        \end{tabular}

    \end{threeparttable}

\end{table*}

The MSA does not provide any information about the individual model uncertainties and confidence intervals. For example, a prediction that is close to the actual value and by that is accurate can have underestimated uncertainties (an overestimated precision) and therefore be many standard deviations away, while a less accurate estimate with larger error bars does include the true value within less standard deviations and by that may provide a better representation of the precision of the planetary parameters. Therefore, we investigated the distance  between injected and retrieved median posterior parameters in terms of standard deviations. We show an overview of these numbers for the different models and activity configurations in Table~\ref{Tab: standard}, while Figs.~\ref{Fig: sigma_diff_period}, ~\ref{Fig: sigma_diff_amplitude}, and \ref{Fig: sigma_diff_ecc} graphically show our results as a function of amplitude and period ratio between the activity and injected planet signal. Examining this metric, we find that while GP Prior III and IV do again perform the best in the majority of cases, there are configurations where a completely unconstrained, or in one case even a Keplerian without a simultaneous GP, do result in the smallest standard deviation distance between injected and retrieved planet parameters.

In particular, the configuration with a random spot distribution and a spot lifetime of five rotation periods stands out. Here, the QP-GP leads to significant deviations between injected and obtained RV semi-amplitude, as the median of the standard deviation distance for the 100 ensembles is as large as 10 or even higher. However, {considering} the derived orbital periods or the eccentricities under the influence of the very same activity configuration, the fits including the GP models are again clearly superior.

    {Nevertheless, the examination of all three investigated planetary parameter results (the distinction between period and amplitude ratio, the scatter of the standard deviation distance and the consistency of the derived standard deviation distance by the number of outliers for a given prior), leads us to conclude that a QP-GP that is constrained in its rotation parameter to the stellar rotation period, but is not constrained in its length scale parameter, has the best performance when it comes to realistic uncertainties of the parameter estimates.}

\section{Discussion}
\label{Sect: Discussion}

\subsection{QP-GP rotation and length scale hyperparameter}

{In the first part, we investigated the relation of the QP-GP rotation period and length scale hyperparameters to the stellar rotation period and spot lifetime using simulated activity data sets. We identified three distinct shapes in the $l_{\textnormal{GP}}$ versus $P_{\textnormal{GP}}$ diagram of the posterior distributions for the rotation period and length scale, which are directly linked to the simulated rotation period and spot lifetime. Further, we found that the spot distribution also influences the determination of the hyperparameters.}

For spot lifetimes of similar magnitude to the rotation period, the posterior distribution contains a large number of solutions that are consistent with any rotation period that is permitted by the prior, but these solutions tend to have a very small length scale. In such cases, we often find that only a small fraction of the posterior solutions are consistent with a higher length scale. Only those solutions that are not part of the plateau as described in \autoref{Sect: alpha_vs_prot} are then also consistent with a {unique} rotation period{, which is, as we later showed, related to the simulated stellar rotation period.}

    {These results suggest that by considering $l_{\textnormal{GP}}$ and $P_{\textnormal{GP}}$ together, a rotation period can be determined by the QP-GP in most cases, even if the star spot lifetime is on the order of the rotation period or less. Consequentially, for the detection of a QP signal in time series data, it can be advantageous to define a lower limit for the QP-GP length scale. The presented empirical approach, to fit first an unconstrained QP-GP and then repeat the fit with a more constrained QP-GP length scale, is versatile and can be used on any time series data with correlated noise. The reason for redoing the fit and excluding the plateau is that the nested sampling algorithm is then more efficient at identifying posterior solutions with higher length scale at the possible rotation period, which otherwise would not have been detected.}

    {However, while based on our simulated data and configurations the simulated stellar rotation period and the rotation period hyperparameter of the QP-GP generally match, there are still some exceptions.} For example, for certain spot distributions, especially in the case of two active longitudes, it is possible that most or all posterior solutions of a QP-GP fit are related to a higher-order harmonic of the rotation period. For the simulations with two active longitudes, the QP-GP has had difficulties in detecting and modeling the fundamental period of the stellar rotation, especially if the spot lifetimes were rather small in comparison to the stellar rotation period, which is a result of the much more dynamic nature of spot formation and decay for such cases.

While the rotation hyperparameter provides a good description of the simulated rotation period, the picture is somewhat different when comparing the QP-GP length scale with the simulated spot lifetime. Here we find a clear {dependence} as longer spot lifetimes result in larger QP-GP length scales, but the absolute value of the length scale does not reflect the true simulated spot lifetime. {Most importantly} we find that the QP-GP length scale is sensitive to the spot distribution and to the quality of the data, which means, jitter and RV error bars have an influence on the estimated median of the spot lifetime. For our simulations, poorer data quality led to higher values for the length scale of the QP-GP.

{Such a varying relation between the determined length scale parameter and the simulated spot lifetime is contradictory to the results from other studies such as \cite{Perger2021} or \cite{Nicholson2022}, who find strong correlations between the simulated spot lifetime and the determined length scale parameter.} There is a clear difference between a parameter that gives the well-defined rotation period (differential rotation left aside) and a parameter that captures something like the ``typical'' lifetime of star spots. {The GP kernel that we use models the lifetime in the sense of the ``decay time'' of the signal (in the sense of 1/e-time), which is different from the setup of our simulations where the lifetime is the time span between the emergence and disappearance of the spot. This leads to about a factor of 2 between the input parameter and the measured decay time of the GP and agrees with the results of \cite{Perger2021}, who use the same configuration to simulate the activity as we do. The factor of 4, which we obtain for the well-behaved cases with spot lifetimes five and ten times the rotation period, results from the additional factor of 2 in the exponent of our kernel. However, with the wide grid of our simulations, we show that this is only the optimal case and that the ratio between the simulated and determined lifetime is not a constant. The positional correlation of star spots, for example, due to active longitudes, can significantly affect the properties of the activity signal and by this the derived QP-GP length scale, as can the uncertainty or jitter of the data set.}

The co-variance function of the QP-GP is not based on an exponential function, but has a Gaussian shape. A number of studies \citep[e.g.,][]{Rajpaul2015, Foreman-Mackey2017, Espinoza2018} argue that this does not have a severe influence on the derived length scale of the QP-GP. This is due to the fact that the two secondary diagonals that are exactly one rotation period away from the main diagonal are especially important regarding the derivation of the length scale parameter; how fast they fall off toward the corners is mostly irrelevant. {But,} this also suggests that not too much physical importance should be given to the length scale parameter.

In contrast to the period hyperparameter, it seems reasonable to regard the length scale hyperparameter as an ``effective model parameter,'' which may be related to stellar physics only very indirectly. One could of course do simulations assuming a certain lifetime for the star spots and then calibrate this parameter \citep[see for example][]{Perger2021}, but it is not clear whether this relationship between QP-GP length scale and lifetime will still be valid if basic assumptions of the model or the data change, which is why we refrain from attempting to perform such a calibration.

A good example are the {simulated} cases with two active longitudes provided {in this work}. Here, the RV variations seem to be stable over all observed rotations of the star, although the simulated spot lifetime is only half of the simulated rotation period. In such cases, an empirically determined lifetime would measure the stability of the magnetic field configuration, or perhaps of active regions controlled by the magnetic field, but not of individual spots. Such cases are not far-fetched; real-world examples for such stars include HD~166435 \citep{Queloz2001} or AD~Leo \citep{Kossakowski2022}. In addition, we found that RV uncertainty and white noise jitter do also affect the derived length scale. It is therefore not easily possible to conclude from the data what importance one should attach to the QP-GP length scale parameter. In most cases, this is not a problem for the field of exoplanets, since we are primarily interested in planet detection and characterization. Therefore, based on our results, we advise against giving too much importance to the absolute value of the QP-GP length scale as it is {not necessarily a reliable measure of} the spot lifetime, even for our simplified simulations consisting of spots that all have exactly the same lifetime.

\subsection{Best practices: Accurate planet parameters from activity-contaminated RV data using QP-GP models}

Based on the MSA and the median eccentricity, we identify several informative key findings. {Not surprisingly, but nevertheless most importantly, adding a GP to the modeling does improve the overall accuracy of the derived planet parameters, independently of the used priors. This is true even in cases where the activity signal is weaker than the planetary signal.} In the majority of our simulated cases, one of the four combinations of a Keplerian model together with a red-noise model in the form of a QP-GP results in smaller MSA compared to a Keplerian model that only included a jitter term assuming white noise.

    {The choice of the GP Prior, however, has only little influence on accuracy and precision of the determined planetary parameters.} Based on our analysis, we find that the best performance of the QP-GP is achieved when the GP is constrained to the stellar rotation period (GP Prior III), sometimes even in combination with the length scale (GP Prior IV). It can therefore be essential to take photometry or other data into consideration to estimate the stellar rotation period {in order to use this as} physical prior knowledge in the modeling of the RV data with the QP-GP. {In contrast to the findings on the stellar activity-only analysis, } we find that a constraint solely on the length scale is not beneficial, but actually reduces the accuracy of the derived planet parameters in comparison to an unconstrained prior. {A reason for this could be} the reduced flexibility of the QP-GP model. The length scale constrained prior (GP Prior II) should therefore only be used to detect rotation periods from time series data, but not for the simultaneous modeling of activity and planetary signals.

\section{Conclusions}
\label{Sect: Conclusion}

In this work we have investigated the properties and advantages of QP-GP models for the detection and characterization of exoplanet signals from RV data. For this purpose, we first generated synthetic RV data that contain stellar rotational signals. In order to do this, we modeled the surface of a rotating star based on different spot distributions and spot lifetimes, and generated the RV data from these models, taking sampling and instrumental precision into account. Based on these simulations and our results, we come to the following conclusions and recommendations: \newline
\begin{enumerate}
    \item  {The rotation period hyperparameter of the QP-GP is in agreement with the stellar rotation period.}
    \item  {The QP-GP length scale hyperparameter correlates with the star's spot lifetime,} but there is no 1:1 connection between the absolute values. In addition, spot distribution and data quality can significantly affect the derived QP-GP length scale. Therefore, we advise against using the QP-GP length scale parameter as an estimate of the typical spot lifetime on the stellar surface.
    \item {To determine an unknown stellar rotation period with the QP-GP kernel, the 2D posterior distribution of the rotation period and the length scale hyperparameter provides crucial information. Constraining the length scale of the QP-GP in the case of a plateau of samples that do not favor any specific rotation period significantly increases the efficiency of detecting the correct value of the rotation period.}
    \item  To derive precise and accurate planet parameters, we strongly recommend fitting for the stellar activity and not ignoring it, even if the stellar activity amplitude is small compared to the planetary amplitude.
    \item  For our simulations, we derived the best results in terms of accuracy and precision regarding the planetary parameters either when the QP-GP was constrained to the rotation period or when it was constrained to the rotation period and had a minimum length scale. However, constraining only the length scale resulted in less accurate planet parameters. This prior should therefore not be used for simultaneous fitting of planet and activity signals, but only for the detection of a rotation period from time series data.
\end{enumerate}

Simulations like those presented in this work are essential for assessing the performance and properties of the models that are used for real data. Especially when it comes to non-parametric noise models such as GP regression, such analyses provide a better understanding of possible pitfalls and considerations that need to be taken into account. While the recommendations based on our astrophysically motivated simulations can certainly help us better understand GPs and the QP-GP in particular, every system can show individual characteristics, not all of which could be considered in the analysis presented in this work. {Nevertheless, in a large majority of our simulations, we were able to correctly identify stellar rotation periods with the QP-GP, and we obtained better planet parameters by using it to model stellar activity.}

\begin{acknowledgements}
    {All authors acknowledge support from the Deutsche Forschungsgemeinschaft (DFG) through the Research Unit FOR2544 ``Blue Planets around Red Stars'', in particular project numbers RE 2694/4-1, RE 2694/8-1 and KU 3625/1-1.
        The authors thank Stefan Dreizler and Manuel Perger for fruitful discussions on Gaussian process regression for radial velocity data.
        \textit{Software:} \texttt{astropy} \cite{AstropyCollaboration.2018}, \texttt{dynesty} \citep{Speagle2019},
        \texttt{scipy} {\citep{Virtanen.2020}}, \texttt{numpy} \citep{Oliphant.2006}, \texttt{matplotlib} \citep{Hunter.2007},  \texttt{pandas} \citep{Thepandasdevelopmentteam.2020}, \texttt{seaborn} \citep{Waskom.2020}, \texttt{radvel} \citep{Fulton2018}, \texttt{george} \citep{Ambikasaran2015}, \texttt{juliet} \citep{Espinoza2018}.}
\end{acknowledgements}

\bibliographystyle{aa} 
\bibliography{references}

\begin{appendix}

    \section{Conversion between $l$ and $\alpha$}
    \label{app:lengthscale}

    The QP-GP kernel that we use in this study is parametrized in \texttt{juliet} according to \cite{Espinoza2018} and is given as

    \begin{equation}
        k(\tau)=\sigma^2_{\textnormal{GP}}\exp{(-\alpha_\textnormal{GP}\tau^2-\Gamma\sin^2{(\pi\tau/P_{\text{rot}}}))},
    \end{equation}where $\sigma_{\text{GP}}$ is the amplitude of the GP component given in parts per million (ppm) for photometric data or $\mathrm{m\,s^{-1}}$ for RV data, $\Gamma$ is the amplitude of the GP sine-squared component and is dimensionless, $\alpha$ is the inverse length scale of the GP exponential component given in d$^{-2}$, $P_{\text{rot}}$ the period of the GP's component given in d, and $\tau$ is the time lag.

    For the fitting of the above kernel, \texttt{juliet} makes use of the python package \texttt{george}, in particular a multiplication of an exponential-squared kernel and an exponential-sine-squared kernel \footnote{See kernel definitions here: \url{https://george.readthedocs.io/en/latest/user/kernels/}}.
    The exponential-squared kernel is defined in \texttt{george} as $\exp(-r^2/2)$, where $r$ corresponds to $\tau$ in the \texttt{juliet} paper. The length scale in \texttt{george} is proportional to $\exp(-r^2/(2\Theta))$ \footnote{See how $\mathrm{k}_1$ is defined in the \texttt{george} tutorial here: \url{https://george.readthedocs.io/en/latest/tutorials/hyper/}}. Within \texttt{juliet}, the inverse of this length scale $\Theta = 1/\alpha$, is defined, so the kernel being fitted by \texttt{juliet} is $\exp(-\alpha \cdot r^2/2)$ and not $\exp(-\alpha \cdot r^2)$ as written in Eq.~8 of \cite{Espinoza2018}. The conversion from the inverse length scale $\alpha$ to $l$ is therefore given as

    \begin{equation}
        l=\sqrt{\alpha}^{-1}.
    \end{equation}

    For the convenience of the reader, we have transformed the parametrization of the GP inverse length scale $\alpha$ given in d$^{-2}$ as provided within \texttt{juliet} to the more intuitive GP length scale $l$ given in days within this work.

    \clearpage

    \section{Model comparisons using the Bayesian evidence}
    \label{app:modelcomparison}
    {In the main body of this work, we examined the connection of the hyperparameters of the QP-GP to the properties of the stellar activity and how the prior choice affects the modeling by the GP. Beyond that, the nested sampling algorithm we used for the fits allows us to investigate in more detail how the prior choice also affects the detection efficiency -- in terms of both stellar activity signals and planetary signals. }

    \subsection{When only stellar activity is present}
    \label{app:modelcomparison_onlyact}

    {First, we tested the GP models against a pure white noise model and a Kepler model, which corresponds to a non-detection and a false detection, respectively. }
    We investigated whether the QP-GP model is the favored model for the stellar {activity} signal in all cases, as one would expect. {Further, we considered the combination of a QP-GP component and a Keplerian component as a method for assessing the presence of additional signals in unexplored data sets, and whether the choice of the GP priors can facilitate false positives in doing so.}

    \subsubsection{Single-model comparison}
    We {compared} the single models GP Prior I, GP Prior II, GP Prior III, GP Prior IV, K11, K22, and W, which are given in Table~\ref{Tab: Priors_GP_RV}. The priors for the Keplerian models assume that there is some information from significant peaks in the periodogram, whose origin, however, whether planetary or due to activity, is not yet determined. With this comparison, we intend to investigate whether the Bayesian evidence is an additional reliable tool for distinguishing between stellar activity and planetary signals in cases of limited information regarding the signals in the data, {for example when there is no known photometric rotation period or no compelling evidence from the stellar activity indicators.}

    \begin{figure}
        \centering
        \includegraphics[width=8cm]{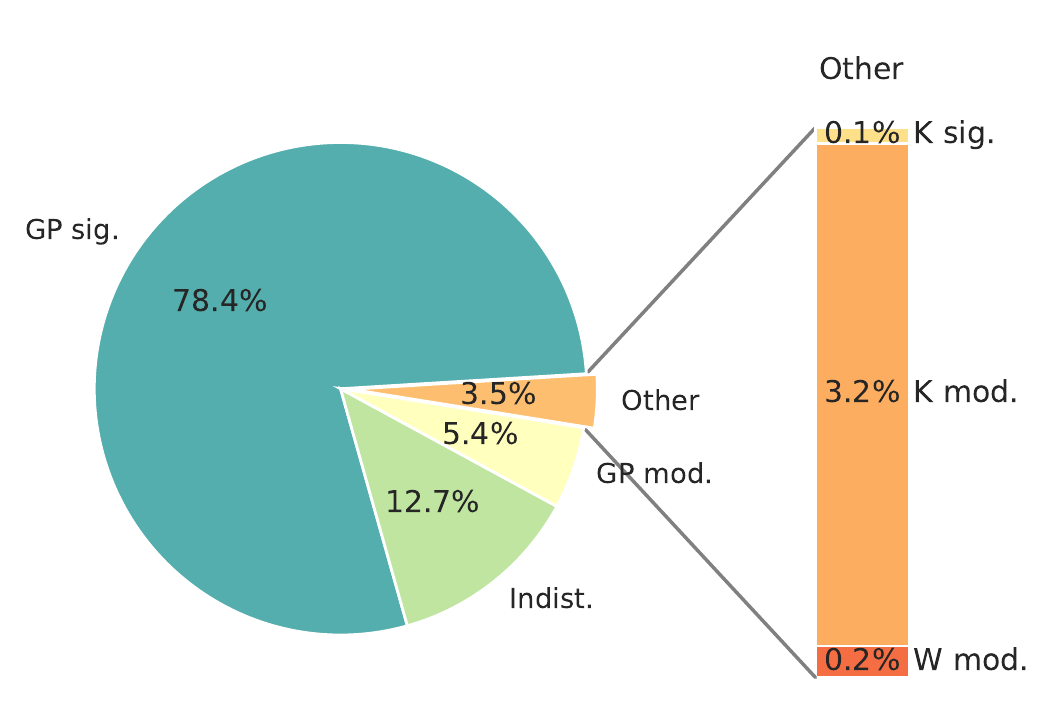}
        \caption{{Single-model comparison of activity-only data.} We show the percentage of significant (sig.), moderately favored (mod.), and indistinguishable (indist.) models based on all ten investigated stellar activity configurations, with each configuration consisting of 100 synthetic RV data sets (a total of \num{1000} data sets).
        }
        \label{Fig:pie_chart_onlyact}
    \end{figure}

    \begin{figure}
        \centering
        \includegraphics[width=\columnwidth]{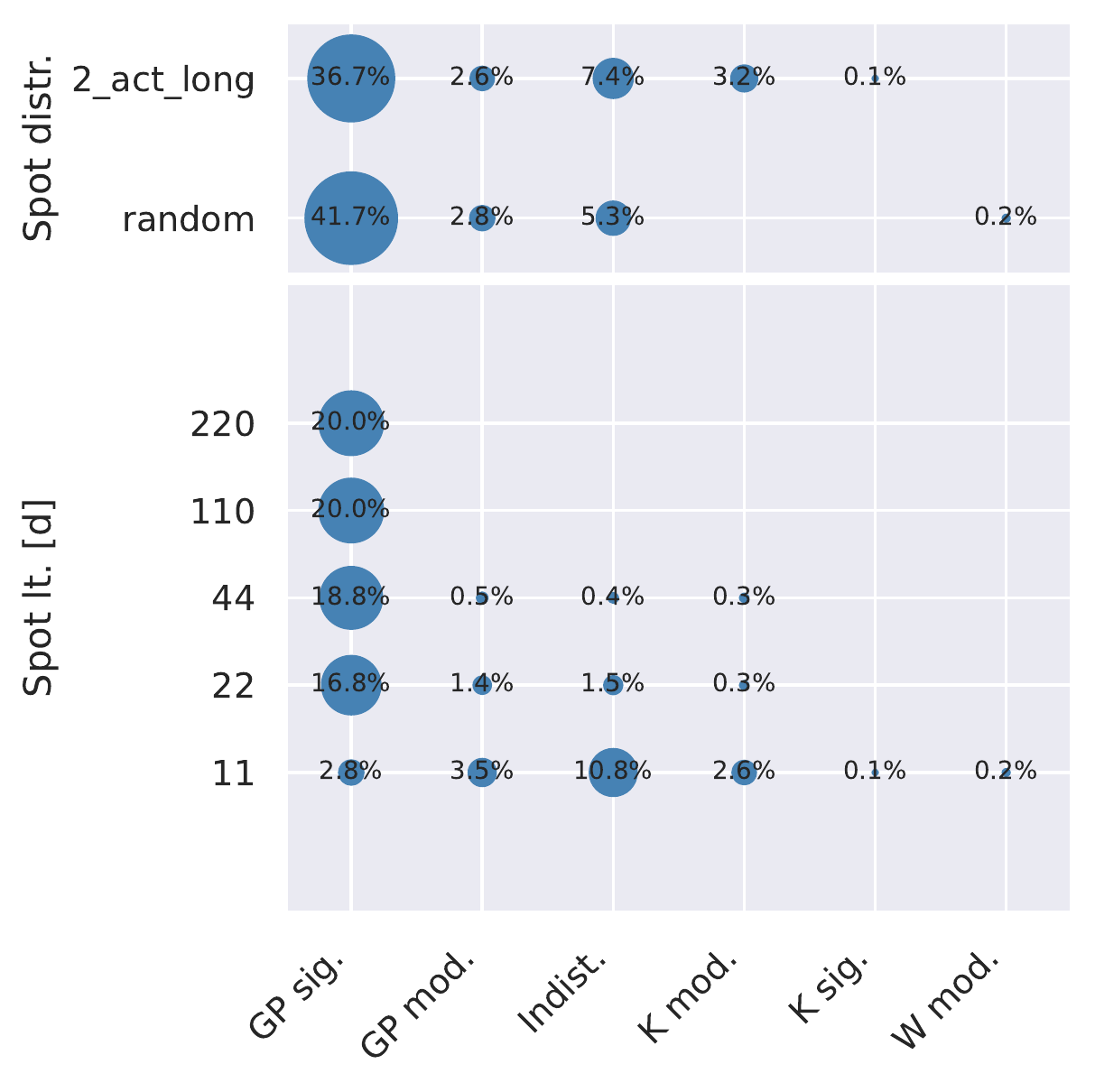}
        \caption{{Single-model comparison of activity-only data. Shown is a breakdown of the values from Fig.\,\ref{Fig:pie_chart_onlyact} into the different activity configurations. The percentages, also indicated by the point sizes, sum up to 100\% for each configuration parameter category highlighted by the gray-shaded areas.}}
        \label{fig:correlationplot_onlyact}
    \end{figure}

    \begin{figure}
        \centering
        \includegraphics[width=\columnwidth]{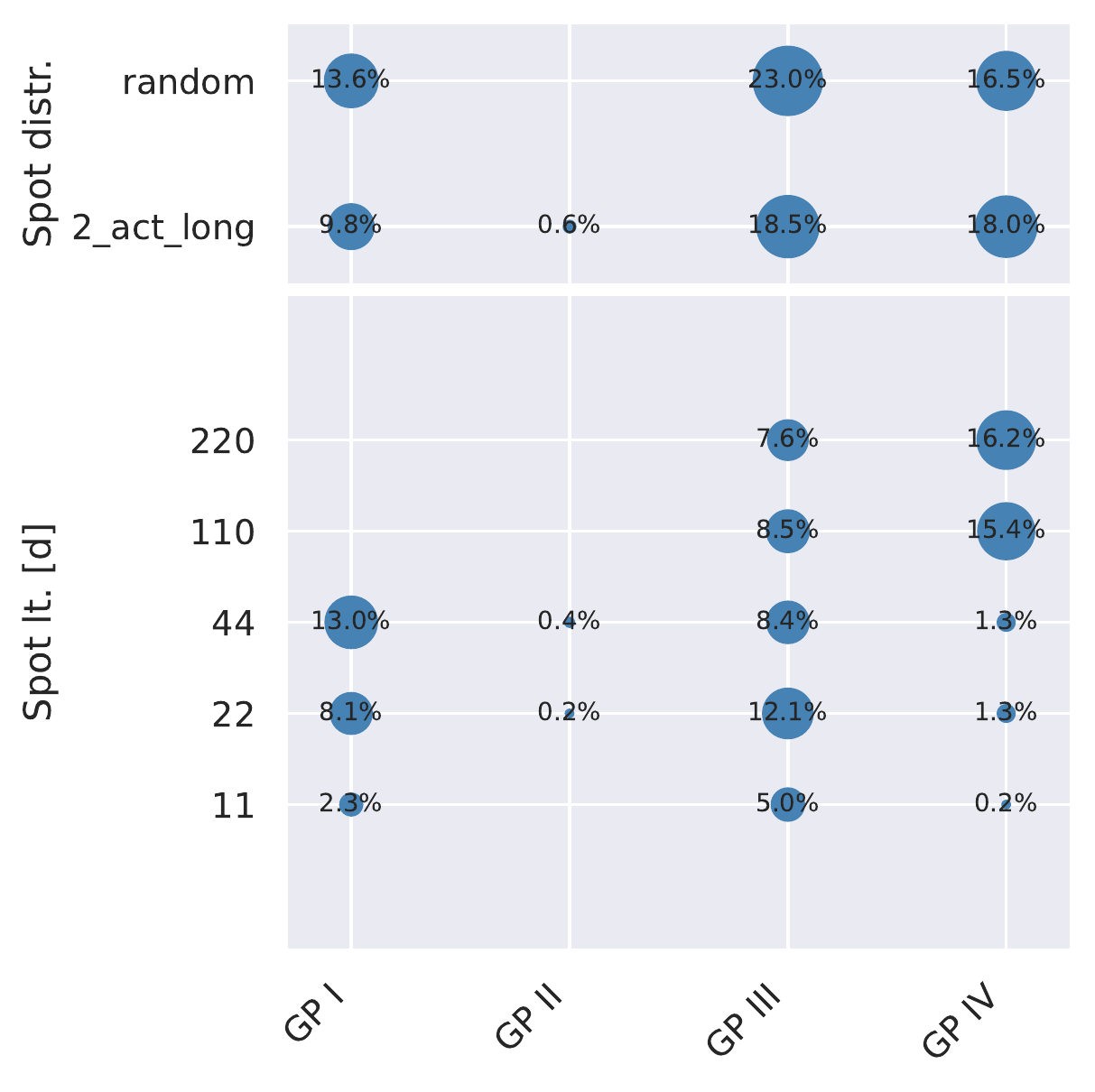}
        \caption{{Single-model comparison of activity-only data. Shown is the breakdown to the individual GP priors for the significant or moderately favored GP models from Fig.\,\ref{Fig:pie_chart_onlyact}. Note that this is only a qualitative comparison based on the highest evidence compared to the best non-GP model. The percentages, also indicated by the point sizes, sum up to 100\% for each configuration parameter category highlighted by the gray-shaded areas.}}
        \label{Fig:prior_comparison_onlyact_single_model}
    \end{figure}{}

    {In doing so, we investigated whether one model family (i.e., GP model, Keplerian model, or white noise model) performed better than another for a synthetic RV data set, independent of the chosen prior distributions. This means that, as soon as a GP model was better than all K models, the GP model was classified as the winning model for the simulation. This is important because the choice of prior affects the derived evidence, which must be taken into account when using Bayesian evidence to compare models.} If two models of different classes performed similarly well ($\Delta\ln \mathcal{Z}<2.5$) they were classified as indistinguishable. The results of this analysis are visualized in the pie chart in Fig.~\ref{Fig:pie_chart_onlyact}.

    For the majority of the data sets, about 78\%, a QP-GP-only model has been the significantly favored model ($\Delta\ln \mathcal{Z}>5$) while for about 5\% it was at least moderately favored ($\Delta\ln \mathcal{Z}>2.5$). In about 13\% of the cases, the two best models of different class were indistinguishable. It is interesting that we find that for about 4\% of the simulated stellar activity time series data sets, a Keplerian or white noise model has been favored for the activity signal by the log-evidence metric. Most of these cases consist of moderately favored Keplerian models.

    Almost all cases where a non-GP model was indistinguishable or performed better
    than the GP model for the activity signal
    belong to simulations where the spot lifetime was small compared to the simulated stellar rotation period. Figure~\ref{fig:correlationplot_onlyact} shows the percentages of the winning models between the different investigated activity configurations. Short spot lifetimes with respect to the rotation period lead to correlated noise on shorter timescales, which can also be well described by a simple white noise model. This explains why many indistinguishable cases and all cases where the white noise model is favored occur for a spot lifetime of 11\,d. Especially in the case of a random spot distribution with spot lifetimes smaller than the rotation period, a W model has often been indistinguishable compared to the QP-GP model.

    In the case of two active longitudes, it has been mainly the Keplerian model that competed with the GP regarding the modeling of the stellar activity, resulting in a higher single-digit percentage of indistinguishable solutions and a few percent of a moderately favored, or in a few cases even significantly favored, Keplerian model with respect to the GP models.
    The fact that a Keplerian model has been indistinguishable, or in a few cases even been moderately favored, compared to a GP model for these activity cases shows that {caution is advised when} using model comparison: Although in this first experiment the evidence has shown to be reliable in the majority of investigated cases, there are still a few cases where the evidence prefers the wrong model. This highlights the importance of additional investigations, auxiliary data, and simulations such as those presented in this work to verify the nature of a signal, in addition to the Bayesian evidence. It is interesting to note that most of the cases where the Keplerian model has been favored were simulations where the $l_{\textnormal{GP}}$ versus $P_{\textnormal{GP}}$ diagram showed a bar shape. Since we used non-eccentric Keplerians, this means that in cases of rather coherent stellar activity a sinusoidal fit can do a decent job at modeling stellar activity. In such cases the flexibility of the QP-GP may not be required.

    For the cases where the GP models were significantly or moderately favored over a Keplerian or white noise model, we investigated which GP prior actually performed best. In Fig.\,\ref{Fig:prior_comparison_onlyact_single_model} we present {a quantitative} breakdown of the {winning} GP models with respect to the activity properties. First, it can be seen that Priors III and IV make up the bulk of the models {that performed better than the Keplerian or white noise models}. Only for shorter spot lifetimes, the unconstrained GP Prior I performs similarly well. This is plausible because the correlated noise on shorter timescales associated with the shorter spot lifetimes can be better modeled by the more flexible GP model. This is also reflected in the generally better performance of the GP Prior III compared to the less flexible GP IV. Solely constraining the length scale hyperparameter as in the GP Prior II, on the other hand, seems to not have any advantage over the other priors and accounts for only a negligible proportion of the winning models.

    \subsubsection{Mixed-model comparison}

    \begin{figure*}
        \centering
        \includegraphics[width=0.75\textwidth]{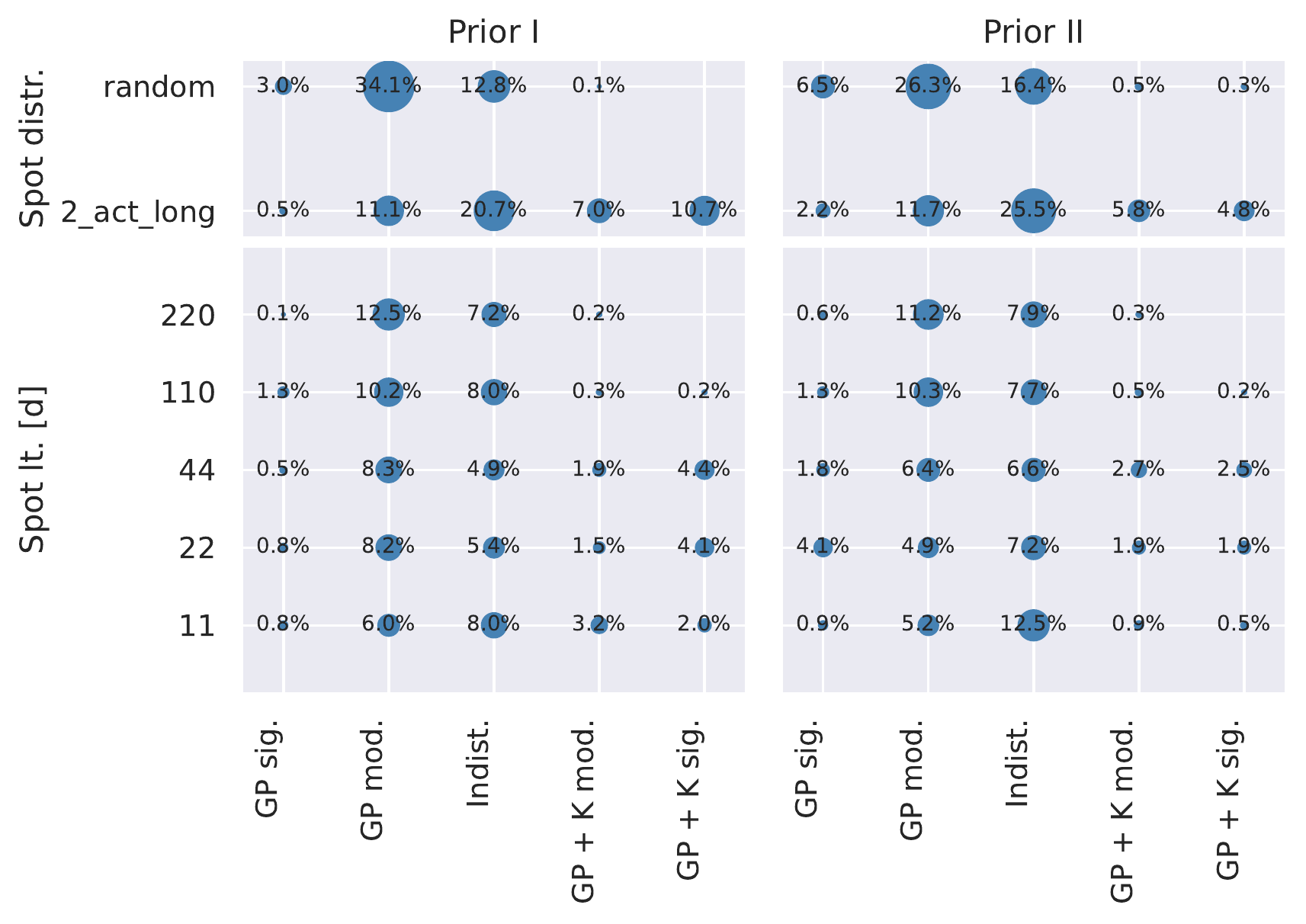}
        \includegraphics[width=0.75\textwidth]{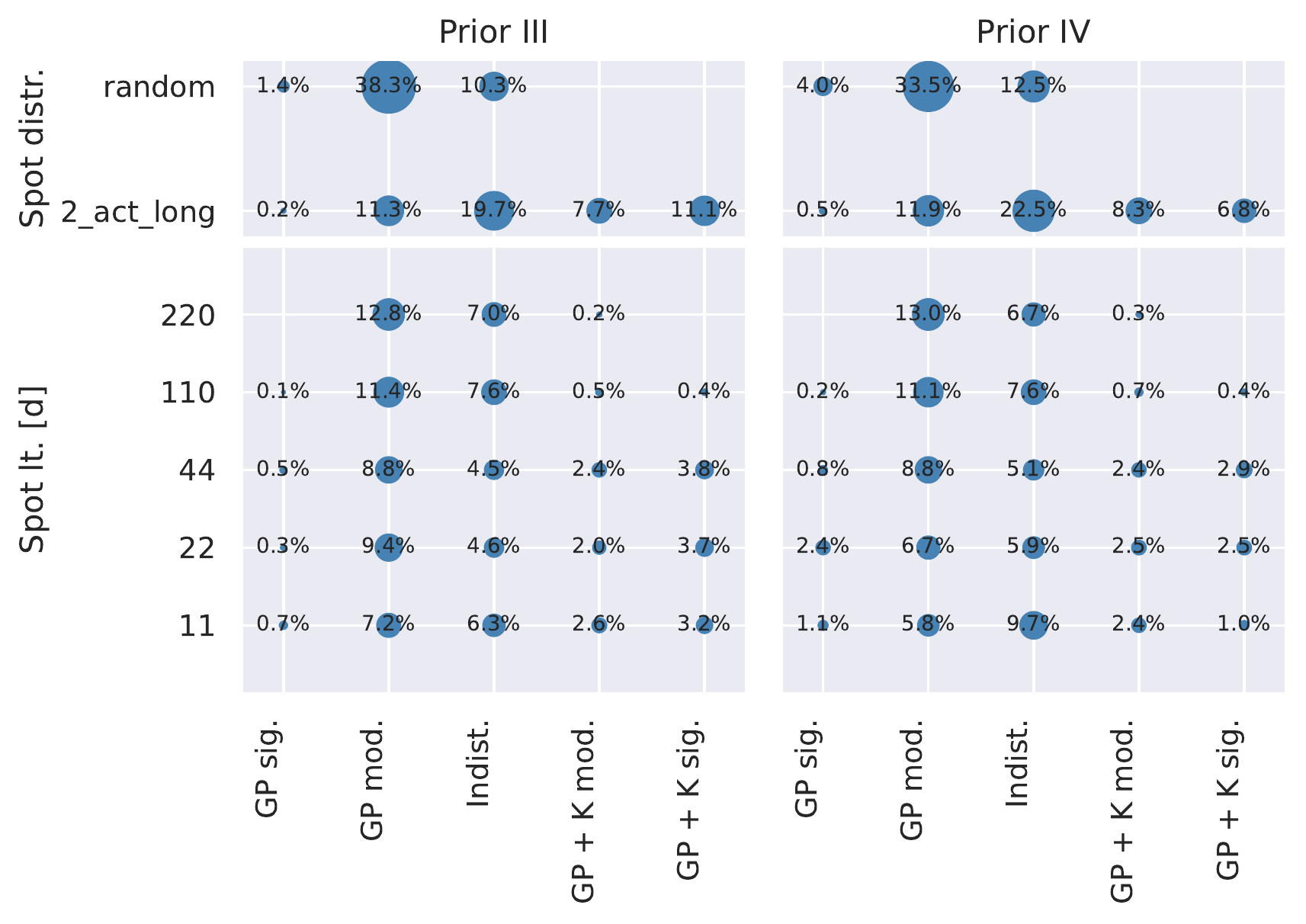}
        \caption{{Mixed-model comparison of activity-only data. Each panel shows the fraction of significant (sig.), moderately favored (mod.), and indistinguishable (indist.) models for one of our GP prior settings. In each plot, the percentages, also indicated by the point sizes, sum up to 100\% for each configuration parameter category highlighted by the gray-shaded areas.}}
        \label{Fig:prior_comparison_onlyact_mixed_model}
    \end{figure*}{}

    The Bayesian evidence is commonly used to justify whether adding an extra Keplerian is better than just fitting for white or red noise -- in other words, whether a bona fide planet has been detected. {Here we compare the QP-GP-only models against mixed models comprising a Keplerian and a QP-GP component (results summarized in Fig.~\ref{Fig:prior_comparison_onlyact_mixed_model}). The goal was to investigate} whether certain prior choices favor the addition of an additional sinusoidal or Keplerian component by the Bayesian evidence, even though only activity signals are included in the data. In particular, we used the four different QP-GP priors motivated before and fitted them together with constrained Keplerian models whose priors are given in Table~\ref{Tab: Priors_GP_RV}. In {the first} scenario, {we assumed that} auxiliary data did not provide any constraints on the rotation period; therefore, a wide uninformative prior for the GP was chosen and the planet hypothesis of a significant peak in the periodogram was checked by the constrained Keplerian (GP I + K) to the rotational signal. Another scenario {we tested} is whether it would be possible within the GP framework to distinguish a planet in a 1:1 or 1:2 spin-orbit resonance, which motivates using the other three more constrained GP priors. It is not entirely clear how reliable the log-evidence metric and the QP-GP kernel are in such complex real-life cases with discrete sampling and various instruments covering a wide wavelength range.

    For all four QP-GP priors studied, we find a number of simulations where an additional Keplerian, respectively sinusoidal, component has been at least moderately favored. This is particularly the case when the simulated configuration has two active longitudes. For this configuration, the K11 model is highly significant when added to the QP-GP model, even if the QP-GP is not constrained to the fundamental period of 22\,d. This shows that while the QP-GP is capable to also model harmonics of the rotation period, it is not very efficient in doing so, which is why an extra component is required by the Bayesian evidence. A similar conclusion was reached by \cite{Perger2021} based on the auto-correlation function, which motivated the authors to use a second periodic component that is added to the standard QP-GP resulting in their QPC. From our investigations, we find that an additional sinusoidal or Keplerian component was not as necessary if the QP-GP length scale was restricted to exclude small values, such that the GP does not include the posterior solutions from the plateau observed in the middle plot of Fig.~\ref{Fig: shapes}.
    Nevertheless, our results regarding the harmonic of the rotation period at 11\,d, and the need for an additional model component to model it, strongly favors the use of a GP in the form of the QPC \citep{Perger2021} or the sum of multiple harmonic oscillator kernels as for example in \cite{Kossakowski2021} It shows that the standard QP-GP in the form that has been applied in this work lacks flexibility to model stellar activity in all its aspects.

    If we leave the K11 model for the second harmonic aside, and concentrate on the K22 model, we still find a number of simulations where an additional Keplerian component at the rotation period is favored by the Bayesian evidence in addition to the QP-GP model. This might indicate that there exists a stable periodic component for certain stellar activity configurations, which can lead to better Bayesian evidence when an extra sinusoidal model fits this stable component while the QP-GP accounts simultaneously for any deviations from it. Our results do definitively show that the improvement of the evidence by the addition of a Keplerian model to a base QP-GP model is not a strong proof that a real planetary signal is present in the data. This has particular relevance for targets such as AD~Leo \citep{Kossakowski2022}.

    \subsection{In the presence of stellar activity and Keplerian signals}
    \label{app:modelcomparison_activity+kepler}
    \begin{figure}[!t]
        \centering
        \includegraphics[width=0.35\textwidth]{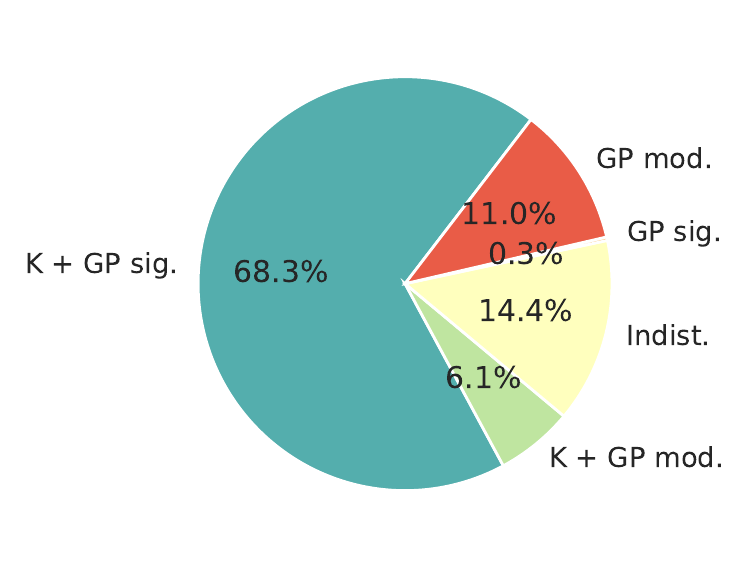}\\
        \caption{{Model comparison on activity data with added Keplerian signals. Shown is the percentage of significant (sig.), moderately favored (mod.), and indistinguishable (indist.) models based on all configurations.}}
        \label{Fig:piechart_addedKep}
    \end{figure}

    \begin{figure}[!t]
        \centering
        \includegraphics[width=\columnwidth]{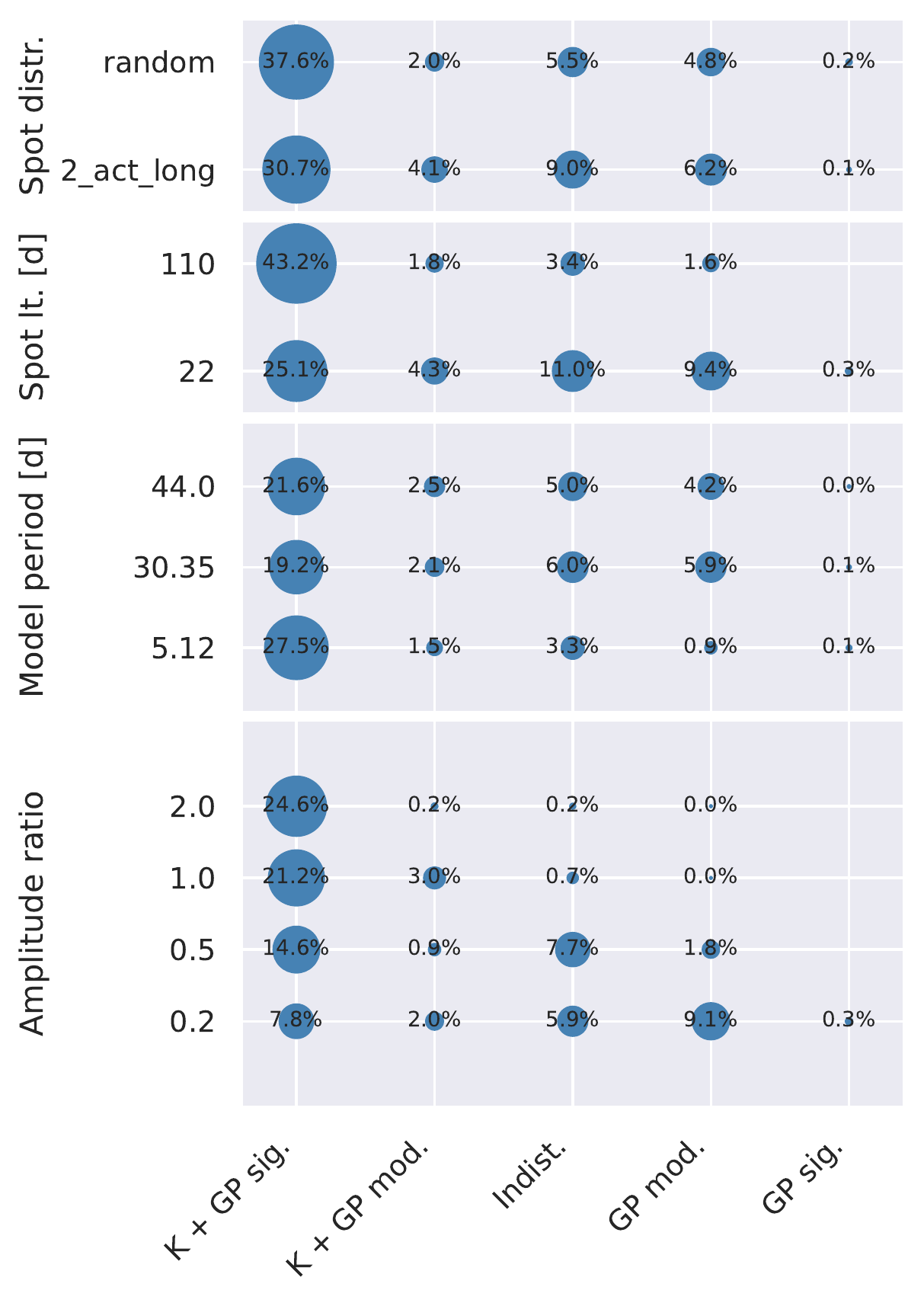}
        \caption{{Model comparison on activity data with added Keplerian signals. Shown is the breakdown of the values from Fig.\,\ref{Fig:piechart_addedKep} into the different activity configurations. The percentages, also indicated by the point sizes, sum up to 100\% for each configuration parameter category highlighted by the gray-shaded areas.}}
        \label{Fig:correlationplot_addedKep}
    \end{figure}

    \begin{figure*}
        \centering
        \includegraphics[width=0.65\textwidth]{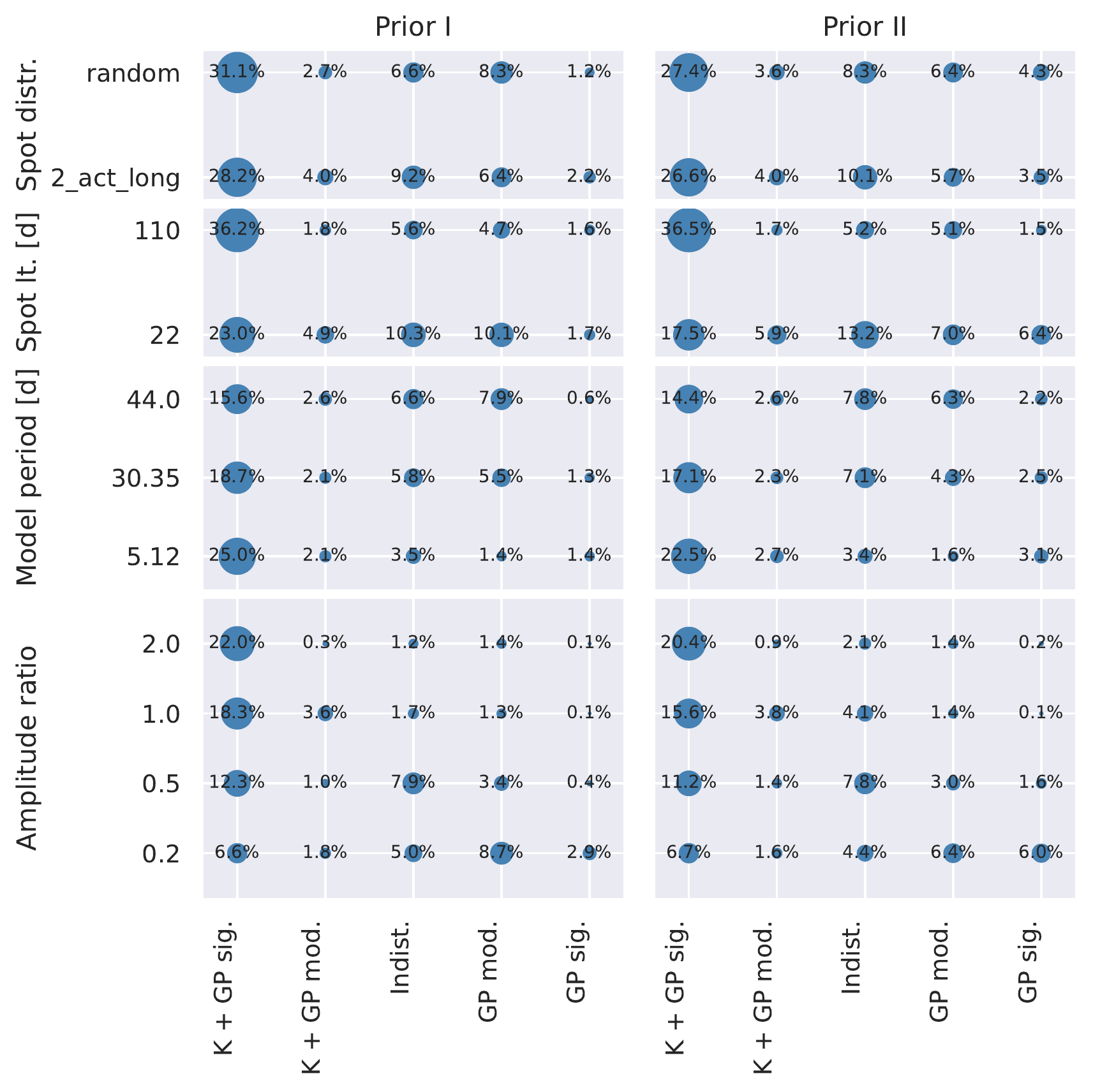}
        \includegraphics[width=0.65\textwidth]{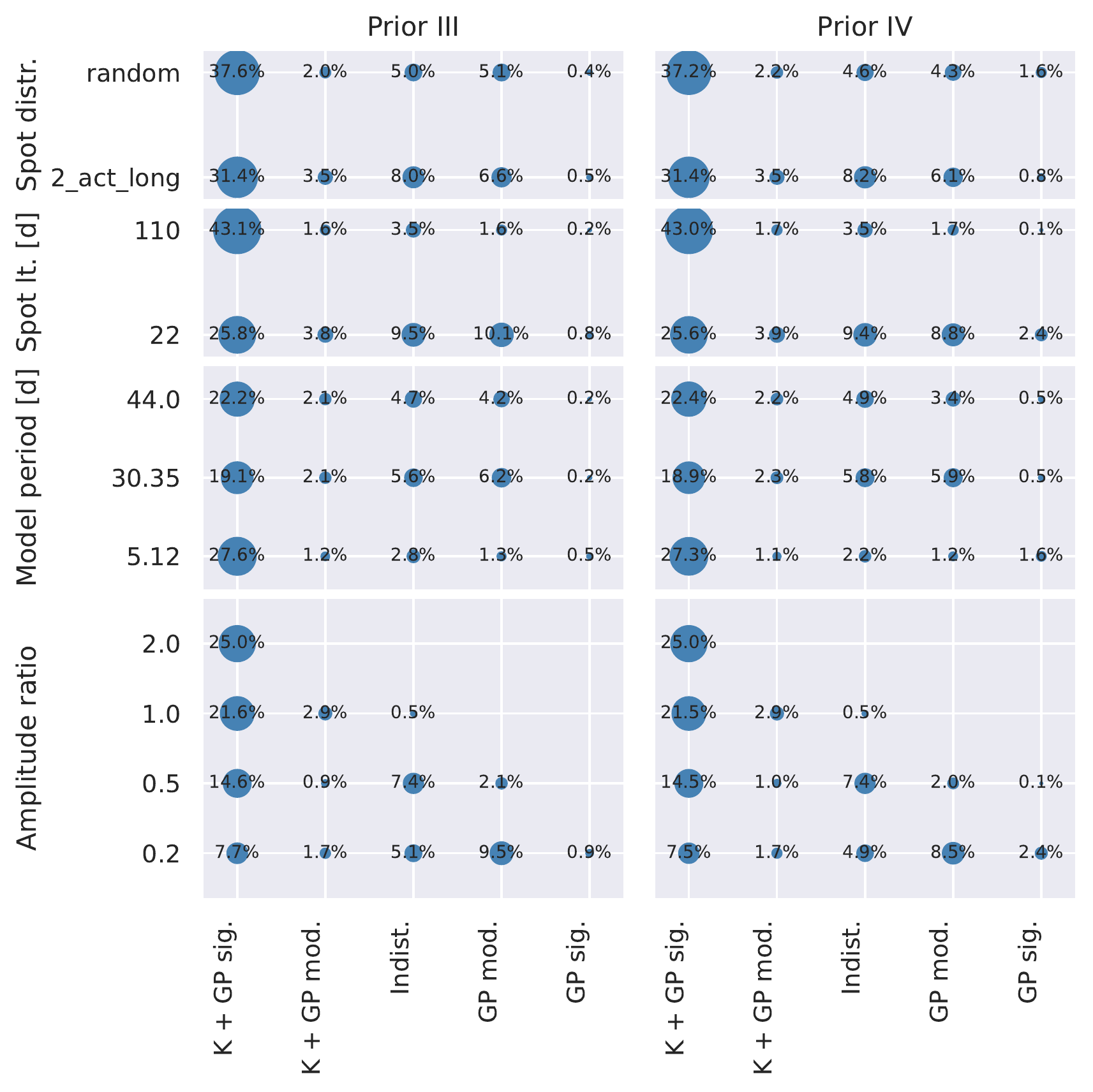}
        \caption{{Model comparison on activity data with added Keplerian signals. Each panel shows the fraction of significant (sig.), moderately favored (mod.), and indistinguishable (indist.) models for one of our GP prior settings. In each plot, the percentages, also indicated by the point sizes, sum up to 100\% for each configuration parameter category highlighted by the gray-shaded areas.}}
        \label{Fig:prior_comparison_addedKep}
    \end{figure*}{}

    Two of the obvious questions are (i) under which conditions Keplerian signals can successfully be recovered from the simulated activity data, and (ii) which assumptions for the QP-GP model provide the {highest detection probability.} To answer them, we explored three kinds of model fits to our grid of activity-plus-Keplerian data. In the first case, we assume GP-only models as the reference for pure activity and thus a non-detection of the Keplerian signal. In addition, we consider Kepler-only and combined Keplerian + GP models based on the GP priors discussed already for the activity only analysis. For the Keplerian components, we assumed a free eccentricity and uniform amplitude between 0 and 40\,ms$^{-1}$, while allowing the period to vary in the range of plus and minus ten percent of the input period. {An overview of the results is provided by the pie chart in Fig.\,\ref{Fig:piechart_addedKep} and the detailed breakdowns into the simulated conditions and used priors in Fig.\,\ref{Fig:correlationplot_addedKep} and Fig.\,\ref{Fig:prior_comparison_addedKep}.}

    First, it is apparent that a simple Keplerian model was never significant compared to the GP-only, or combined Keplerian+GP models, even in the case of an amplitude ratio in favor of the injected planetary signal. We therefore {discuss} here only the direct comparison of GP-only versus combined Keplerian + GP models. In only 68\% of the cases, combined Keplerian + GP models performed significantly better than the pure activity models and thus provided a clear detection of the Keplerian signal. For almost 15\% of the cases the combined models are indistinguishable from the best GP-only models, and in 11\% of the cases a GP-only model even wins.

        {Comparing the results for the different priors in Fig.\,\ref{Fig:prior_comparison_addedKep}, }most of the significant detections of the Keplerian signal are found in combination with the GPs that had the rotation period constrained (GP Prior III), or both, the period and length scale, constrained (GP Prior IV). A sole constraint on the length scale (GP Prior II), however, performed {apparently} worse, also compared to the unconstrained GP Prior I. These results do not differ between the two simulated spot patterns and are in good agreement with our findings for the activity only analysis in Sect.\,\ref{app:modelcomparison_onlyact}.

    {With respect to the different configurations of simulated activity and added Kepler signals, the amplitude of the simulated Kepler signal has, as expected, the largest influence on the proportion of false positives, since a large proportion of the models in which the Kepler signal could not be clearly detected are associated with an amplitude ratio between Kepler and activity of 0.2 or 0.5.} The second-biggest factor seems to be the spot lifetime, since a larger number of non-detections is obtained when the lifetime is 22 days. This is plausible since a shorter lifetime of active regions means a less coherent activity signal and therefore more random dispersion of the RVs. In contrast, the simulated spot pattern does not make a strong difference, even though the random distribution would be expected to also lead to a less coherent activity signal. The simulated period of the Keplerian signal seemed to have no impact on the detection probability.

    \clearpage

    \onecolumn
    {\section{Detailed results of Sect.~\ref{Sect: Results2} }}

    \begin{figure}[!h]
        \centering
        \includegraphics[width=11cm]{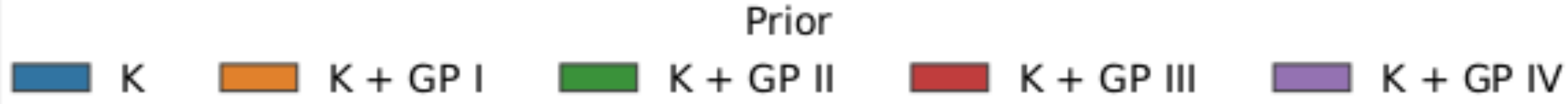}\\
        \includegraphics[width=18cm]{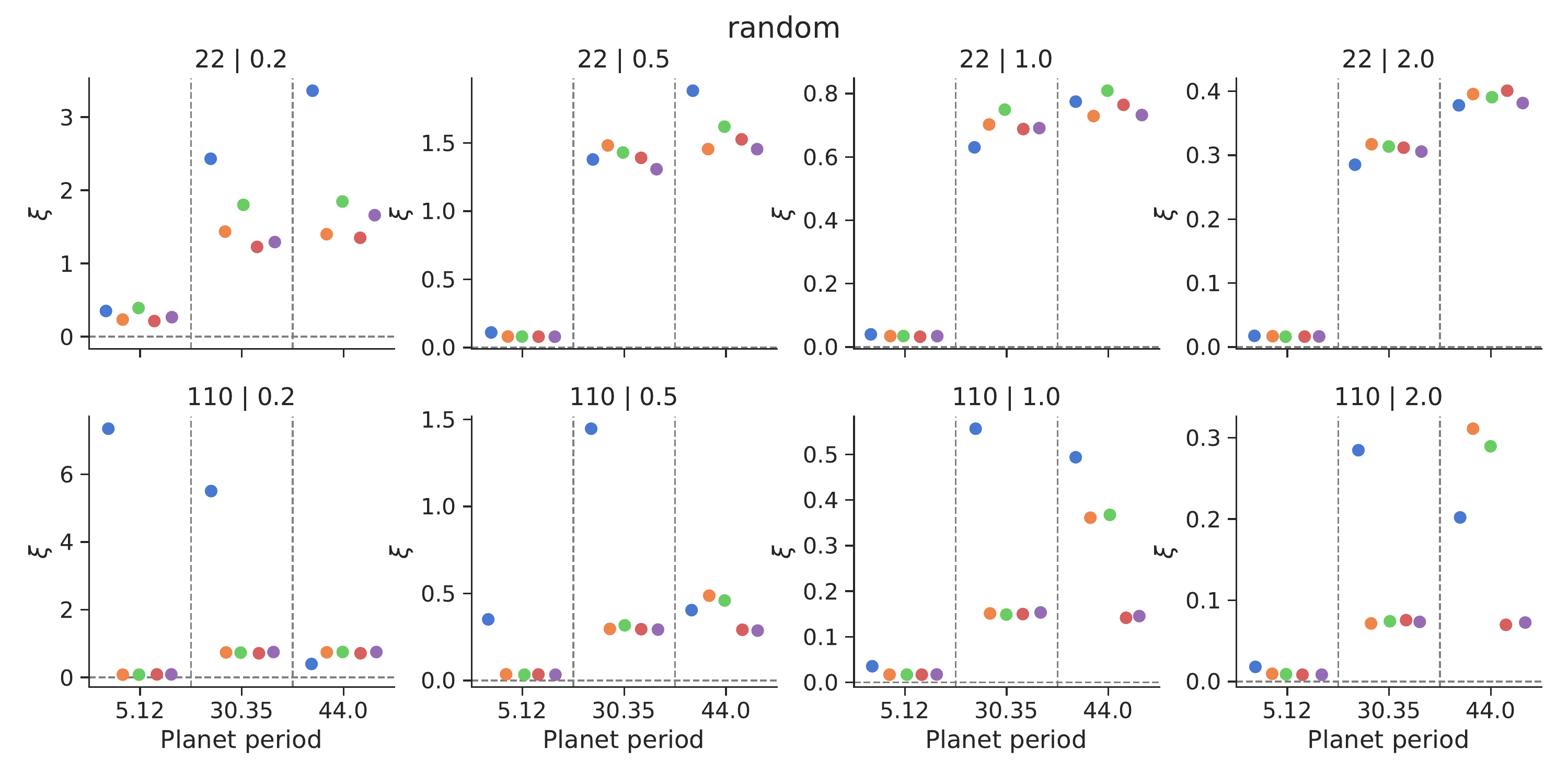}
        \includegraphics[width=18cm]{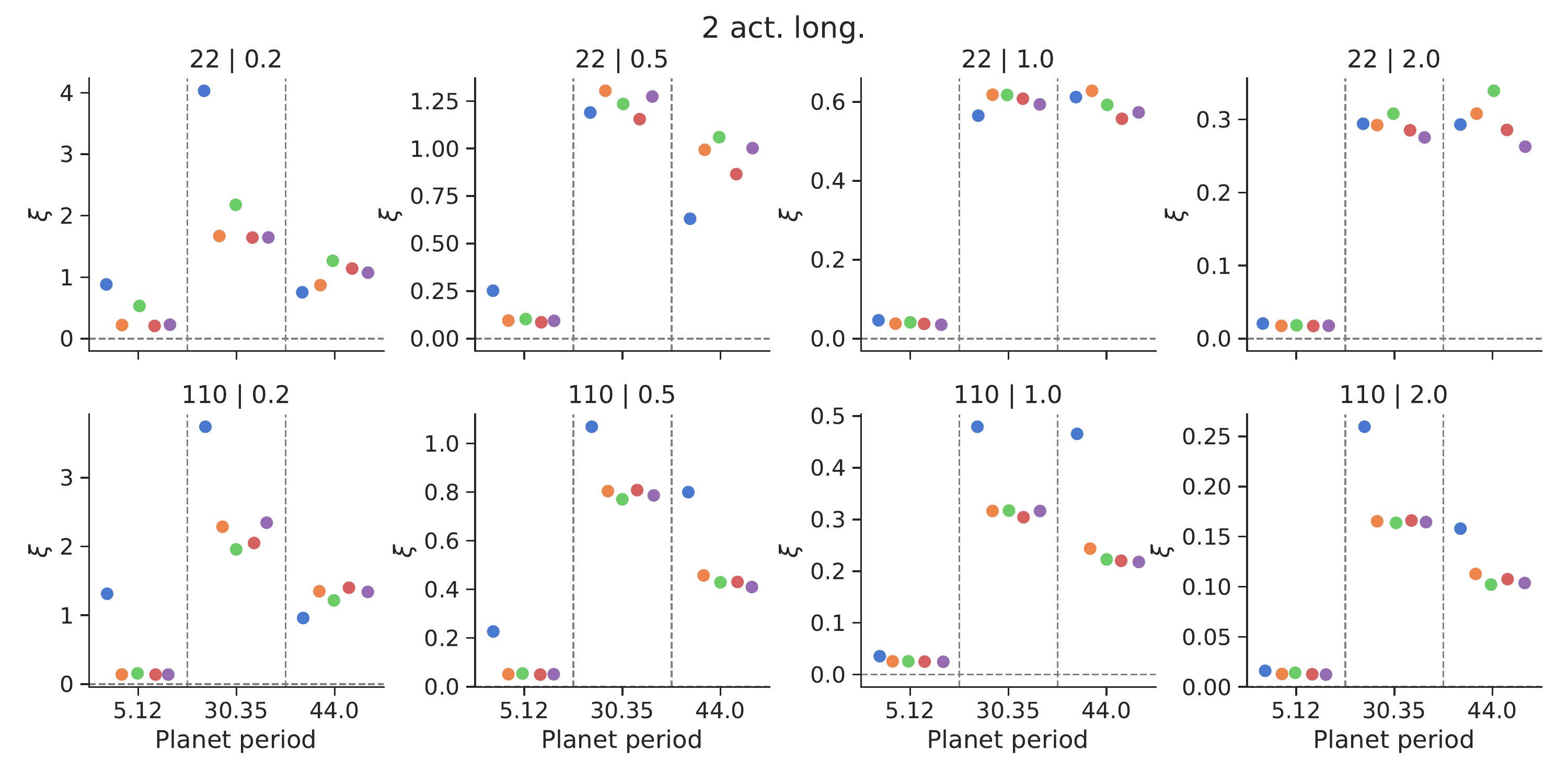}
        \caption{MSA ($\xi$) of the planetary orbital periods based on 100 ensembles, distinguishing between three differently injected orbital periods (separated by the vertical dashed gray lines) and between the five differently applied models (distinguishable by the different color of the dots). The ratio of the injected planet amplitude to the modeled stellar rotational signal increases in favor of the planet signal from the left columns to right columns. The top eight plots show the results for a random spot distribution, and the lower eight plots show the results for two active longitudes on the stellar surface. These two cases are further separated into the top row, which shows the results for a simulated spot lifetime of 22\,d, and the bottom row, which shows the results for a simulated spot lifetime of 110\,d. }
        \label{Fig: msa_periods}
    \end{figure}{}

    \begin{figure*}
        \centering
        \includegraphics[width=11cm]{figures/figures_part2/legend.pdf}\\
        \includegraphics[width=18cm]{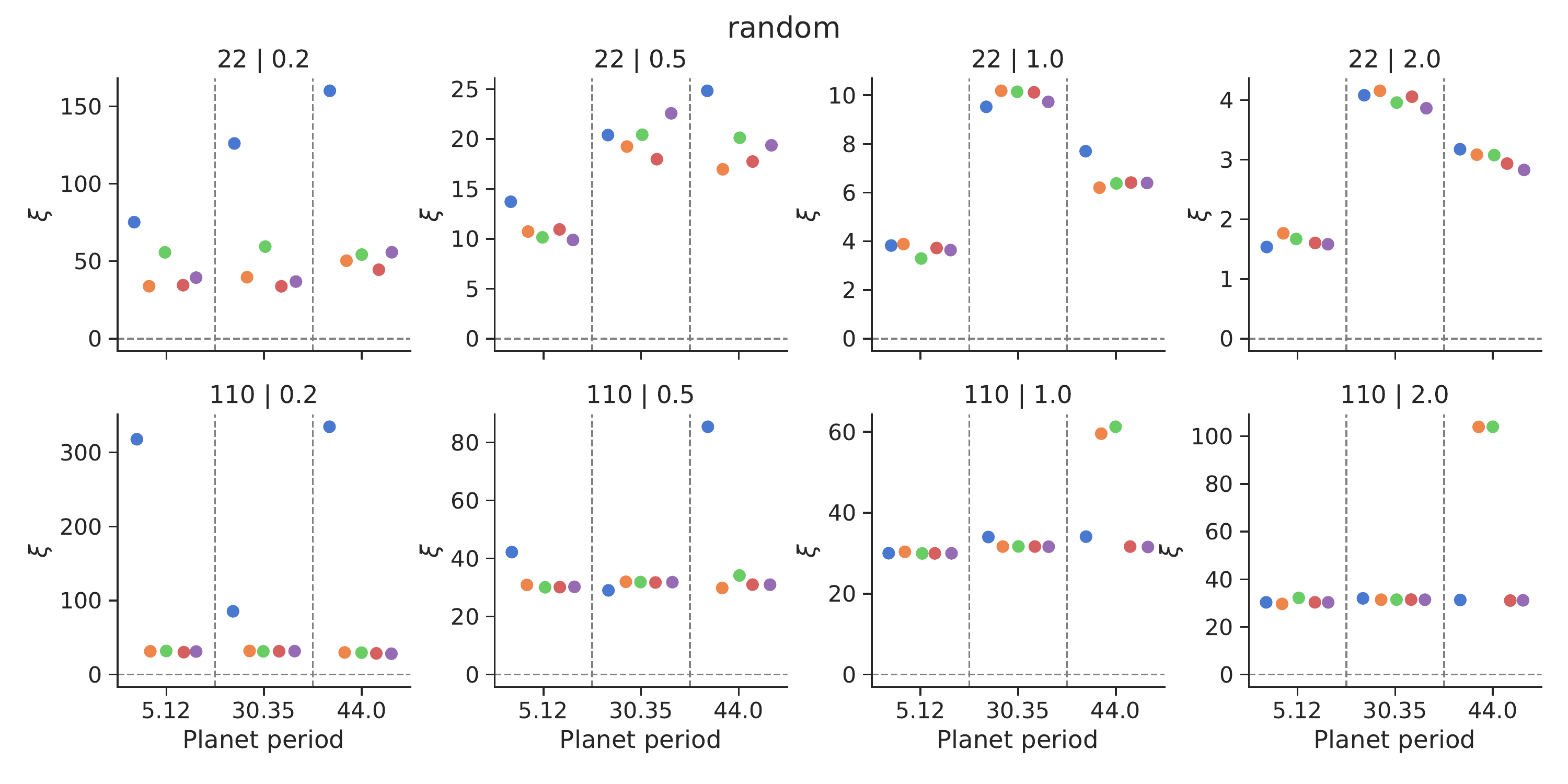}
        \includegraphics[width=18cm]{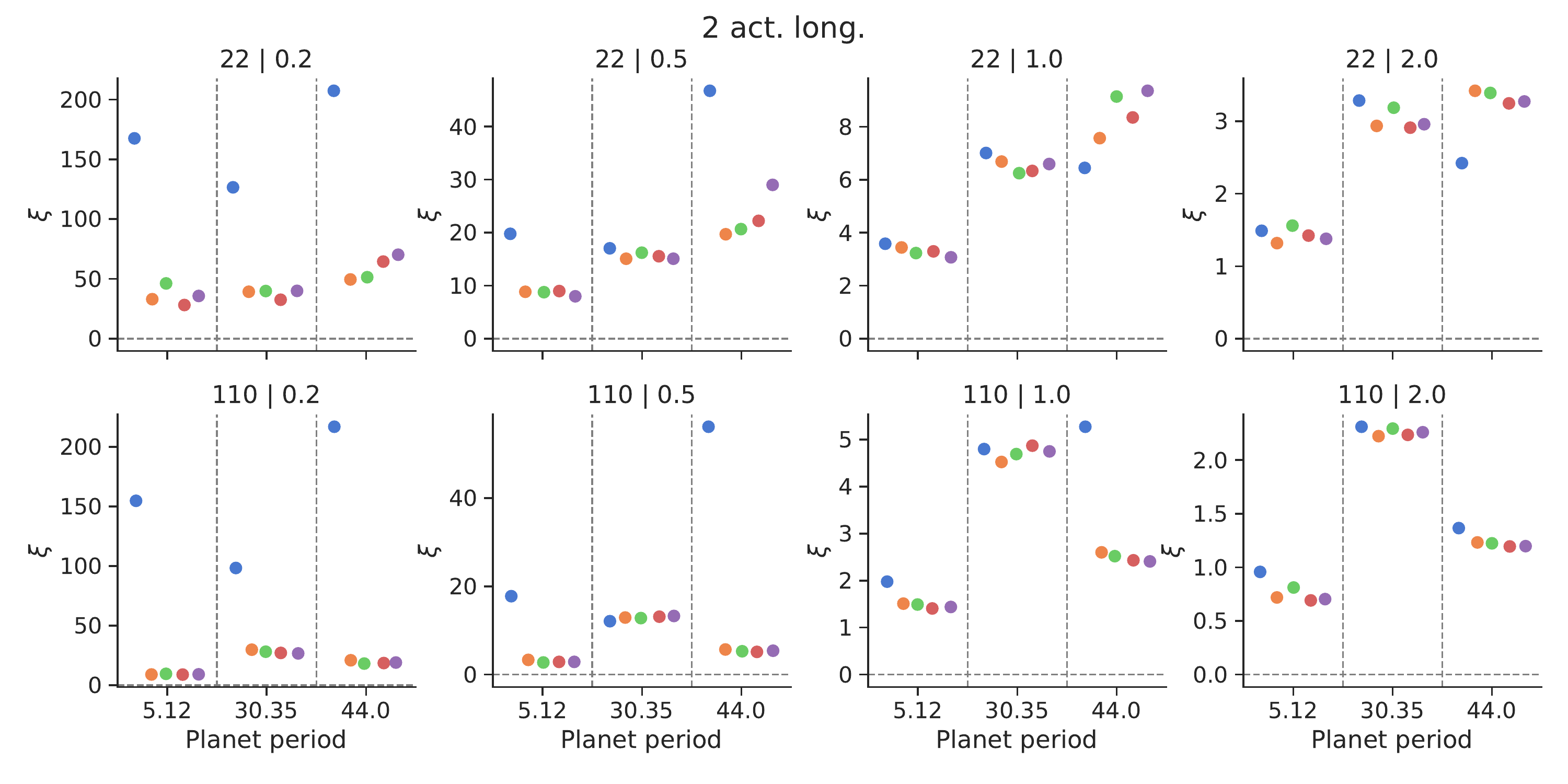}
        \caption{Same as Fig.~\ref{Fig: msa_periods} but for the planetary RV semi-amplitudes.}
        \label{Fig: msa_amplitudes}
    \end{figure*}{}

    \begin{figure*}
        \centering
        \includegraphics[width=11cm]{figures/figures_part2/legend.pdf}\\
        \includegraphics[width=18cm]{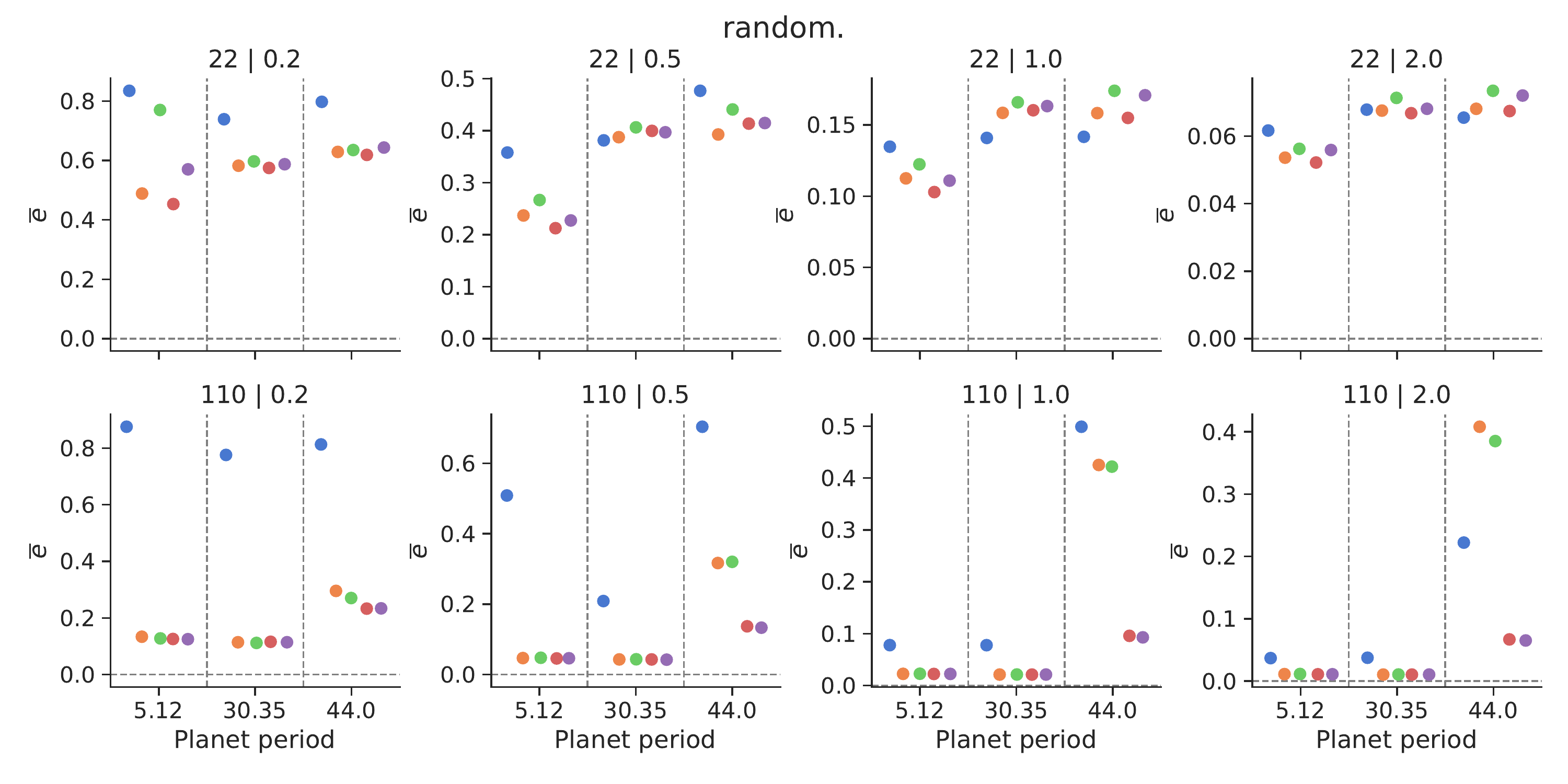}
        \includegraphics[width=18cm]{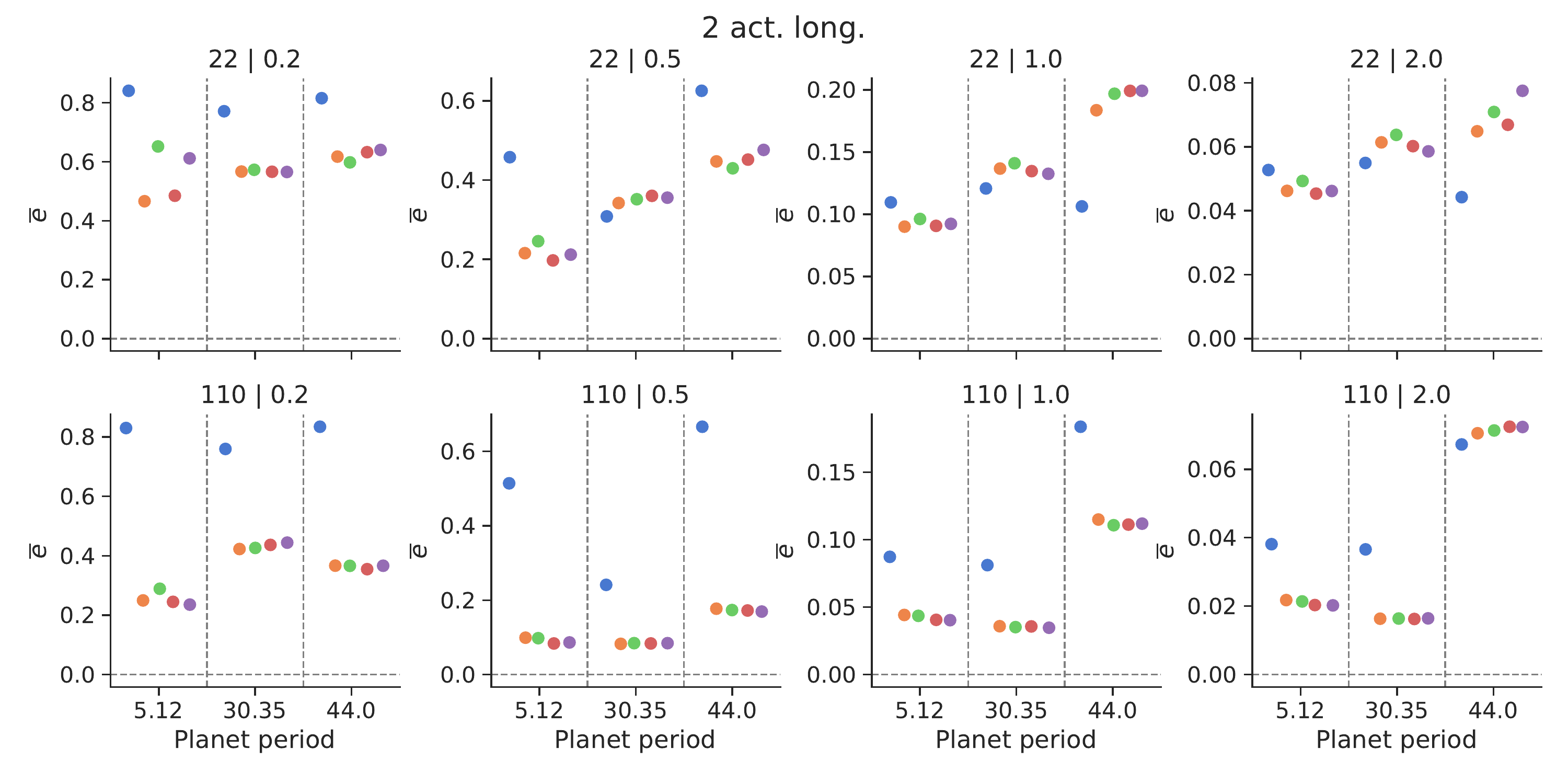}
        \caption{Same as Fig.~\ref{Fig: msa_periods} but instead of the MSA, the median of the derived eccentrics for the 100 ensembles is shown.}
        \label{Fig: med_ecc}
    \end{figure*}{}

    \begin{figure*}
        \centering
        \includegraphics[width=11cm]{figures/figures_part2/legend.pdf}\\
        \includegraphics[width=18cm]{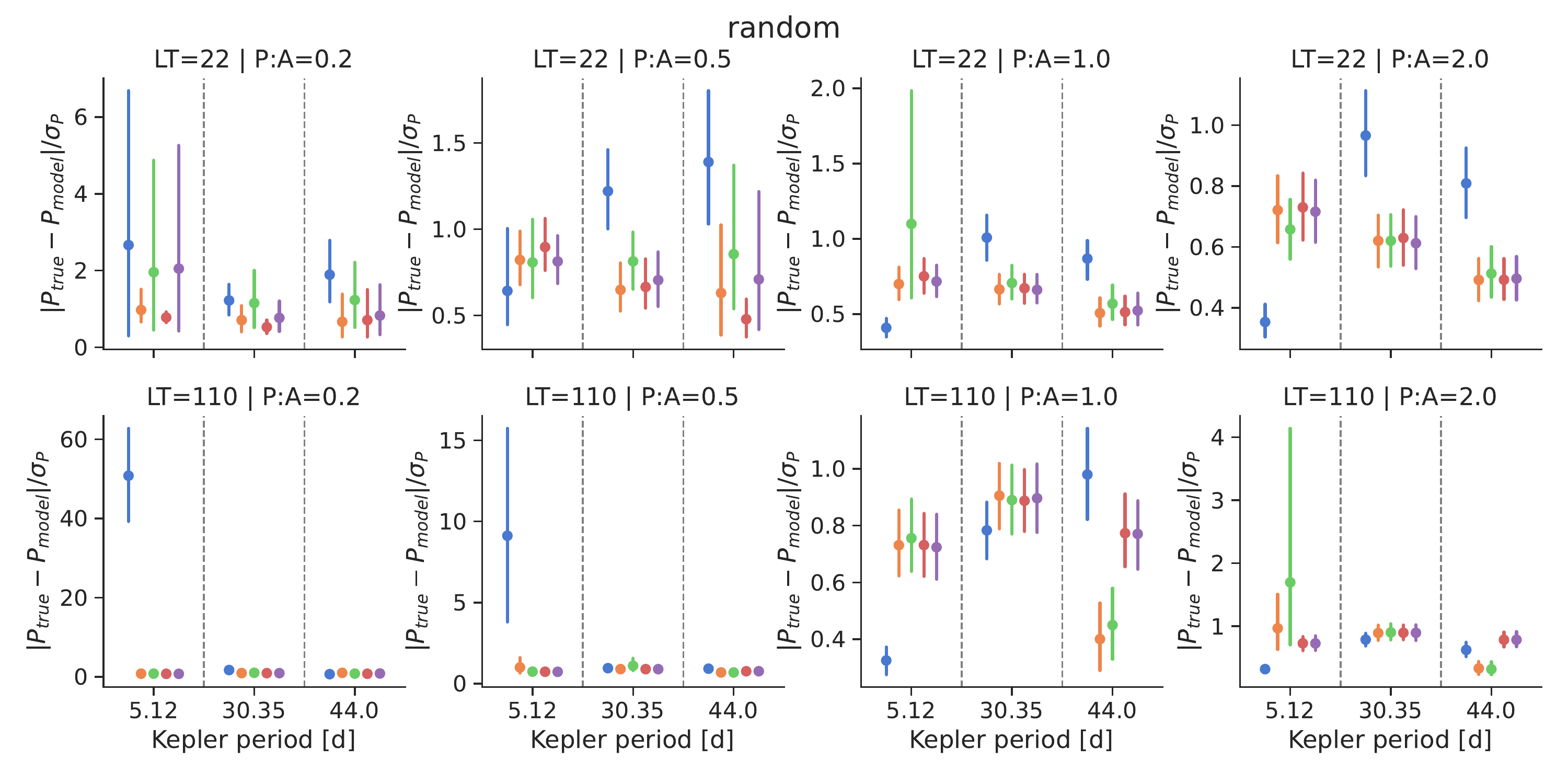}
        \includegraphics[width=18cm]{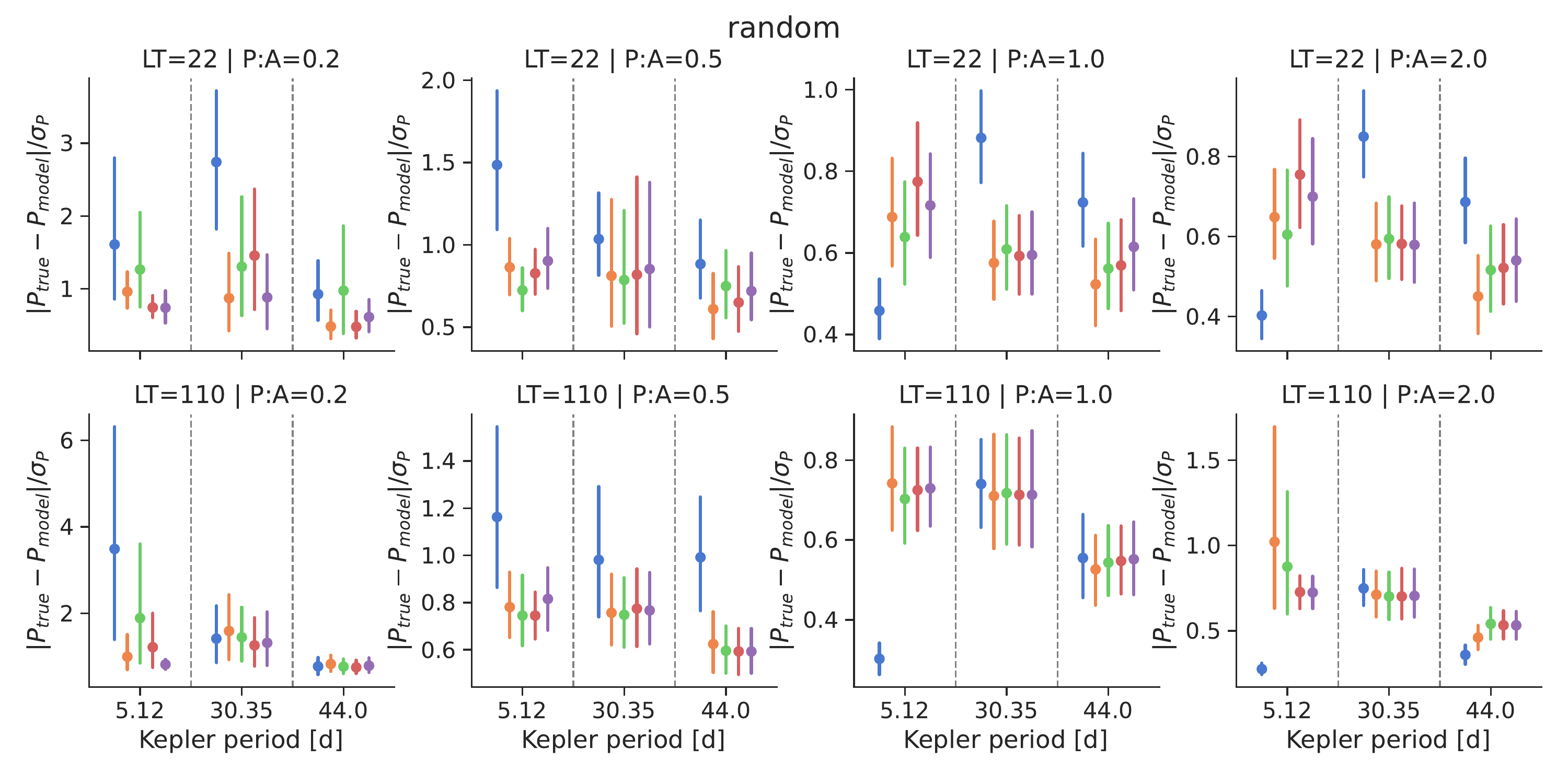}
        \caption{Median of the absolute standard deviation distance between the injected and retrieved planetary orbital periods based on the median posteriors of 100 ensembles each. Each plot distinguishes between three different orbital periods (separated by the vertical dashed gray lines) and between the five differently applied models (distinguishable by the different color of the dots). The error bars show the 0.16 and 0.84 quartiles. The ratio of the injected planet amplitude to the modeled stellar rotational signal increases in favor of the planet signal from the left columns to right columns. The top eight plots show the results for a random spot distribution, and the lower eight plots show the results for two active longitudes on the stellar surface. These two cases are further separated into the top row, which shows the results for a simulated spot lifetime of 22\,d, and the bottom row, which shows the results for a simulated spot lifetime of 110\,d.}
        \label{Fig: sigma_diff_period}
    \end{figure*}{}

    \begin{figure*}
        \centering
        \includegraphics[width=11cm]{figures/figures_part2/legend.pdf}\\
        \includegraphics[width=18cm]{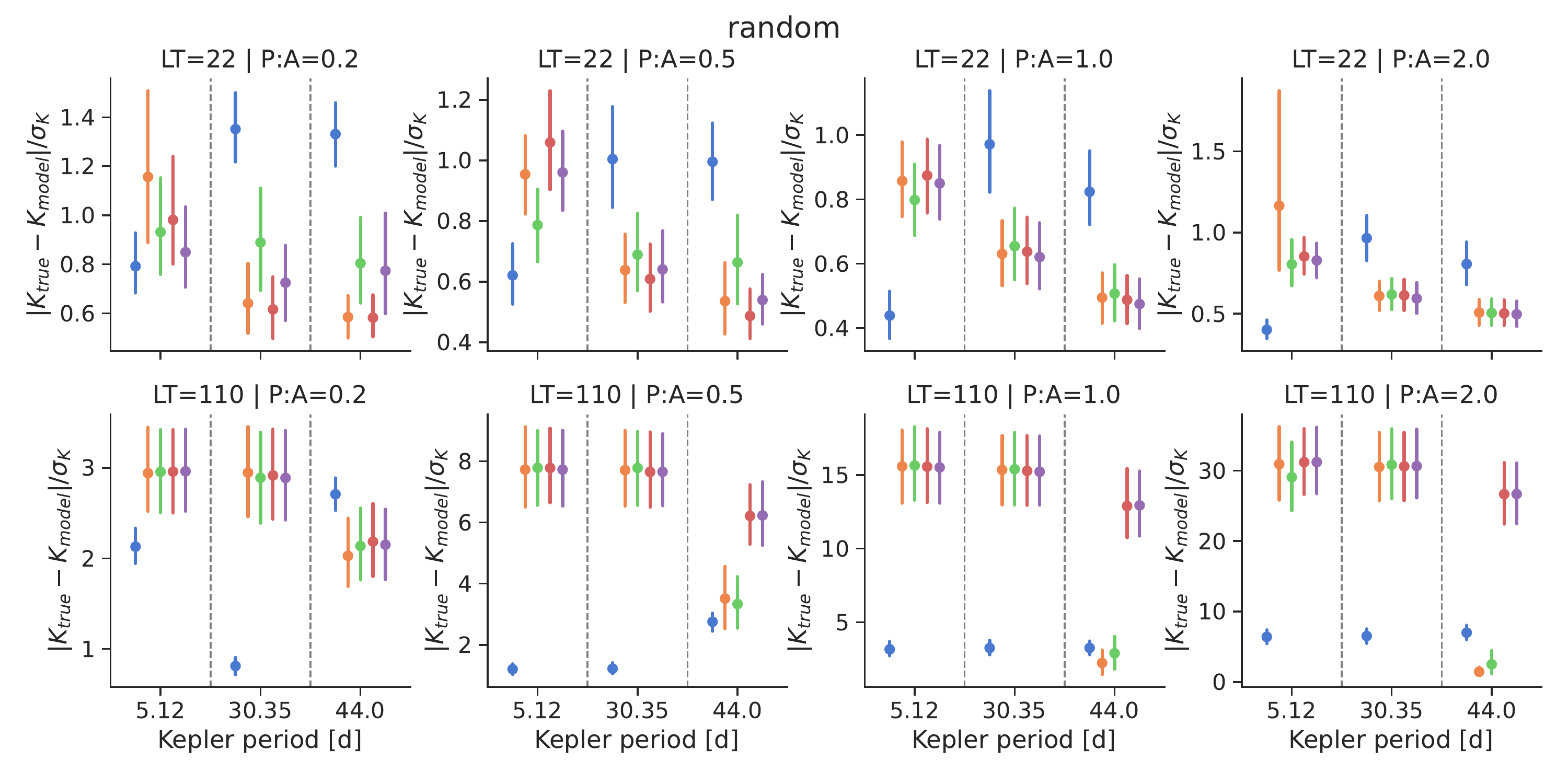}
        \includegraphics[width=18cm]{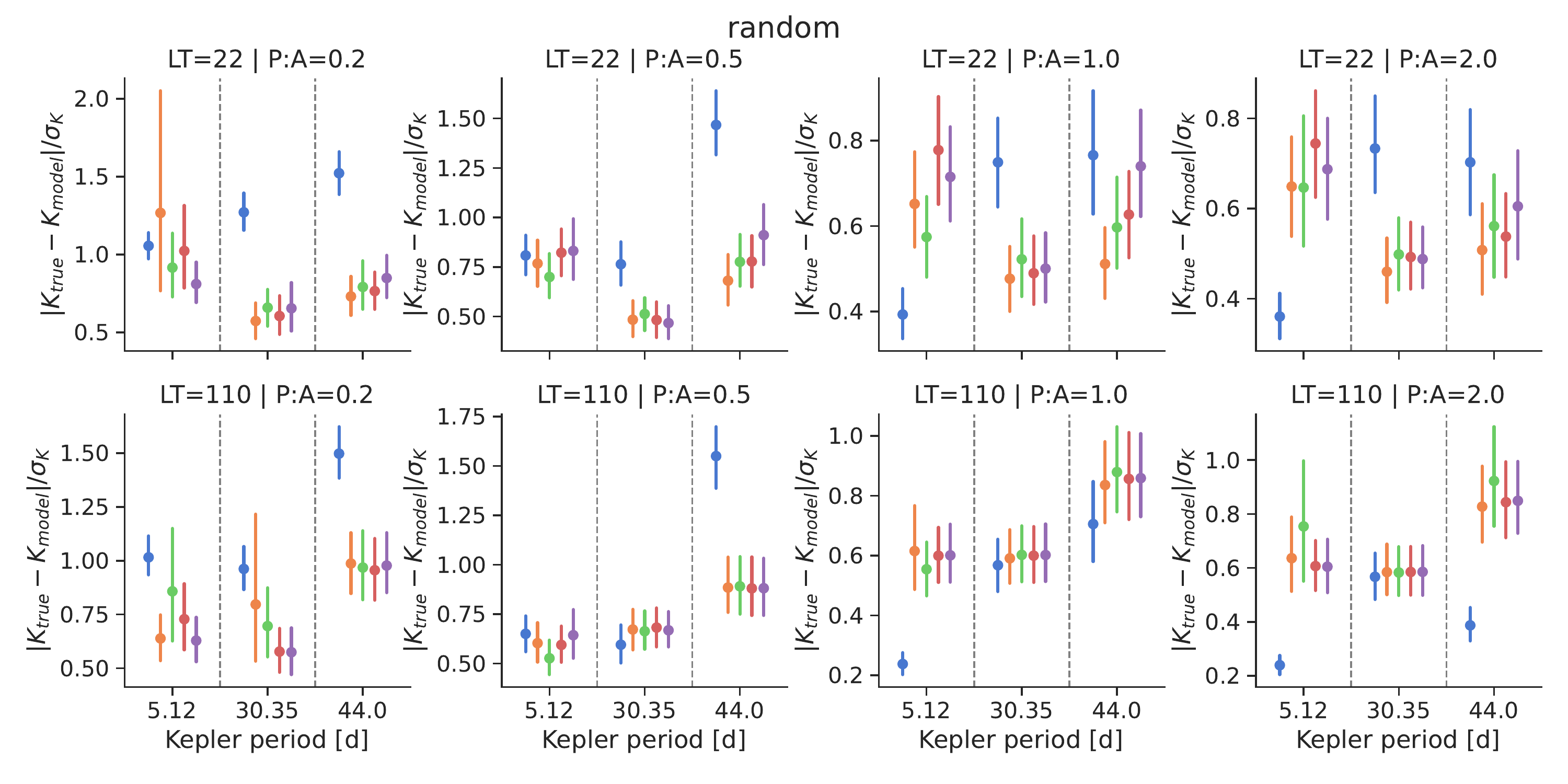}
        \caption{Same as Fig.~\ref{Fig: sigma_diff_period} but for the planetary RV semi-amplitudes.}
        \label{Fig: sigma_diff_amplitude}
    \end{figure*}{}

    \begin{figure*}
        \centering
        \includegraphics[width=11cm]{figures/figures_part2/legend.pdf}\\
        \includegraphics[width=18cm]{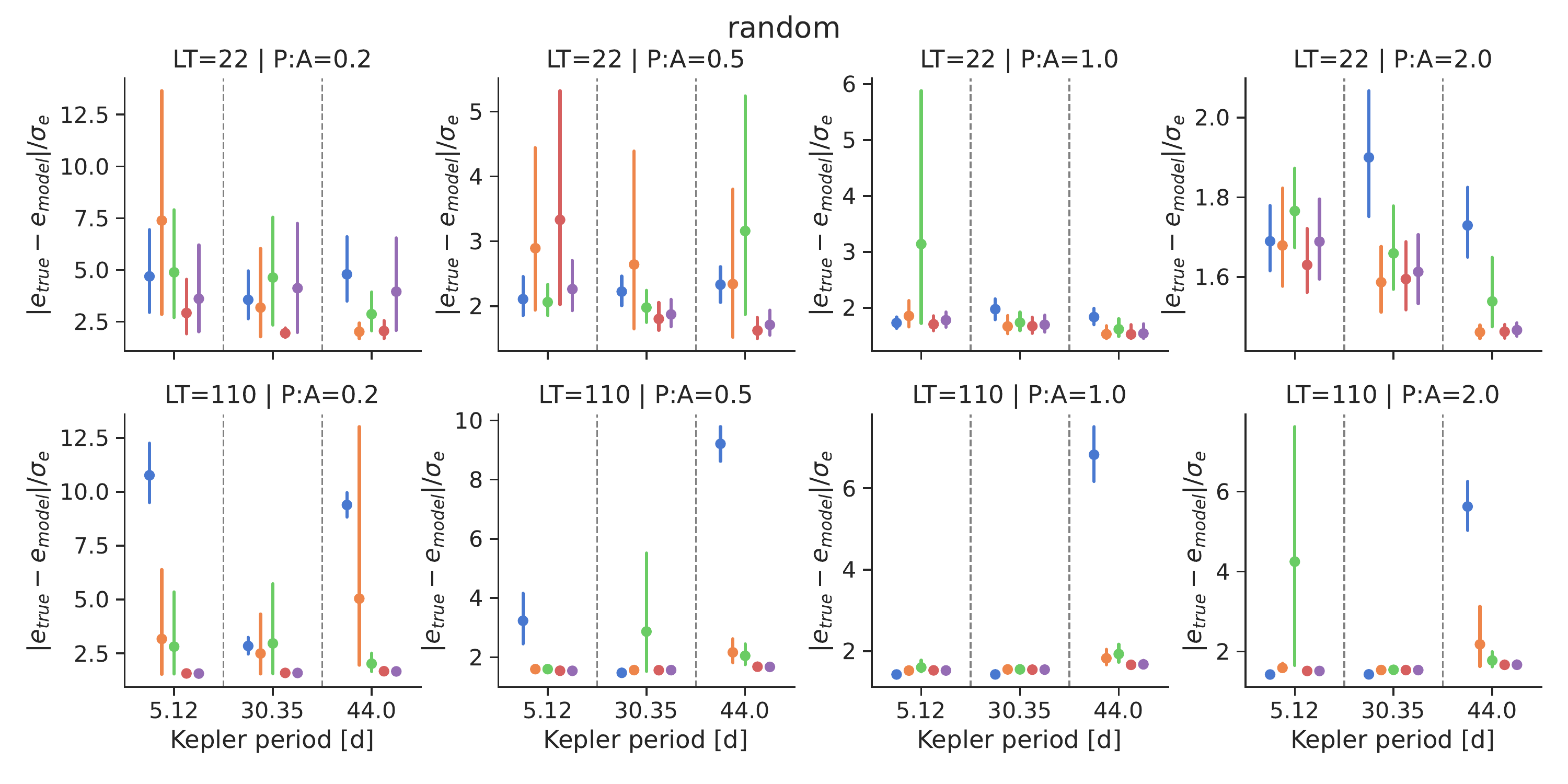}
        \includegraphics[width=18cm]{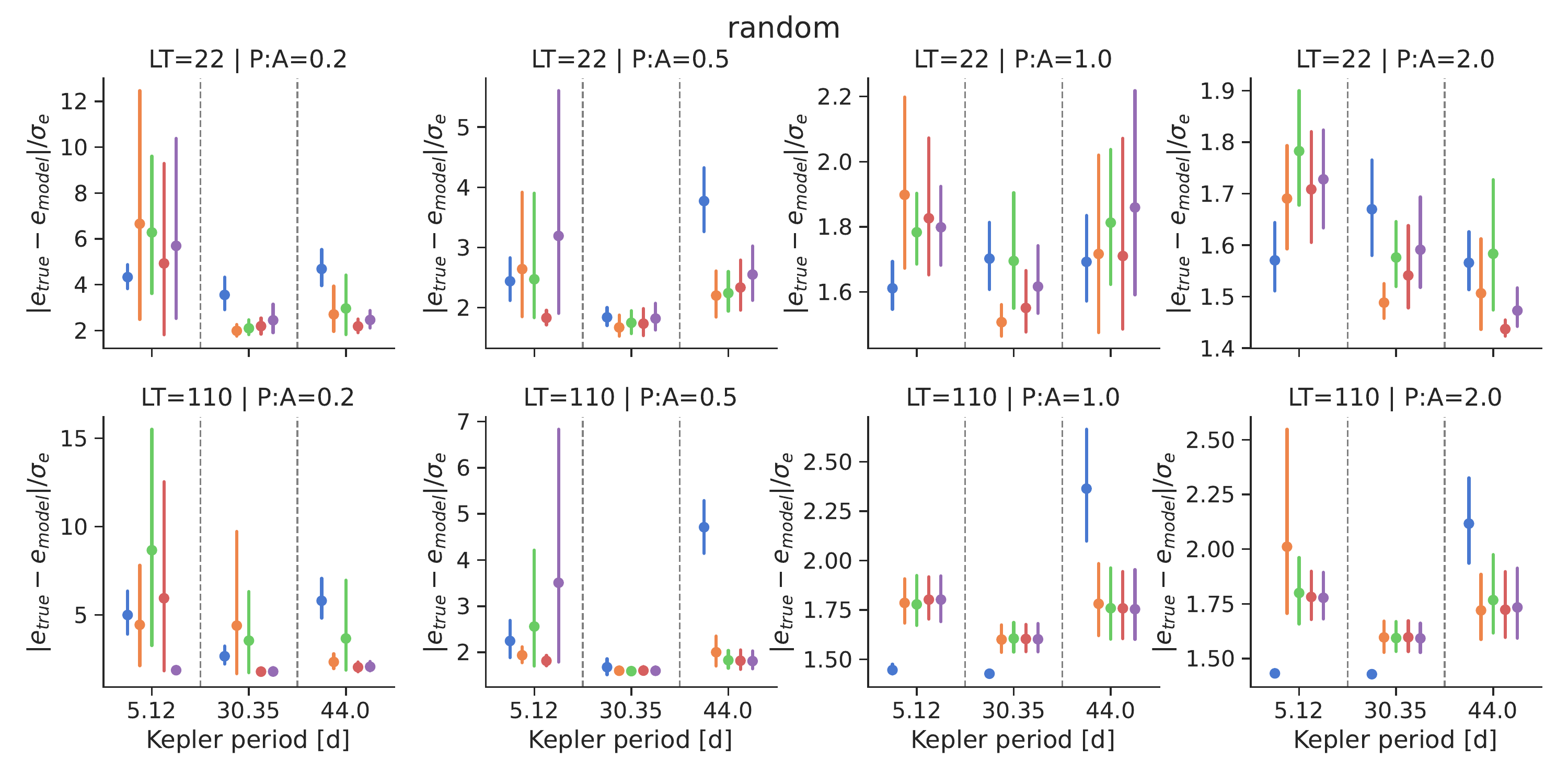}
        \caption{Same as Fig.~\ref{Fig: sigma_diff_period} but for the planetary RV semi-amplitudes.}
        \label{Fig: sigma_diff_ecc}
    \end{figure*}{}

\end{appendix}
\end{document}